\documentclass[article,onecolumn,superscriptaddress,nofootinbib,floatfix,onecolumn]{revtex4}

\usepackage{tabularx}
\usepackage[T1]{fontenc}

\usepackage{dcolumn}
\usepackage{bm}
\usepackage{xspace}
\usepackage{diagbox}
\usepackage{pdflscape}
\usepackage{float}
\usepackage{epsfig}
\usepackage{longtable}
\usepackage{rotating}
\usepackage{ulem}

\usepackage{subfigure}
\usepackage{amssymb}
\usepackage{amsmath}
\usepackage{txfonts}
\usepackage{upgreek}
\usepackage{color}
\usepackage{mathtools}
\usepackage{tabularx}
\usepackage{booktabs}
\usepackage{appendix}
\usepackage{latexsym}
\usepackage{hyperref}
\usepackage{graphics}
\usepackage{multirow}
\usepackage{makecell}

\usepackage{soul} 

\usepackage{comment}

\usepackage[capitalize]{cleveref}
\definecolor{nicered}{rgb}{.7,.1,.1}
\definecolor{nicegreen}{rgb}{.1,.5,.1}
\definecolor{darkblue}{rgb}{0,0,.5}
\hypersetup{colorlinks, citecolor=nicegreen,linkcolor=nicered, urlcolor=darkblue}
\usepackage{multirow}
\usepackage{slashed}
\usepackage{verbatim}

\definecolor{kkbrown}{rgb}{0.82,0.41,0.12}

\def\mean#1{\ensuremath{\left<#1\right>}}
\def\ttt#1{\texttt{\small #1}}

\providecommand{\qqbar}{\rm q\overline{q}}

\providecommand{\twoccbar}{\rm cc\overline{cc}}
\providecommand{\bbbar}{\rm b\overline{b}}
\providecommand{\ttbar}{\rm t\overline{t}}
\providecommand{\lele}{\ell^{+}\ell^{-}}

\newcommand{\pp}{p-p}
\newcommand{\ppbar}{p-$\overline{\mathrm{p}}$}

\providecommand{\pA}{p-A}

\providecommand{\pPb}{p-Pb}
\providecommand{\AaAa}{A-A}
\providecommand{\PbPb}{Pb-Pb}
\providecommand{\AuAu}{Au-Au}

\newcommand{\epem}{\mathrm{e}^+\mathrm{e}^-}
\newcommand{\ellell}{\ell^+\ell^-}
\newcommand{\mumu}{\mu^+\mu^-}
\newcommand{\tautau}{\tau^+\tau^-}

\newcommand{\pleptonium}{(\ell^+\ell^-)_0}
\newcommand{\ppositronium}{(\mathrm{e}^+\mathrm{e}^-)_0}
\newcommand{\pdimuonium}{(\mu^+\mu^-)_0}
\newcommand{\pditauonium}{(\tau^+\tau^-)_0}
\newcommand{\gaga}{\gamma\gamma}

\newcommand{\etacOneS}{\mathrm{\eta_\mathrm{c}(1\mathrm{S})}}
\newcommand{\etacTwoS}{\mathrm{\eta_\mathrm{c}(2\mathrm{S})}}

\newcommand{\chicZero}{\mathrm{\chi_\mathrm{c0}}}
\newcommand{\chicTwo}{\mathrm{\chi_\mathrm{c2}}}
\newcommand{\chicOne}{\mathrm{\chi_\mathrm{c1}}}

\newcommand{\UpsnS}{\Upsilon(\mathrm{nS})}

\newcommand{\etabOneS}{\mathrm{\eta_\mathrm{b}(1\mathrm{S})}}
\newcommand{\etabTwoS}{\mathrm{\eta_\mathrm{b}(2\mathrm{S})}}
\newcommand{\etabnS}{\mathrm{\eta_\mathrm{b}(n\mathrm{S})}}

\newcommand{\chibZero}{\mathrm{\chi_\mathrm{b0}}}
\newcommand{\chibTwo}{\mathrm{\chi_\mathrm{b2}}}

\newcommand{\etatOneS}{\mathrm{\eta_{t}(1\mathrm{S})}}
\newcommand{\etatTwoS}{\mathrm{\eta_{t}(2\mathrm{S})}}
\newcommand{\etatnS}{\mathrm{\eta_{t}(n\mathrm{S})}}

\newcommand{\Apipi}{\mathrm{A}_{2\pi}}
\newcommand{\pipifree}{\pi^+\pi^-}
\newcommand{\AKK}{\mathrm{A}_{2\mathrm{K}}}

\newcommand{\DD}{\mathrm{A}_{2\mathrm{D}}}
\newcommand{\DsDs}{\mathrm{A}_{2\mathrm{D_{s}}}}
\newcommand{\BB}{\mathrm{A}_{2\mathrm{B}}}
\newcommand{\BcBc}{\mathrm{A}_{2\mathrm{B_{c}}}}

\newcommand{\baryonium}{(\mathrm{h\overline{h}})_0}
\newcommand{\protonium}{(\mathrm{p\overline{p}})_0}
\newcommand{\Sigmaonium}{(\mathrm{\Sigma^+\Sigma^-)}_0}

\newcommand{\overbar}[1]{\mkern 1.5mu\overline{\mkern-1.5mu#1\mkern-1.5mu}\mkern 1.5mu}

\newcommand{\Xionium}{(\mathrm{\Xi^-\overbar{\Xi}^+})_0}
\newcommand{\Omegaonium}{(\mathrm{\Omega^-\overbar{\Omega}^+})_0}
\newcommand{\Lambdaconium}{(\mathrm{\Lambda_\mathrm{c}^+\overbar{\Lambda}_\mathrm{c}^-})_0}
\newcommand{\Xiconium}{(\mathrm{\Xi_\mathrm{c}^+\overbar{\Xi}_\mathrm{c}^-})_0}

\newcommand{\Xibonium}{(\mathrm{\Xi_\mathrm{b}^-\overbar{\Xi}_\mathrm{b}^+})_0}

\newcommand{\Omegabonium}{(\mathrm{\Omega_\mathrm{b}^-\overbar{\Omega}_\mathrm{b}^+})_0}

\newcommand{\jpsi}{\mathrm{J}/\psi}

\newcommand{\shat}{\hat{s}_{\gamma\gamma}}
\newcommand{\sqrtshat}{\!\sqrt{\hat{s}_{\gamma\gamma}}}
\newcommand{\sqrts}{\!\sqrt{s}}
\newcommand{\sgg}{s_{_{\gamma\,\gamma}}}

\newcommand{\sqrtsnn}{\!\sqrt{s_{_\text{NN}}}}
\newcommand{\snn}{s_{_\text{NN}}}
\newcommand{\alphas}{\alpha_\mathrm{s}}
\newcommand{\pT}{p_\text{T}}

\newcommand{\BR}{\mathcal{B}}
\newcommand{\Lumi}{\mathcal{L}}

\newcommand{\LumiInt}{\mathcal{L}_{\mathrm{int}}}

\newcommand{\gammaUPC}{\ttt{gamma-UPC}}

\usepackage{xspace}
\newcommand*{\eg}{e.g.,\@\xspace}
\newcommand*{\ie}{i.e.,\@\xspace} 
\newcommand*{\cm}{c.m.\@\xspace}

\begin{document}

\title{Exclusive photon-fusion production of even-spin resonances and exotic QED atoms\\ in high-energy hadron collisions}

\author{David~d'Enterria}\email{david.d'enterria@cern.ch}
\affiliation{CERN, EP Department, CH-1211 Geneva, Switzerland}
\author{Karen Kang}\email{kkang25@amherst.edu}
\affiliation{Amherst College, Amherst, MA 01002, USA}

\begin{abstract}
\noindent 
The cross sections for the single exclusive production of (pseudo)scalar and (pseudo)tensor hadrons, as well as of even-spin QED bound states formed by pairs of opposite-charge leptons or hadrons, are estimated for photon-fusion processes in ultraperipheral collisions (UPCs) of proton-proton, proton-nucleus, and nucleus-nucleus at the RHIC, LHC and FCC colliders, as well as in proton-air interactions at the highest energies reached by cosmic-rays impinging on Earth. The UPC cross sections are computed in the equivalent photon approximation with realistic photon fluxes from the charged form factors of proton, lead, gold, and nitrogen ions. The production of four types of even-spin systems are considered: quarkonium (spin-0,\,2,\,4 meson bound states, from the lightest $\pi^0$ meson up to toponium), exotic hadrons (including candidate multiquark states), leptonium (positronium, dimuonium, and ditauonium), as well as mesonium (pionium, kaonium, D-onium, and B-onium) and baryonium (notably, protonium) QED atoms. The expected yields at the different colliders are presented for about 50 such even-spin composite resonances, for which the ALICE and LHCb experiments have potential reconstruction capabilities at the LHC. The impact of the diphoton decays of such even-spin states is also discussed as resonant backgrounds in the measurement of light-by-light scattering ($\gaga\to\gaga$) over $m_{\gaga} \approx 0.1$--15~GeV masses in \PbPb\ UPCs at the LHC.
\end{abstract}

\maketitle

\vspace{-0.4cm}
\tableofcontents

\section{Introduction}

The electric field created by a charged particle accelerated to high energies can be interpreted, in the Weizs\"{a}cker--Williams (WW) equivalent photon approximation (EPA)~\cite{vonWeizsacker:1934nji,Williams:1934ad}, as a flux of quasireal photons whose energies $E_{\gamma}$ and number densities $N_{\gamma}$ grow proportionally to the Lorentz relativistic factor ($E_{\gamma}\propto \gamma_L$) and squared charge ($N_{\gamma}\propto Z^2$) of the beam particles~\cite{Brodsky:1971ud,Budnev:1975poe}. Such quasireal photon beams have been exploited for decades to study high-energy photon-photon ($\gaga$) interactions at particle colliders~\cite{Morgan:1994ip,Whalley:2001mk,Bertulani:2005ru,Baltz:2007kq,deFavereaudeJeneret:2009db}. Research on $\gaga$ interactions at multi-GeV energies was first realized in the laboratory in $\epem$ collisions at DESY PETRA in the 1980s~\cite{Morgan:1994ip} and at CERN LEP in the 1990s~\cite{Whalley:2001mk}, and has received a significant experimental and theoretical boost at hadron colliders in the last twenty years thanks to the large EPA $\gamma$ energies and luminosities accessible at the BNL Relativistic Heavy-Ion Collider (RHIC)~\cite{Bertulani:2005ru} and at the CERN Large Hadron Collider (LHC)~\cite{Baltz:2007kq,deFavereaudeJeneret:2009db}. At hadron colliders, photon-photon processes can be studied in particularly clean conditions in the so-called ultraperipheral collisions (UPCs), where the colliding hadrons interact with transverse separations larger than their matter radii, \ie\ without hadronic overlap, and thereby survive their purely electromagnetic interaction. Such UPCs provide the means to study the exclusive production of a single neutral object, or a pair of opposite-charge objects, at central rapidities in an otherwise empty detector~\cite{Bertulani:1987tz}. 

The quantity of interest in a $\gaga$ collision of charges A and B is the effective two-photon luminosity, $\mathcal{L}_{\gaga}^{(\mathrm{AB})}$, obtained from the integral of the EPA photon fluxes of the colliding charges. By denoting as $f(x)\mathrm{d}x$ the number of photons carrying a fraction between $x$ and $x+\mathrm{d}x$ of the energy of the charge $Z$, \ie\ $x=E_\gamma/E_\text{beam}$, the two-photon luminosity as a function of the fractional center-of-mass (\cm) energy squared $\tau = \shat/s$ (where $s$ and $\shat$ are the squared \cm\ energy of the colliding hadronic and $\gaga$ system, respectively) can be written\footnote{Natural units, $\hbar=c=1$, are used throughout the paper.} as~\cite{Baur:1988tu,Cahn:1990jk},
\begin{equation}
\frac{\mathrm{d} \mathcal{L}_{\gaga}^{(\mathrm{AB})}}{\mathrm{d} \tau} =  \int_{\tau}^{1} \mathrm{d}x_1  \mathrm{d}x_2 f(x_1)f(x_2)\delta(\tau-x_1x_2) = \int_{\tau}^{1} \frac{\mathrm{d}x}{x}f(x)f(\tau/x),
\label{eq:L_gaga}
\end{equation}
where the last equality assumes 
that the colliding charges are identical (and of opposite momentum) and that their $\gamma$ fluxes factorize\footnote{Simple factorization of photon fluxes is not fully realistic for quantitatively accurate estimates of $\gaga$ cross sections in UPCs, 
as discussed below, but it is a good approximation for illustrative purposes here.} as a function of $x$. The inclusive photon-photon cross section for any final state X in a  given $\mathrm{A} \mathrm{B} \xrightarrow{\gaga} \mathrm{A~X~B}$ collision can then be obtained from the corresponding elementary $\gaga$ cross section, $\hat{\sigma}_{\gaga \to\mathrm{X}}$, via
\begin{equation}
\sigma(\mathrm{A} \mathrm{B} \xrightarrow{\gaga} \mathrm{A~X~B}) = \int \mathrm{d}\tau \frac{\mathrm{d} \mathcal{L}_{\gaga}^{(\mathrm{AB})}}{\mathrm{d} \tau} \hat{\sigma}_{\gaga \to\mathrm{X}}(\shat).
\label{ref:sigma_gaga_UPC}
\end{equation}
Once the EPA flux of the colliding charges $f(x)$ is known, one can compute any arbitrary $\gaga$ cross section in high-energy collisions. If the $\gamma$ source is an electron (with mass $m_\mathrm{e}$), the EPA flux depends on the photon virtuality $Q^2$, and reads
\begin{equation}
f_{\gamma/\mathrm{e}}(x) = \frac{Z^2\alpha}{\pi x}\int_{x^2m_\mathrm{e}^2}^{\infty} \frac{\mathrm{d}Q^2}{Q^2} \approx \frac{\alpha}{\pi x}\ln(s/m_\mathrm{e}^2),
\end{equation}
where $\alpha$ is the fine structure constant, and the last approximation, obtained by setting $Z=1$ and the upper limit of integration to $s$, agrees with the usual WW form: $f(x) = \frac{\alpha}{\pi x}\ln(s/(4m_\mathrm{e}^2))\frac{1}{2}[1+(1-x)^2]$.
The EPA photon flux for a hadronic beam A, with nucleon (proton) mass $m_\mathrm{N}=0.9315~(0.9383)$~GeV, has much smaller virtualities, constrained by the form factor of the matter distribution $F_\mathrm{A}(Q^2)$, and reads
\begin{equation}
f_{\gamma/\mathrm{A}}(x) = \frac{Z^2\alpha}{\pi x}\int_{x^2m_\mathrm{N}^2}^{\infty} \frac{\mathrm{d}Q^2}{Q^2} F_\mathrm{A}(Q^2)^2 \left(1-\frac{x^2m_\mathrm{N}^2}{Q^2}\right) \approx \frac{Z^2\alpha}{\pi x}\ln\left(\frac{Q_0^2}{(xm_\mathrm{N})^2}\right),
\label{eq:A_photonflux}
\end{equation}
with $\alpha = 1/137.036$, and where the last approximation, valid for not too large $x$ values, takes a maximum virtuality $Q_0\approx 1/R_\mathrm{A}$ given by the inverse of the transverse radius of the hadron $R_\mathrm{A}$. 
Indeed, the coherent photon emission from the full hadron charge distribution forces the photons to be (quasi) real, \ie\ (almost) on-mass shell, limiting their virtuality to very low values $Q^{2} < 1/R_\mathrm{A}^{2}$, namely $Q_0^2\approx 0.08$~GeV$^2$ for protons (with $R_\mathrm{A}\approx 0.7$~fm), and $Q_0^2\approx 4\cdot10^{-3}$~GeV$^2$ for nuclei (with $R_\mathrm{A}\approx 1.2\,A^{1/3}$~fm, for mass number $A>14$). The $f_{\gamma/\mathrm{A}}(x)\propto 1/x$ behavior of Eq.~(\ref{eq:A_photonflux}) shows that the longitudinal photon energies have a typical $E_{\gamma}^{-1}$ bremsstrahlung-like spectrum, up to energies of the order of $E_{\gamma}^\text{max}\approx\gamma_\mathrm{L}/R_\mathrm{A}$ (beyond which the photon flux is not zero, but decreases much more steeply) where $\gamma_\mathrm{L}=\sqrtsnn/2m_\mathrm{N}$ is the Lorentz gamma factor of the beam. Plugging the photon flux (\ref{eq:A_photonflux}) into Eq.~(\ref{eq:L_gaga}) and integrating over photon fractional energies, one obtains a simple approximate parametrization for the effective two-photon luminosity in hadronic UPCs as a function of fractional $\gaga$ \cm\ energy~\cite{Baur:1988tu,Cahn:1990jk},
\begin{equation}
\frac{\mathrm{d} \mathcal{L}_{\gaga}^{(\mathrm{AB})}}{\mathrm{d} \tau} \approx \left(\frac{Z^2\alpha}{\pi}\right)^2 \frac{16}{3\tau} \ln^3\left(\frac{\gamma_\mathrm{L}}{ m_\mathrm{N} R_\mathrm{A}}\right), 
\label{eq:L_gaga_UPC}
\end{equation}
which illustratively provides intuitive parametric dependencies of the \cm-fractional photon-photon luminosities in UPCs: they scale as $Z^4$ and as $\ln^3(\!\sqrtsnn)$. The fourth power on the charge $Z$ enhances the $\gaga$ cross sections by a factor of about $50\cdot 10^6$ in \PbPb\ compared to $\epem$ or \pp\ collisions, although larger $\sqrt{\sgg}$ can be reached with charges with smaller radii given the $E_{\gamma}^\text{max}\propto1/R_\mathrm{A}$ dependence. Since the $\gaga$ luminosity increases as the cube of the logarithm of the beam energy, 
UPCs are anticipated to play an even bigger role at the upcoming Future Circular Collider (FCC), 
with \cm\ energies about one order of magnitude larger than at the LHC, $\sqrtsnn= 39$--100~TeV~\cite{FCC:2018vvp}, and even more in proton-air (p-air, mostly p-nitrogen) collisions at the maximum energies observed in interactions of primary cosmic-ray protons with air nuclei in the upper atmosphere\footnote{Atmospheric nuclei are not fully stripped of their electrons, at variance with nuclei at colliders, and therefore their full charge (and associated ``target'' $\gamma$ flux) is not visible to the ``projectile'' $\gamma$ flux unless the p-air interaction happens at small impact parameters ($b$) below the first electron shell, \ie\ smaller than the Bohr radius, $b<r_\text{Bohr}\approx 53\cdot10^3\mbox{ fm}/Z\approx 3800$~fm for nitrogen ($Z=14$), which is the case for most of the systems produced in the $\gaga$ collisions considered~here (except, maybe, for a fraction of the positronium yields).}, the so-called Greisen--Zatsepin--Kuzmin (GZK) cutoff~\cite{Greisen:1966jv,Zatsepin:1966jv} corresponding to \cm\ energies of $\sqrtsnn\approx 400$\,TeV~\cite{dEnterria:2011twh}.
Table~\ref{tab:1} summarizes the typical parameters for \pp, \pA, and \AaAa\ UPCs at RHIC, LHC and FCC energies, as well as in p-air collisions at the GZK cutoff. For the latter fixed-target collisions, the Lorentz factor of the proton projectile and \cm\ systems are related via $\gamma_\mathrm{L} = \sqrt{(\gamma_\mathrm{p}+1)/2}$, and we consider a photon energy of the nitrogen nucleus at rest of $\mathcal{O}(10$~MeV), typical of collective nuclear excitations.

\begin{table}[htpb!]
\tabcolsep=2.mm
\centering
\caption[]{Summary of the generic characteristics of photon-photon collisions in ultraperipheral proton and nuclear collisions at RHIC, HL-LHC~\cite{Bruce:2018yzs,dEnterria:2022sut} and FCC~\cite{FCC:2018vvp,Dainese:2019gab} energies, and in fixed-target cosmic-ray collisions at GZK-cutoff energies~\cite{dEnterria:2011twh}. For each colliding system, we quote its (i) nucleon-nucleon (NN) \cm\ energy $\sqrtsnn$, (ii) nominal total integrated luminosity per experiment $\LumiInt$ (for \pp, we quote in parentheses the estimated values collected under low-pileup conditions), (iii) beam energies $E_\text{beam}$, (iv) Lorentz factor $\gamma_\mathrm{L}=E_\text{beam}/m_\mathrm{N}$, (v) effective charge radius $R_\mathrm{A}$, (vi) photon ``maximum'' energy $E_{\gamma}^\text{max} \approx \gamma_\mathrm{L}/R_\mathrm{A}$, 
and (vii) ``maximum'' photon-photon \cm\ energy $\sqrt{s_{\gaga}^\text{max}} = \sqrt{4E_{\gamma,1}E_{\gamma,2}}$.}
\label{tab:1}
\vspace{0.2cm}
\begin{tabular}{lc cccccc}
\hline
System & $\sqrtsnn$ & $\LumiInt$ & $E_\text{beam1}+E_\text{beam2}$ & $\gamma_\mathrm{L}$ & $R_\mathrm{A}$ & $E_{\gamma}^\text{max}$ & $\sqrt{s_{\gaga}^\text{max}}$
\\\hline
\AuAu\ & 200~GeV & 10~nb$^{-1}$ & $100 + 100$~GeV & 107 & 6.9 fm & 3.1~GeV & 6.2~GeV \\ \hline
\PbPb\ & 5.52~TeV & 10~nb$^{-1}$ & 2.76 + 2.76~TeV & 2960 & 7.1 fm & 80~GeV & 160~GeV \\
\pPb\ & 8.8~TeV & 1~pb$^{-1}$ & 7.0 + 2.76~TeV & 7450, 2960 & 0.7,~7.1 fm & 2.45~TeV, 130~GeV & 1.1~TeV \\ 
\pp\ & 14~TeV & 3~ab$^{-1}$ (1~fb$^{-1}$) & 7.0 + 7.0~TeV & 7450 & 0.7 fm & 2.45~TeV& 4.5~TeV \\ \hline 
\PbPb\ & 39.4~TeV & 110~nb$^{-1}$ & 19.7 + 19.7~TeV & 21\,100 & 7.1 fm & 600~GeV & 1.2~TeV \\ 
\pPb\ & 62.8~TeV & 29~pb$^{-1}$ & 50. + 19.7~TeV & 53\,300, 21\,100 & 0.7,~7.1 fm & 15.2~TeV, 600~GeV & 6.0~TeV \\ 
\pp\ & 100~TeV & 30~ab$^{-1}$ (10~fb$^{-1}$) & 50. + 50.~TeV & 53\,300 & 0.7 fm & 15.2~TeV & 30.5~TeV \\ \hline
p-air & 400~TeV & -- & ${\approx}10^{8}$~TeV$\,+\,{\approx}0$ & ${\approx}10^{11},~1$ & 0.7,~2.9~fm & ${\approx}3\cdot 10^{10}$\,GeV, ${\approx}10$\,MeV & ${\approx}35$~TeV \\ \hline
\end{tabular}
\end{table}

In photon-photon collisions, any singly produced particle $\gaga\to X$ must have $\rm J^{P C}$ quantum numbers (representing total angular momentum J, parity P, and charge conjugation C) that respect basic conservation rules. In particular, the production of vector (spin-1) particles is forbidden in $\gaga$ collisions because real (massless) photons cannot combine to form a vector particle as per the Landau--Yang theorem~\cite{Landau:1948kw,Yang:1950rg}, and only even-spin resonance states with positive C-parity, such as $\rm J^{PC}=0^{-+},0^{++},2^{-+},2^{++},\dots$, are allowed. Such a selection rule makes of photon-photon collisions a particularly clean environment for the study of $\rm J=0$ (pseudo)scalar and $\rm J=2$ (pseudo)tensor particles~\cite{Morgan:1994ip,Krauss:1997vr,Whalley:2001mk}. Higher even-spin $\rm J=4, \ldots$ resonances can also be theoretically produced, but none of the few presently known spin-4 hadrons has a clearly established $\gaga$ decay width~\cite{ParticleDataGroup:2024cfk}, and they remain unobserved in such a production mode. Interestingly, $\gaga$ processes can also produce a pair of opposite-charged particles that can subsequently form short-lived spin- and C-even bound states under their common quantum electrodynamics (QED) interaction. Photon-photon collisions provide thereby ideal conditions to produce and investigate exotic QED atoms such as leptonium $(\lele)_0$, for $\ell^\pm=\rm e^\pm, \mu^\pm, \tau^\pm$, and ``hadronium'' atoms of two sorts: ``mesonium'' $\rm A_\mathrm{2h}$ (for $\rm h = \pi^\pm, K^\pm, D^\pm, B^\pm$ mesons), and ``baryonium'' $(\rm h\overline{h})_0$ (for $\rm h = baryon$, where the '0' subindex indicates their spin-0 para-state). 

\begin{figure}[htpb!]
  \centering
  \includegraphics[width=0.95\linewidth]{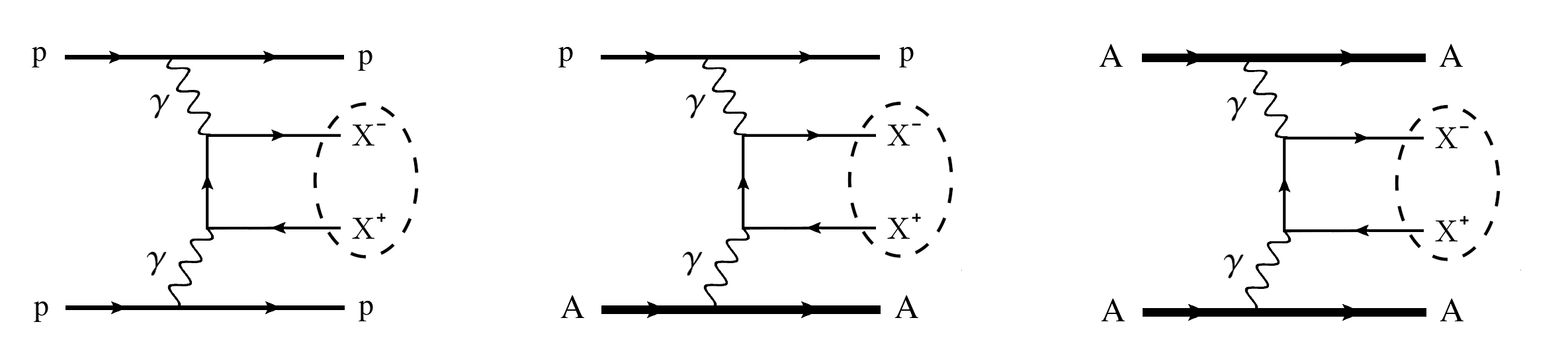}
  \caption{Schematic diagrams of the photon-photon production of a pair of opposite-charge particles $\rm X^+X^-$ followed by the formation of an $\rm (X^+X^-)$ onium-like bound state in \pp\ (left), proton-nucleus (center), and nucleus-nucleus (right) collisions. The $\rm (X^+X^-)$ states considered in this work are even-spin states formed by pairs of quarks ($\rm X= q$) bound by their QCD interaction, or of leptons ($\rm X = e, \mu, \tau$) or hadrons ($\rm X = mesons$, or baryons) bound by their QED interaction.}
  \label{fig:gaga_diags}
\end{figure}

The purpose of this paper is to study the two-photon production of even-spin systems in UPCs of protons and/or ions via the processes shown in Fig.~\ref{fig:gaga_diags}. The elementary cross-section for the production of a resonance X of mass $m_\mathrm{X}$, even spin J, total width $\Gamma_\mathrm{tot}$, and two-photon width $\Gamma_{\gaga}$, in photon-photon collisions at a center-of-mass energy $\sqrtshat = \sqrt{4E_{\gamma1}E_{\gamma2}}$ is given by Low's formula~\cite{Low:1960wv},
\begin{equation}
\hat{\sigma}_{\gaga \to\mathrm{X}}(\shat)= 8 \pi^2(2 J+1) \frac{\Gamma_{\gaga}\Gamma_\mathrm{tot}}{\left(\shat-m_\mathrm{X}^2\right)^2+\left(m_\mathrm{X}\Gamma_\mathrm{tot}\right)^2} 
= 4 \pi^2(2 J+1) \frac{\Gamma_{\gaga}}{m_\mathrm{X}^2} \delta\left(\shat-m_\mathrm{X}^2\right),
\label{eq:low}
\end{equation}
where the last equality holds in the narrow-width approximation, $\Gamma_\mathrm{tot}\ll m_\mathrm{X}$, and the delta function ensures total 4-momentum conservation as physical particles can only be produced on their mass shell. Since the $\gaga$ decay and production modes use the same matrix elements, Eq.~(\ref{eq:low}) provides a simple and useful expression that allows relating the diphoton width to the $\gaga$-fusion cross section, with proper phase-space and polarization summation factors. From this last expression and Eq.~(\ref{ref:sigma_gaga_UPC}), one can derive the ``master formula'' used in the remainder of this work to compute the production of any given C-even resonance $\mathrm{X}$ via photon-photon collisions in a generic UPC of charged hadrons A and B at nucleon-nucleon \cm\ energy $\sqrtsnn$~\cite{Budnev:1975poe},
\begin{equation}
\sigma(\mathrm{A} \mathrm{B} \xrightarrow{\gaga} \mathrm{A~X~B}) = \left.4 \pi^2(2 J+1) \frac{\Gamma_{\gaga}}{m_\mathrm{X}^2} \frac{\mathrm{d} \mathcal{L}_{\gaga}^{(\mathrm{AB})}}{\mathrm{d} m_{\gaga}}\right|_{m_{\gaga}=m_\mathrm{X}},
\label{eq:sigma_X_master}
\end{equation}
where $\left.\frac{\mathrm{d} \mathcal{L}_{\gaga}^{(\mathrm{AB})}}{\mathrm{d}m_{\gaga}}\right|_{m_{\gaga}=m_\mathrm{X}}$ is the value of the photon-photon luminosity at the resonance mass $m_\mathrm{X}$.


The $\gaga$ production of even-spin resonances, and/or QED bound states and other exotic atoms in UPCs (Fig.~\ref{fig:gaga_diags}) has been first considered in Refs.~\cite{Krauss:1997vr, Natale:1994nb, Baur:1998ay, Nystrand:1998hw,Baur:2001jj,Bertulani:2001zk,Ginzburg:1998df,Kotkin:1998hu} as well as in more recent works~\cite{Moreira:2016ciu,Goncalves:2018hiw,Azevedo:2019hqp, Esposito:2021ptx,Goncalves:2021ytq,Shao:2022cly,Niu:2022cug, Biloshytskyi:2022dmo,dEnterria:2022ysg,Francener:2021wzx,Fariello:2023uvh,Dai:2024imb}. In our paper, we extend these previous studies by (i) including multiple new hadronic resonances and exotic atoms not considered previously, (ii) using improved photon-photon luminosities for UPCs, (iii) with a proper propagation of theoretical uncertainties to their production cross sections, and also (iv) adding predictions for current and future colliders, such as the FCC-hh, as well for cosmic-rays interactions at GZK cutoff energies. According to Table~\ref{tab:1}, UPCs at RHIC can produce even-spin particles with masses $m_\mathrm{X}\lesssim 6$~GeV, whereas UPCs at the LHC, FCC, and GZK-cutoff energies can produce any resonance with masses up to hundreds or thousands of~GeV. In our study, we will present the production cross sections for all C-even states (with known diphoton width) between the lowest-mass (positronium) and the highest-mass (toponium) objects currently known.\\

The paper is organized as follows. In Section~\ref{sec:th} the basic theoretical ingredients are presented, including realistic effective two-photon luminosity functions as a function of $\gaga$ invariant mass obtained with the \gammaUPC\ Monte Carlo code~\cite{Shao:2022cly}, which allow the determination of the production cross sections of any given C-even resonance in UPCs by means of the Low's formula (Section~\ref{sec:th_xsec}), as well as a concise overview of the generic properties of QED bound states that can be produced in two-photon collisions (Section \ref{sec:QED_onium}).
In Sections \ref{sec:quark}, \ref{sec:leptonium}, and \ref{sec:hadronium}, we present, respectively, the theoretical UPC cross sections computed for quarkonium, leptonium, and hadronium final states for all colliding systems shown in the diagrams of Fig.~\ref{fig:gaga_diags}. In Section~\ref{sec:LbL}, we assess the impact of the diphoton decays of such objects as resonant backgrounds for the measurement of light-by-light scattering ($\gaga\to\gaga$) over $m_{\gaga} \approx 0.1$--15~GeV in \PbPb\ UPCs at the LHC.
The main findings are summarized in Section~\ref{sec:summ}. Finally, the Appendix provides in tabulated form all cross sections and yields obtained in the study.

\section{Theoretical ingredients}
\label{sec:th}

The effective $\gaga$ luminosities used in our cross section calculations, as well as a discussion of the basic formulas to obtain relevant properties (mass, Bohr radius, and diphoton width) of the exotic QED atoms studied in this work, are presented in this section.

\subsection{Effective photon-photon collision luminosities in UPCs}
\label{sec:th_xsec}

Often in the literature, approximate expressions for the effective $\gaga$ luminosities such as Eq.~(\ref{eq:L_gaga_UPC}) have been used to estimate cross sections for the production of even-spin resonances in UPCs. Those formulas are valid in the limit where the hadrons are described with a simplistic form factor, \eg\ a point-like charge or a ``hard sphere'' of radius $R_\mathrm{A}$, and actually include interactions where their matter distributions overlap and produce final states that are not distinguishable from standard hadronic interactions. In this work, we employ more realistic expressions based on the effective two-photon luminosity implemented in \gammaUPC~\cite{Shao:2022cly}:
\begin{equation}
\frac{\mathrm{d} \mathcal{L}_{\gaga}^{(\mathrm{AB})}}{\mathrm{d} \shat}=\frac{2 \shat}{s_{\mathrm{NN}}} \int \frac{\mathrm{d} E_{\gamma_1}}{E_{\gamma_1}} \frac{\mathrm{d} E_{\gamma_2}}{E_{\gamma_2}} \delta\left(\frac{\shat}{s_{\mathrm{NN}}}-\frac{4 E_{\gamma_1} E_{\gamma_2}}{s_{\mathrm{NN}}}\right) \frac{\mathrm{d}^2 N_{\gamma_1 / \mathrm{Z}_1, \gamma_2 / \mathrm{Z}_2}^{(\mathrm{AB})}}{\mathrm{d} E_{\gamma_1} \mathrm{~d} E_{\gamma_2}},
\label{eq:gagalumi}
\end{equation}
where 
\begin{equation}
\frac{\mathrm{d}^2N^{(\mathrm{AB})}_{\gamma_1/\mathrm{Z}_1,\gamma_2/\mathrm{Z}_2}}{\mathrm{d}E_{\gamma_1}\mathrm{d}E_{\gamma_2}} =  \int{\mathrm{d}^2\pmb{b}_1\mathrm{d}^2\pmb{b}_2\,P_\text{no\,inel}(\pmb{b}_1,\pmb{b}_2)\,N_{\gamma_1/\mathrm{Z}_1}(E_{\gamma_1},\pmb{b}_1)N_{\gamma_2/\mathrm{Z}_2}(E_{\gamma_2},\pmb{b}_2)},
\label{eq:2photonintegral}
\end{equation}
is derived from the convolution of the two-photon number densities $N_{\gamma_i/\mathrm{Z}_i}(E_{\gamma_i},\pmb{b}_i)$ with energies $E_{\gamma_{1,2}}$ at impact parameters $\pmb{b}_{1,2}$ from hadrons A and B, respectively (the vectors $\pmb{b}_{1}$ and $\pmb{b}_{2}$ have their origins at the center of each hadron, and, therefore, $|\,\pmb{b}_{1}-\pmb{b}_{2}|$ is the impact parameter between them); and $P_\text{no\,inel}(\pmb{b}_1,\pmb{b}_2)$ encodes the probability of hadrons A and B to remain intact after their interaction, which depends on their relative impact parameters. 
For the $P_\text{no\,inel}(b)$ probability to have no inelastic hadronic interaction at impact parameter $b$, the standard opacity (optical density, also known as ``eikonal Glauber''~\cite{Glauber:1970jm} ) expressions are used:
\begin{eqnarray}
P_\text{no\,inel}\left(b\right)&=& \left\{\begin{array}{ll} 
e^{-\,\sigma^{\mathrm{NN}}_{\text{inel}}\cdot T_{\mathrm{AB}}(b)}, & \text{for nucleus-nucleus UPCs}\\
e^{-\,\sigma^{\mathrm{NN}}_{\text{inel}}\cdot T_\mathrm{A}(b)}, & \text{for proton-nucleus UPCs}\\
\left|1-\Gamma(s_{_\text{NN}},b)\right|^2,\; \mbox{ with }\;\Upgamma(s_{_\mathrm{NN}},b)\propto e^{-b^2/(2b_0)} & \text{for \pp\ UPCs.}\\
\end{array}\right.
\label{eq:Psurv}
\end{eqnarray}
Here $\rm T_\mathrm{A}(b)$ and $\rm T_{\mathrm{AB}}(b)$ are the nuclear thickness and overlap functions, respectively, commonly derived from the hadron transverse density profiles via a Glauber MC model~\cite{Loizides:2017ack}, $\sigma^{\mathrm{NN}}_{\text{inel}} \equiv \sigma^{\mathrm{NN}}_{\text{inel}}(\!\sqrtsnn)$ is the inelastic NN scattering cross section at the hadronic \cm\ energy $\sqrtsnn$, and $\Upgamma(s_{_\mathrm{NN}},b)$ is the Fourier transform of the \pp\ elastic scattering amplitude modeled by an exponential function~\cite{Frankfurt:2006jp} with inverse slope $b_0 \equiv b_0(\!\sqrtsnn)$ dependent on the NN \cm\ energy. The $\sigma^{\mathrm{NN}}_{\text{inel}}$ and $b_0$ parameters dependent on $\sqrtsnn$ that are used in \gammaUPC\ are obtained from fits of the available experimental data~\cite{dEnterria:2020dwq}.

\begin{figure}[htpb!]
\centering
\includegraphics[width=0.49\textwidth]{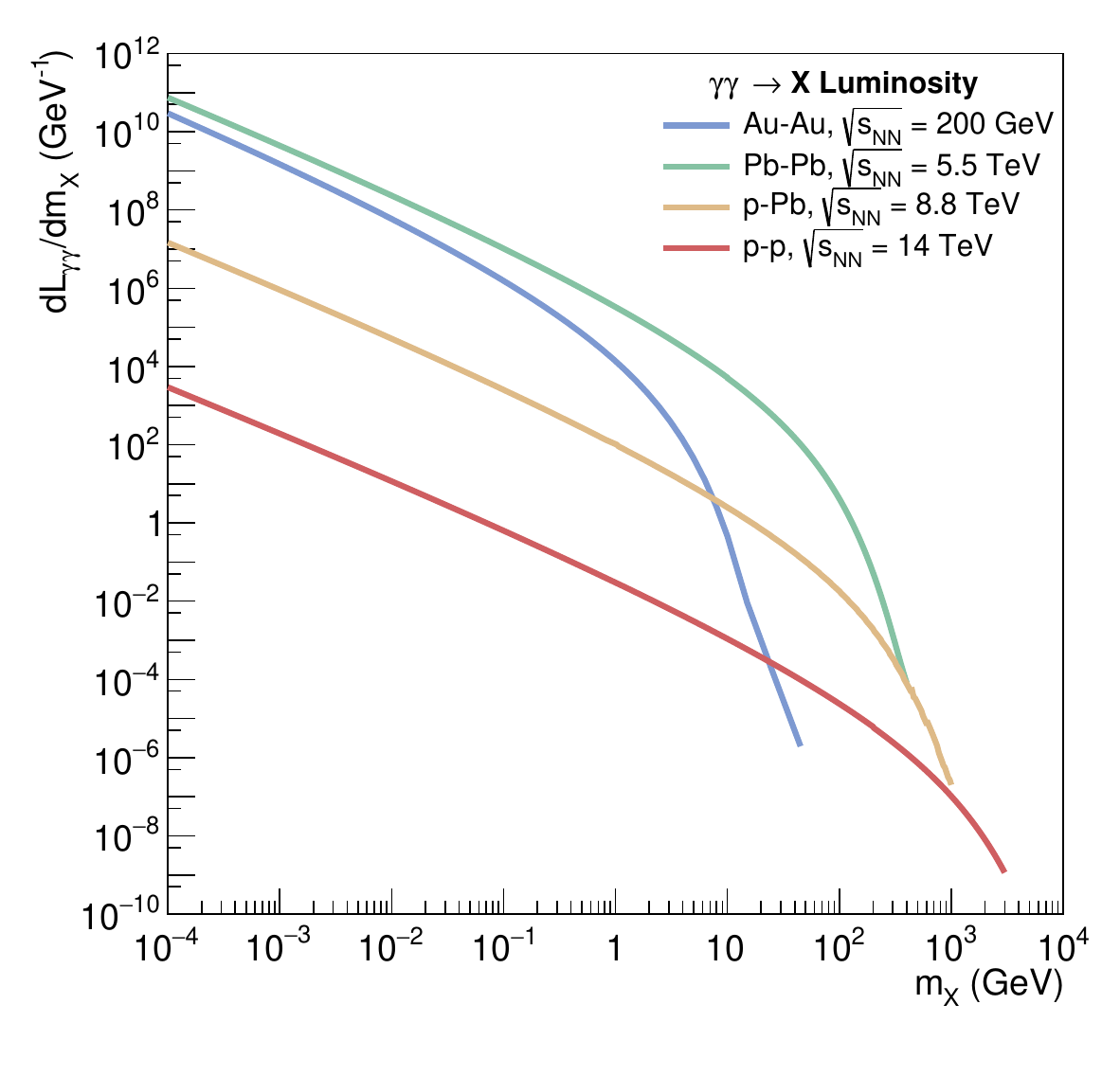}
\includegraphics[width=0.49\textwidth]{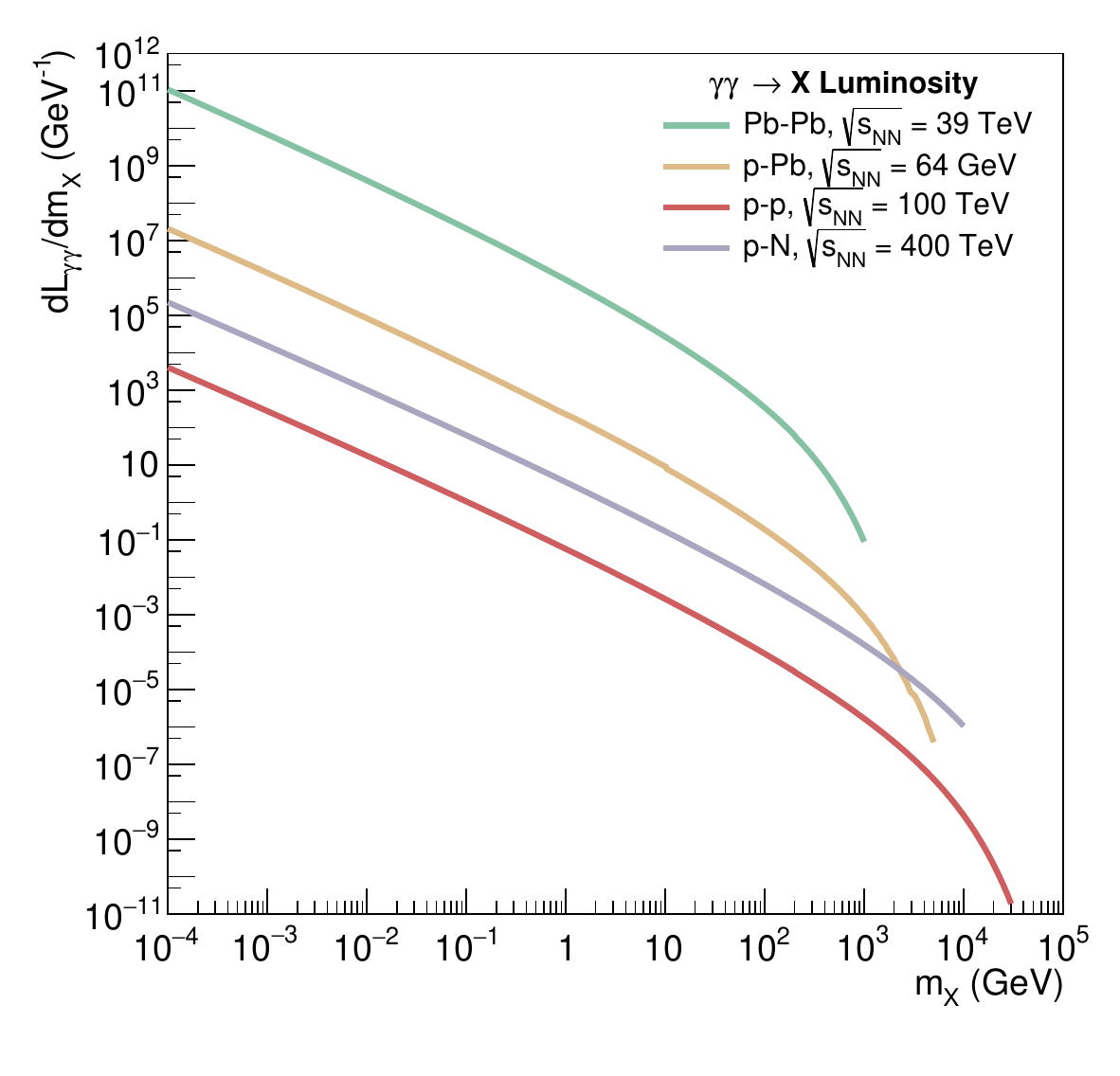}
\caption{Effective photon-photon luminosities as a function of $m_\mathrm{X}$, $\mathrm{d}{\Lumi}_{\gaga}/\mathrm{d}m_\mathrm{X}$ given by Eqs.~(\ref{eq:gagalumi})--(\ref{eq:Psurv}), for ultraperipheral \AuAu\,(200\,GeV), \PbPb\,(5.5\,TeV), \pPb\,(8.8\,TeV), and \pp\,(14\,TeV) collisions at RHIC and LHC (left), and \PbPb\,(39\,TeV), \pPb\,(62.8\,TeV), and \pp\,(100\,TeV) at FCC, and p-air\,(400\,TeV) for cosmic rays at the GZK cutoff (right), obtained with the \gammaUPC\ code with $\gamma$ photon fluxes derived from the corresponding ion charged form factors.}
\label{fig:dLdWUPC}
\end{figure}

In order to compute the $\gaga$ cross sections via Eq.~(\ref{eq:sigma_X_master}) for a variety of final states and for multiple colliding systems at RHIC, LHC, FCC, and GZK-cutoff energies, we employ the effective photon-photon luminosities obtained through Eqs.~(\ref{eq:gagalumi})--(\ref{eq:Psurv}) with the $\gammaUPC$ code, using the photon fluxes $N_{\gamma_i/\mathrm{Z}_i}(E_{\gamma_i},\pmb{b}_i)$ derived from the charged form-factors of protons and ions~\cite{Shao:2022cly}. The ion charged form-factor is more realistic than the electric dipole form factor commonly used in the literature, which diverges at low impact parameters and depends on a cutoff radius. The charged form factor reproduces better the precision $\gaga\to\lele$ measurements performed at the LHC, leading to theoretical uncertainties associated to the photon flux in the few percent (and neglected hereafter)~\cite{Shao:2024dmk}.
The corresponding $\gaga$ luminosities $\mathrm{d}{\Lumi}_{\gaga}/\mathrm{d}m_\mathrm{X}$ as a function of $m_\mathrm{X}$ are plotted in Fig.~\ref{fig:dLdWUPC} for \AuAu\ UPCs at RHIC as well as for \PbPb, \pPb, and \pp\ UPCs at the LHC (left), and for  \PbPb, \pPb, and \pp\ UPCs at FCC and p-air collisions at the GZK cutoff (right). The plotted luminosities cover 12--14 orders-of-magnitude from $m_\mathrm{X} = 0.1$~MeV up to the high-mass tails, and they approximately follow, except in the tails, a power-law decrease with a dependence of the form $\mathrm{d}{\Lumi}_{\gaga}/\mathrm{d}m_\mathrm{X}\propto m_\mathrm{X}^{-n}$, with exponent $n=1.25$--1.8 depending on the system and \cm\ energy. Employing such $\gaga$ luminosity curves and Eq.~(\ref{eq:sigma_X_master}), we can compute the production cross section for any arbitrary system existing between the lowest-mass (positronium) and the highest-mass (toponium) C-even particles currently known. 

\subsection{Basic properties of QED bound states}
\label{sec:QED_onium}

For the predictions of the production cross section of onium states and exotic atoms in photon-photon collisions, it is useful to review basic properties of QED bound states such as their mass, Bohr radius, and diphoton width. An onium system formed by two identical opposite-charge particles X$^\pm$, of 
mass $m_\mathrm{X}$ and electric charge $Z$, bound by their common QED interaction, has nonrelativistic momenta and, at first approximation, can be
described by the nonrelativistic Schr\"odinger equation for the wavefunction $\psi(\vec{r})$ of a hydrogen-like atom,
\begin{equation}
\left(-\frac{\nabla^2}{2 m_\text{red}} + V(r)\right)\psi(\vec{r}) =\left(-\frac{\nabla^2}{m_\mathrm{X}} + V(r)\right)\psi(\vec{r}) = E_n\psi(\vec{r})\,,
\label{eq:Schro}
\end{equation}
where $m_\text{red} = \frac{m_1m_2}{m_1+m_2}$ is the reduced mass of the system, which simplifies to $m_\text{red} = m_\mathrm{X}/2$ (second expression) in the symmetric case of two opposite-charge constituents of equal mass ($m_\mathrm{X} = m_1 = m_2$), $E_n$ is the energy of the state with principal quantum number $n$, and $V(r)$ is the one-photon-exchange Coulomb potential as a function of the radial distance $r$,
\begin{equation}
V(r)=-\frac{Z\alpha}{r}~,
\end{equation}
with $\alpha$ the fine structure constant evaluated at $m_\mathrm{X}$. The average distance between the constituents of such an onium system is given by their Bohr radius
\begin{equation}
r_\text{Bohr}=\frac{n^2}{Z\alpha m_\text{red}} = \frac{2n^2}{Z\alpha m_\mathrm{X}},
\label{eq:Rbohr}
\end{equation}
(where again the last equality assumes equal-mass constituents). Since, apart from meson resonances, we will be focusing on QED bound states, this formula is useful to check that the average distance between any pair of hadronic objects is much larger than the range of the strong interactions, \ie\ $r_\text{Bohr}\gg 1$~fm, so that the pure QED formulas below are applicable.


For bound states in a central potential, it is convenient to decompose the Schr\"odinger wavefunction $\psi(\vec{r})$ of Eq.~(\ref{eq:Schro}) into radial $R_{nl}$ and angular $Y_{l}^m$ parts (with $l$ and $m$ being the orbital angular momentum and its projection) as
$\psi_{n l m}(r, \theta, \phi)=R_{n l}(r) \cdot Y_{l}^m(\theta, \phi)$.
A simple Coulomb model for the binding force implies that for the S-wave level (\ie\ $l=0$) $n$ state ($n = 1$ is the ground state, and $l=0,\,m=0$),
the radial part $R_{n 0}(r)$ depends on the principal quantum number $n$ with probability at the origin $\left|R_{n 0}(0)\right|^2 = 4n^3/r_\text{Bohr}^3 = (\alpha\,m)^3/n^3$, where the $n^3$ term comes from the Laguerre polynomial with $l=0$, and the spherical harmonic is constant: $Y_0^0(\theta, \phi)=\frac{1}{\sqrt{4 \pi}}$. Thus, all the nonperturbative information about the formation of a given bound state is contained in the amplitude of  the radial wavefunction at the origin $|R_{n0}(0)|$ (and its derivatives, for non-zero angular momentum states). In the case of an onium resonance formed by two fermions, the ground state 
has two states with total angular momentum $\rm J=0$~and~1, depending on the relative (opposite or parallel) orientation of its constituent particles, known as para- ($\rm J^\mathrm{PC}=0^{-+}$) and ortho- ($\rm J^\mathrm{PC} = 1^{--}$) states. 
Since, as aforementioned, only the para-state is producible in photon-photon collisions, we will focus on this state. The wavefunction at the origin of a QED para-onium bound state, $\psi_\mathrm{para(X^+X^-)}(r=0)$, amounts to
\begin{equation}
\left|\psi_\mathrm{para(X^+X^-)}(r=0)\right|^2 = \left|R_{n0}(0)\cdot Y^0_0(\theta,\phi)\right|^2 = \frac{1}{4\pi}\left|R_{n0}(0)\right|^2 = \frac{1}{\pi r_\text{Bohr}^3}=\frac{\left(Z\alpha m_\mathrm{X}\right)^{3}}{8\pi n^3},
\label{eq:para_WF}
\end{equation}

where the binding energy of the ground ($n=1$) and excited states are given, respectively, by:
\begin{equation}
    E_{\text{bind},\,n=1}=-\frac{1}{4}m_\mathrm{X}\left(Z\alpha\right)^{2},\;\mbox{ and } E_{\text{bind},\,n} = \frac{E_{n=1}}{n^2}.
\label{eq:Ebind}
\end{equation}
Higher-order corrections to the Coulomb potential can be computed in the nonrelativistic QED framework. The first $\mathcal{O}(\alpha^3)$ correction to the binding energy is given by the Lamb shift~\cite{Lamb47,bethe1957quantum}, which can be derived using the nonrelativistic Uehling potential, characterizing difermion insertions into the Coulomb’s photon propagator, yielding (for the $n=1$ ground state)~\cite{dEnterria:2022alo,HS2025}
\begin{equation}
 \delta E_{\mathrm{Lamb}, n=1}=-\frac{(Z\alpha)^2}{18 \pi r_\text{Bohr} u}\left[u\left(-46+21 \pi \sqrt{1-u^2}-12 u^2\left(-2+\pi \sqrt{1-u^2}\right)\right)+{3\left(3-9 u^2+4 u^4\right) \log \frac{1-u}{1+u}}\right],
 \label{eq:E_Lamb}
\end{equation}
with $u = \sqrt{1 - (r_\text{Bohr}m_f)^2}$, where $m_f$ is the mass of charged fermions circulating in the loop. 
This  equation can be explicitly written as an $\mathcal{O}(\alpha^3)$ expression, taking into account the $\alpha^{-1}$ dependence of the Bohr radius, and retaining the dominant contribution from the electron loop alone (\ie\ setting $m_f=m_e$). In this limit, the Lamb shift energy correction can be simplified to
\begin{equation}
 \delta E_{\mathrm{Lamb}, n=1}=-\frac{(Z\alpha)^3 m_\mathrm{X}}{\pi}\left( a + b\,m_\mathrm{X}  \right),
 \mbox{ with } a = 0.68 \mbox{ and } b = 3.55 \cdot 10^{-5}\,\mathrm{MeV}^{-1},
 \label{eq:E_Lamb2}
\end{equation}
where $a$ and $b$ are numerically fitted constants that reproduce the full expression Eq.~(\ref{eq:E_Lamb}), including all difermion loops, to within $\pm20\%$.
The next correction to the binding energy, of order $\mathcal{O}(\alpha^4)$, can be obtained in the nonrelativistic Breit potential approach~\cite{Eides:2000xc}, taking into account scattering and annihilation channels, and reads
\begin{equation}
\delta E_\mathrm{Breit}=-\frac{1}{2m_\mathrm{X}}\left(E^{2}-2E\left\langle V\right\rangle +\left\langle V^{2}\right\rangle \right)=-\frac{21}{64}m_\mathrm{X}\left(Z\alpha\right)^{4} \frac{1}{n^3}~.
\label{eq:E_Breit}
\end{equation}
Combining Eqs.~(\ref{eq:Ebind}), (\ref{eq:E_Lamb2}), and (\ref{eq:E_Breit}), the mass of any given QED para-onium\footnote{At leading order, the wavefunction of the spin-1 and spin-0 bound states are equal, $\psi_\mathrm{ortho(X^+X^-)}(0) = \psi_\mathrm{para(X^+X^-)}(0)$, since they satisfy the same Schr\"odinger equation, but the ortho state is slightly more massive as~\cite{Efimov:2010ih}: $m_{\psi_\mathrm{ortho(X^+X^-)}} = m_{\psi_\mathrm{para(X^+X^-)}} + \frac{7}{12} (Z\alpha)^4 m_\mathrm{X}$.} ground state $(n=1)$, $\psi_\mathrm{para(X^+X^-)}$, can be obtained from
\begin{equation}
m_{\psi_\mathrm{para(X^+X^-)}}=2m_\mathrm{X}+E_{\text{bind},\,n=1}+\delta E_{\text{Lamb},n=1} +\delta E_\mathrm{Breit,n=1} = m_\mathrm{X}\left[2-\frac{1}{4}(Z \alpha)^2-\frac{\mathcal{O}(1)}{\pi}(Z\alpha)^3-\frac{21}{64}(Z \alpha)^4\right],
\label{eq:m_para}
\end{equation}
where the approximate value of the $\mathcal{O}(1)$ prefactor of the $(Z\alpha)^3$ correction can be estimated from Eq.~(\ref{eq:E_Lamb2}). 

As a last result, we lay out the general formula to compute the diphoton decay width of an even-spin QED-onium state. Fermi's golden rule provides the means to calculate the transition rate (decay width) from an initial quantum state $|i\rangle$ to a final state $|f\rangle$ under the influence of a perturbing Hamiltonian $H_{\text{int}}$:
\begin{equation}
   \Gamma=2 \pi\left|\left\langle f\left|H_{\mathrm{int}}\right| i\right\rangle\right|^2 \rho(E_f),
\end{equation}
where $\left\langle f\left|H_{\text{int}}\right| i\right\rangle$ is the matrix element of the interaction Hamiltonian between the initial and final states, and $\rho(E_f)$ is the density of final states with energy $E_f$.
The general form for the two-photon decay width involves the annihilation cross-section described by the matrix element $|\mathcal{M}|$, their relative velocity, and their probability density at the origin:
\begin{equation}
    \Gamma= \int \frac{|\mathcal{M}|^2}{ m_{\psi_\mathrm{para}}^2}|\psi_{\mathrm{para}}(0)|^2 \delta(E-|\vec{k_1}|-|\vec{k_2}|)\, \mathrm{d}^3 k_1 \mathrm{d}^3 k_2.
\end{equation}
For a pair of bound charged fermions, such as positronium\footnote{The two-photon annihilation of positronium is a standard result found in QED textbooks such as \eg\ Sec.\,89 of~\cite{Berestetskii:1982qgu}, or p.\,282 of~\cite{Jauch:1976ava}.} (see Section~\ref{sec:leptonium}), the matrix elements and phase space integration give:
\begin{equation}
\Gamma_{\psi_\mathrm{para(X^+X^-)}\to\gaga} = \frac{(Z\alpha)^5 m_\mathrm{X}}{2n^3}, \quad\mbox{for X = fermion}.
\label{eq:Gamma_gaga_fermion}
\end{equation}
A pair of bound charged bosons form a scalar state, such as a $(\pi^{+}\pi^{-})$ ``pionium''\footnote{Hadronic systems bound by their QED interaction are treated perturbatively neglecting their strong interaction, which is justified as long as their Bohr radius is $r_\text{Bohr}\gg 1$~fm, which is the case for the scalar pionium bound state~\cite{Palfrey:1961kt}.} (see Section~\ref{sec:hadronium}), for which the symmetric wavefunction introduces an additional $1/2$ factor, resulting in a diphoton width: 
\begin{equation}
\Gamma_{\psi_\mathrm{scalar(X^+X^-)}\to\gaga} = \frac{(Z\alpha)^5 m_\mathrm{X}}{4n^3}, \quad\mbox{for X = boson}.
\label{eq:Gamma_gaga_boson}
\end{equation}

\section{Photon-fusion production of even-spin hadron resonances}
\label{sec:quark}

The cross section for the single exclusive production of any given C-even meson X through $\gaga$ fusion in an UPC can be computed through Eq.~(\ref{eq:sigma_X_master}), and is completely determined from its spin $\rm J=0,2,4,...$, two-photon width $\Gamma_{\gaga}$, and the photon-photon effective luminosity of the colliding system at the particle mass (curves of Fig.~\ref{fig:dLdWUPC}). 
In the following, we collect the cross sections results for the UPC production of even-spin meson resonances formed by light (u,~d,~s) quarkonium in Section~\ref{sec:light_onia}, and heavy (c,~b,~t) quarkonium  in Section~\ref{sec:heavy_onia}. We provide first the meson properties, and then the computed cross sections and expected yields in UPCs at current and future hadron colliders. In Section~\ref{sec:multiquark}, we also compute the theoretical cross sections for the production of exotic hadron states (including multiquark candidate states). The corresponding cross sections and yields are listed in the Appendix sections~\ref{app:lightmeson},~\ref{app:heavyquarkonium}, and~\ref{app:exotichadrons}, respectively.

\subsection{Production of light meson resonances}
\label{sec:light_onia}

Table~\ref{tab:light_mesons} lists the relevant properties of all experimentally known spin-0,\,2,\,4 mesons formed by light quarks (u, d, s) included in the 2024 PDG review, with known diphoton width~\cite{ParticleDataGroup:2024cfk}. For each particle we list its mass, total and partial diphoton widths, and dominant decay mode (with branching fraction $\mathcal{B}$). We include only established particles, which is around 80\% of the PDG catalog. Their quoted diphoton widths are either those experimentally measured (and quoted in the PDG) or, in some cases, theoretically computed as explained below and/or in the provided references. 

\begin{table}[htpb!]
\centering
\caption{Properties of C-even light quark (u,d,s) resonances (spin-0,\,2,\,4 mesons). For each particle, we quote its $\rm J^{PC}$ quantum numbers, mass $m_\mathrm{X}$, full width $\Gamma$, diphoton partial width $\Gamma_{\gaga}$ and branching fraction $\BR(\rm X\to\gaga)$, from measurements~\cite{ParticleDataGroup:2024cfk} or theoretical predictions (with the indicated reference), and dominant decay modes (and their branching fraction $\BR$). 
\label{tab:light_mesons}}
\vspace{0.2cm}
\resizebox{\textwidth}{!}{
\begin{tabular}{lc cccccc}
\toprule
Resonance & $\rm J^{PC}$ & $m_\mathrm{X}$ (MeV) & $\Gamma_\mathrm{tot}$ (MeV) & $\Gamma_{\gaga}$ (keV) & $\BR(\rm X\to\gaga)$ & Dominant decay ($\BR$) \\
\midrule
$\pi^{0}$ & $0^{-+}$ & $134.9768 \pm 0.0005$ & $(7.808 \pm 0.120) \cdot 10^{-6}$ & $(7.716 \pm 0.119) \cdot 10^{-3}$ & $(98.823 \pm 0.034)$\% & $\gaga$ \\ 
$\eta$ & $0^{-+}$ & $547.862 \pm 0.017$ & $(1.31\pm 0.05)\cdot 10^{-3}$ & $0.515\pm 0.018$ & $(39.36 \pm 0.18)\%$ & $\gaga;\,3\pi^0$ ($32.57 \pm 0.21\%$) \\ 
$\sigma/f_0(500)$ & $0^{++}$ & $449_{-16}^{+22}$ & $550 \pm 24$ & $0.33 \pm 0.07$~\cite{Cappiello:2021vzi}
& $(6.0 \pm 1.3)\cdot10^{-7}$ & $\pi\pi, \gaga$ seen  \\ 
$\eta'$ & $0^{-+}$ & $957.78 \pm 0.06$ & $0.188 \pm 0.006$ & $4.34 \pm 0.14$ & $(2.307 \pm 0.033)\%$ & $\pi^{+} \pi^{-} \eta$ ($42.5 \pm 0.5\%$) \\ 
$f_0(980)$ & $0^{++}$ & $990 \pm 20$ & 10--100 & $0.29_{-0.06}^{+0.11}$ & $(2.9 \pm 0.6)\cdot 10^{-6}$ & $\rm \pi\pi, K \overline{K}, \gaga$ seen  \\ 
$a_0(980)$ & $0^{++}$ & $980 \pm 20$ & 50--100 & $0.30 \pm 0.10$ \footnote{Using $\Gamma_{\gaga} \mathcal{B}(\eta \pi)=(0.24 \pm 0.08)$~keV~\cite{Amsler:1997up}.} & $(3.0 \pm 1.0)\cdot 10^{-6}$ & $\rm \eta\pi, K \overline{K}, \eta'\pi, \gaga$ seen  \\ 
$f_2(1270)$ & $2^{++}$ & $1275.4 \pm 0.8$ & $186.6 \pm 2.3$ & $2.6 \pm 0.5$ & $(1.42 \pm 0.24) \cdot 10^{-5}$ & $\pi\pi$ ($84.3_{-0.9}^{+2.9}\%$) \\ 

$a_2(1320)$ & $2^{++}$ & $1318.2 \pm 0.6$ & $107 \pm 5$ & $1.00 \pm 0.06$ & $(9.4 \pm 0.7) \cdot 10^{-6}$ & $3\pi$ ($70.1 \pm 2.7\%$) \\ 

$a_0(1450)$ & $0^{++}$ & $1439 \pm 34$ & $258 \pm 14$ & $(4.65\pm 0.12)\cdot 10^{-3}$ \footnote{Using $\Gamma_{\gaga} \mathcal{B}(\pi \eta)=(432 \pm 6_{-256}^{+1073})$~keV~\cite{Belle:2009xpa}.} & $(1.8 \pm 0.2)\cdot 10^{-5}$ & $\pi\eta$ $(9.3\pm 2.0\%$) \\ 

$f_2' (1525)$ & $2^{++}$ & $1517.4 \pm 2.5$ & $86 \pm 5$ & $0.082 \pm 0.009$ & $(1.12 \pm 0.15) \cdot 10^{-6}$ & $\rm K \overline{K}$ ($87.6 \pm 2.2\%$) \\ 
$f_2(1565)$ & $2^{++}$ & $1571 \pm 13$ & $132 \pm 23$ & $0.70 \pm 0.14$ 
& $(5.3 \pm 1.7)\cdot10^{-6}$ & $\rm K \overline{K}$ ($87.6 \pm 2.2\%$) \\ 


$a_2(1700)$ & $2^{++}$ & $1706 \pm 14$ & $380_{-50}^{+60}$ & $0.30 \pm 0.05$ 
& $(7.9 \pm 1.7) \cdot 10^{-7}$ & $\eta\pi$ ($2.5 \pm 0.6\%$) \\ 

$f_0(1710)$ & $0^{++}$ & $1733_{-7}^{+8}$ & $150_{-10}^{+12}$ & $(3.33\pm 0.68)\cdot 10^{-5}$ \footnote{Using $\Gamma_{\gaga} \mathcal{B}(\rm K \overline{K})=(12_{-2}^{+3}\,_{-8}^{+227})~\text{keV}$~\cite{Belle:2013eck}.} & $(2.2 \pm 0.5) \cdot 10^{-7}$ & $\rm K\overline{K}~(36 \pm 12\%$)\\ 


$\eta_2(1870)$ & $2^{-+}$ & $1842 \pm 8$ & $225 \pm 14$ & $4.53 \pm 0.29 \pm 0.51$~\cite{CrystalBall:1991zkb} & $(2.0 \pm 0.3)\cdot 10^{-5}$ & $\eta \pi\pi, a_2(1320) \pi, f_2(1270) \eta,a_0(980) \pi$ \\ 

$f_4(2050)$ & $4^{++}$ & $2018 \pm 11$ & $237 \pm 18$ & $(1.36 \pm 0.02)\cdot 10^{-4}$ \footnote{Using $\Gamma_{\gaga}\mathcal{B}(\pi\pi) = (23.1_{-3.3}^{+3.6}{}_{-15.6}^{+70.5})$~keV~\cite{Belle:2009ylx}.} & $(5.7 \pm 0.6)\cdot 10^{-7}$ & $\pi\pi$ $(17.0 \pm 1.5\%)$ \\ 
\bottomrule
\end{tabular}
}
\end{table}

Scalar mesons decay dominantly into pairs of pseudoscalar mesons ($\rm \pi\pi, K\overline{K}, \pi\eta, \eta\eta$ or $\eta\eta'$) and, since broad overlapping states decaying into the same final state interfere, this complicates the determination of their masses and widths. The scalar resonances of light mesons are particularly difficult to resolve experimentally because they can have large decay widths~\cite{pdg2024scalarmeson}. Their mass ($m$) and width ($\Gamma$) can be theoretically estimated from the position of the nearest pole in the process amplitude (S- or T-matrix) at an unphysical sheet of the complex energy plane, traditionally labeled $\sqrt{s_{\text{pole}}}=m-i \Gamma / 2$. One such case is the $f_0(500)$ particle, also known as the $\sigma$ meson, the scalar partner of the SU(3) chiral meson nonet, which decays into $\pi\pi$ and $\gaga$ final states and for which the Breit--Wigner parameterization fails~\cite{Gardner:2001gc}. 
The quark/gluon/meson composition of such a broad state has been subject to discussions for many years, and its measurement in UPCs would provide useful information. 
The same holds true for the $a_0(980)$, $f_0(980)$ scalars, which have been often considered as four-quark states. Furthermore, the $0^{++}$ ground-state glueball expected below 2 GeV, will mix with the isoscalar $0^{++}$ $\qqbar$ states. The study of all these states in UPCs would therefore provide useful discriminating information about their nature and properties. 

The two-photon width of the $f_0(500)$ cannot be easily extracted from data because of its large width, but it has been estimated in the narrow width approximation as
$\Gamma_{\gaga} \simeq \alpha^2 |g_\gamma|^2 /\left(4 \text{Re}\sqrt{s_\mathrm{pole}}\right) = \alpha^2\left| g_\gamma^2\right|/(4 m)$,
where $g_{\gamma}$ is the residue at the pole to two photons~\cite{Morgan:1990kw}. 
Over the last decade, multiple calculations using dispersive techniques have been performed of the diphoton decay width of the $f_0(500)$ meson, but the interpretation of the results remains inconclusive,
yielding $\mathcal{O}(1.2$--3.1)~keV values~\cite{Pelaez:2015qba}. 
A more recent dispersive analysis of the $\gaga\to\pi\pi$ data~\cite{Cappiello:2021vzi} suggests $\Gamma_{\gaga} = 0.33 \pm 0.07$~keV (for a mass at $m = 471\pm 23$~MeV), which we use in our narrow width approximation here.
The diphoton widths of the $f_2(1565)$ and $a_2(1700)$ mesons are obtained from analyses of L3 data at LEP~\cite{Shchegelsky:2006et}. 

Table~\ref{tab:light_mesons_unknown} lists 16 even-spin hadrons with no precise value of their diphoton partial in the latest PDG review. 
There are two isoscalar $0^{++}$ mesons, $f_0(1370)$ and $f_0(1500)$, which are expected to mix with the $f_0(1710)$ meson. Both $f_0(1500)$ and $f_0(1710)$ have been proposed as glueball candidates. Among the signatures naively expected for glueballs is a reduced $\gaga$ coupling (although photon couplings of glueballs are sensitive to glue mixing with $\qqbar$ mesons). The observation of the production of any such resonance in UPCs (in any of their dominant decay modes listed in the last column of the table) would allow determining their partial diphoton decays, by simply inverting Eq.~(\ref{eq:sigma_X_master}), namely, by using their measured cross section $\sigma_\text{exp}(\mathrm{A} \mathrm{B} \xrightarrow{\gaga} \mathrm{A~X~B})$ divided by the expected two-photon luminosity at their mass, via
\begin{equation}
 \Gamma_{\rm X\to\gaga} = \frac{\sigma_\text{exp}(\mathrm{A} \mathrm{B} \xrightarrow{\gaga} \mathrm{A~X~B})}{4 \pi^2(2 J+1)}\,m_\mathrm{X}^2 \left[ \frac{\mathrm{d} \mathcal{L}_{\gaga}^{(\mathrm{AB})}}{\mathrm{d}m_{\gaga}}\mathrel{\bigg|}_{m_{\gaga}=m_\mathrm{X}}\right]^{-1}.
 \label{eq:diphoton_width_from_sigmaUPC}
\end{equation}
The two spin-4 resonances known: $a_4(1970)$ (without established $\gaga$ decay mode to date) and $f_4(2050)$, have theoretical $\gaga$ cross sections enhanced by factors of $(2J+1)=9$ compared to spin-0 mesons of similar mass, and therefore should be produced with an order-of-magnitude larger probability than any lower-spin counterpart with equal mass and diphoton width.

\begin{table}[htpb!]
\centering
\caption{Properties of C-even light quark (u,d,s) resonances (spin-0, -2, -4 mesons) without known 
diphoton decay width. For each particle, we quote its $\rm J^{PC}$ quantum numbers, mass $m_\mathrm{X}$, full width $\Gamma$, upper limit of the diphoton partial width $\Gamma_{\gaga}$ (if observed), and dominant decay modes from measurements~\cite{ParticleDataGroup:2024cfk}. 
\label{tab:light_mesons_unknown}}
\vspace{0.2cm}
\resizebox{\textwidth}{!}{
\begin{tabular}{lc ccrrr}
\toprule
Resonance & $\rm J^{PC}$ & $m_\mathrm{X}$ (MeV) & $\Gamma_\mathrm{tot}$ (MeV) & $\Gamma_{\gaga}$ (keV) & Dominant decays  \\
\midrule
$\eta (1295)$ & $0^{-+}$ & $1294 \pm 4$ & $55 \pm 5$ & $< 0.066$ & $\eta \pi^{+} \pi^{-}, a_0(980) \pi, \eta \pi \pi,  \sigma \eta, {\rm K\overline{K}} \pi, \gaga$ seen \\ 
$\pi(1300)$ & $0^{-+}$ & $1300 \pm 100$ & 200--600 & $< 0.8\cdot 10^{3}$~\cite{Shchegelsky:2006et} & $\rho \pi, \pi(\pi\pi)$ seen  \\ 

$f_0(1370)$ & $0^{++}$ & $1370 \pm 170$ & 200--500 & seen & $\pi\pi,\,4\pi, \eta\eta, \rm K\overline{K}, \gaga$ seen \\ 

$\eta(1405)$ & $0^{-+}$ & $1408.8 \pm 2.0$ & $50.3 \pm 2.5$ & $<1.78$ & $ {\rm K \overline{K}}, \eta \pi\pi, a_0(980) \pi,  f_0(980) \eta, 4\pi, \rho^0 \gamma, \rm K^*K$ seen \\  
$f_0(1500)$ & $0^{++}$ & $1522 \pm 25$ & $108 \pm 33$ & no obs.\ $\gaga$ decay & $\pi\pi$ ($34.5 \pm 2.2\%$) \\ 

$\eta_2(1645)$ & $2^{-+}$ & $1617 \pm 5$ & $181 \pm 11$ & no obs.\ $\gaga$ decay & $a_2(1320) \pi,\rm K \overline{K} \pi, K^* \overline{K}, \eta \pi^{+} \pi^{-},a_0(980) \pi$ seen \\ 

$\pi_2(1670)$ & $2^{-+}$ & $1670.6_{-1.2}^{+2.9}$ & $258_{-9}^{+8}$ & $<0.072$ & $3\pi$ ($95.8 \pm 1.4\%$) \\ 

$\pi (1800)$ & $0^{-+}$ & $1810_{-11}^{+9}$ & $215_{-8}^{+7}$ & no obs.\ $\gaga$ decay & \makecell[r]{$\pi^{+} \pi^{-} \pi^{-}, f_0(500,980,1370,1500) \pi^{-},\eta \eta \pi^{-}, a_0(980) \eta$,\\ $\eta \eta^{\prime}(958) \pi^{-},\rm K_0^*(1430) K^{-}$} \\ 
$\pi_2(1880)$ & $2^{-+}$ & $1874_{-5}^{+26}$ & $237_{-30}^{+33}$ & no obs.\ $\gaga$ decay & $\eta \eta \pi^{-}, a_0(980) \eta,a_2(1320) \eta, f_0(1500) \pi, f_1(1285) \pi, \omega \pi^{-} \pi^0$ \\
$a_4(1970)$ & $4^{++}$ & $1967 \pm 16$ & $324_{-18}^{+15}$ & no obs.\ $\gaga$ decay & $\rm K \overline{K}, \pi^{+} \pi^{-} \pi^0, \rho \pi, f_2(1270) \pi, \omega \pi^{-} \pi^0,\omega \rho, \eta \pi, \eta^{\prime}(958) \pi$ \\ 
$f_2(2010)$ & $2^{++}$ & $2011_{-80}^{+60}$ & $202 \pm 60$ & no obs.\ $\gaga$ decay & $\rm \phi\phi, K\overline{K}$ seen\\ 
$f_0(2020)$ & $0^{++}$ & $1982_{-3}^{+54.1}$ & $436 \pm 50$ & no obs.\ $\gaga$ decay & $\rho \pi\pi, \pi^0 \pi^0, \rho\rho, \omega \omega, \eta \eta, \eta' \eta'$ seen \\ 
$f_2(2300)$ & $2^{++}$ & $2297 \pm 28$ & $149 \pm 40$ & seen & $\rm \phi\phi, K\overline{K}, \gaga,\Lambda \overline{\Lambda}$ seen \\ 
$f_2(2340)$ & $2^{++}$ & $2346_{-10}^{+21}$ & $331_{-18}^{+27}$ & no obs.\ $\gaga$ decay & $\rm \phi\phi, \eta\eta, \eta'\eta'$ seen \\
\bottomrule
\end{tabular}
}
\end{table}

\begin{figure}[htpb!]
    \centering
    \includegraphics[width=0.6\linewidth]{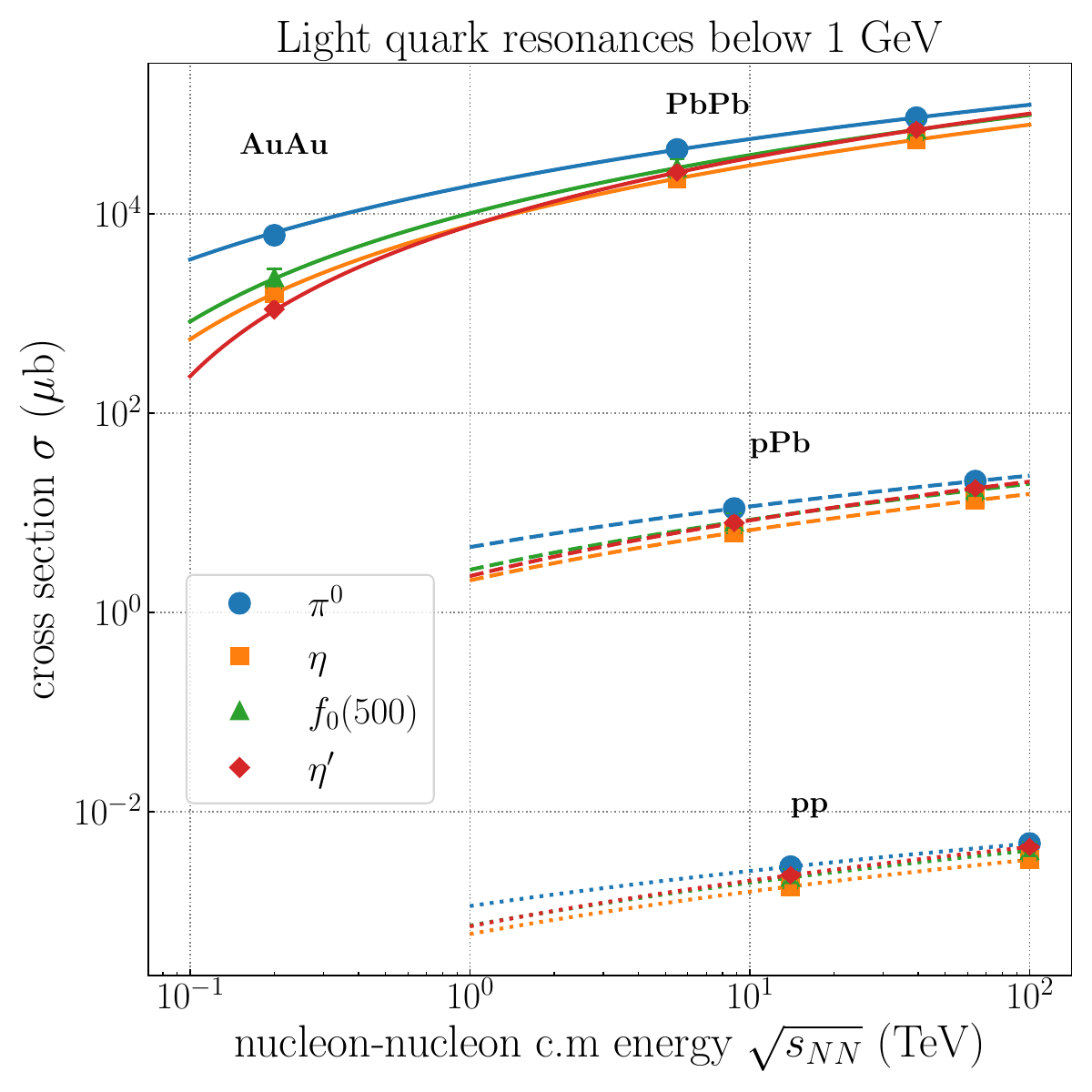}
    \caption{Cross sections for the $\gaga$ production of even-spin light mesons (with $m_\mathrm{X}\lesssim 1$~GeV and known $\gaga$ widths) as a function of nucleon-nucleon \cm\ energy $\sqrtsnn$, in \PbPb\ or \AuAu\ (solid curves), \pPb\ (dashed curves), and \pp\ (dotted curves) UPCs. The curves are $\ln ^3(\snn)$ fits to guide the eye.}
    \label{fig:sigma_vs_sqrts_1GeV}
\end{figure}

\begin{figure}[htpb!]
    \centering
    \includegraphics[width=0.6\linewidth]{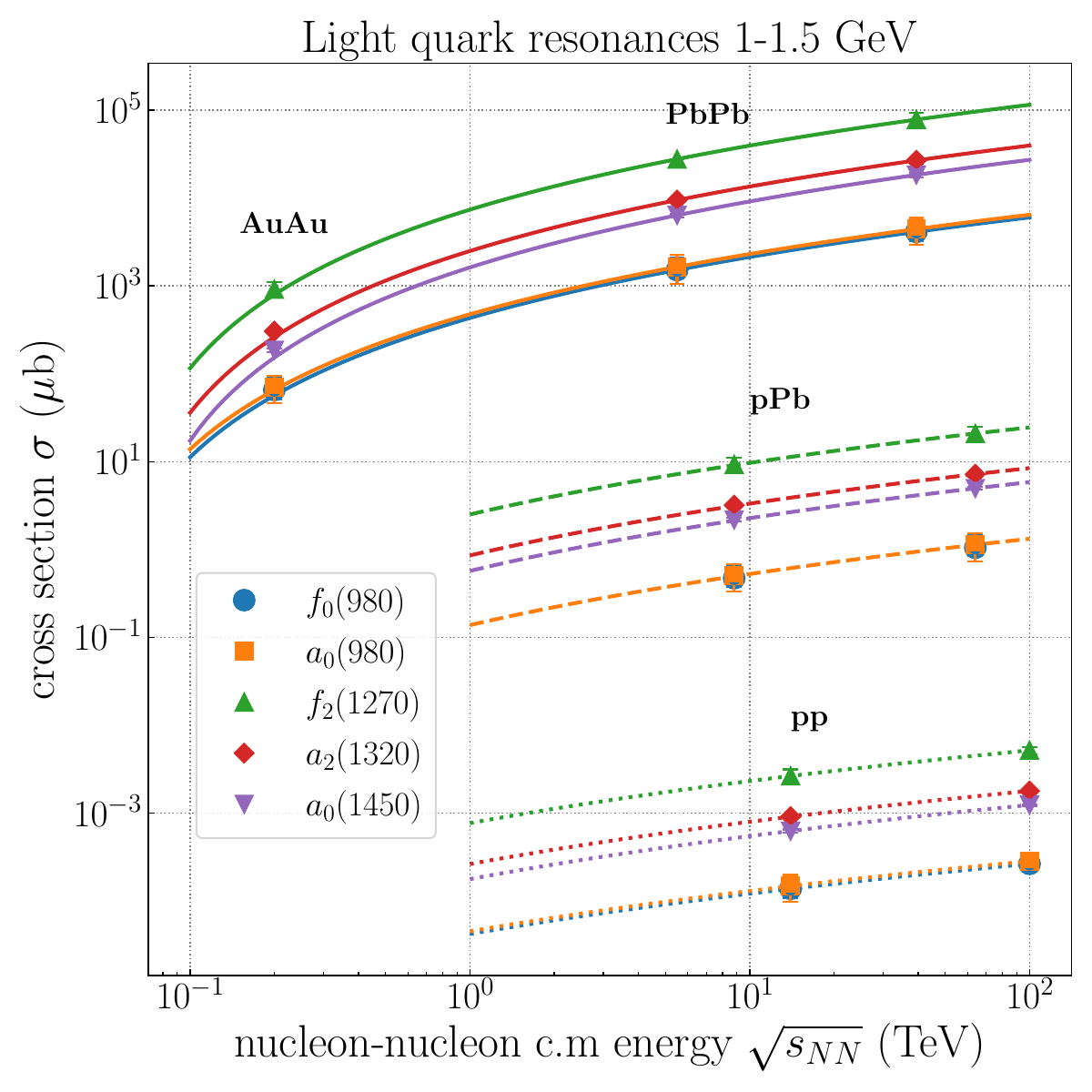}
    \caption{Cross sections for the $\gaga$ production of even-spin light mesons (with $m_\mathrm{X}= 1$--1.5~GeV and known $\gaga$ widths) as a function of nucleon-nucleon \cm\ energy $\sqrtsnn$, in \PbPb\ or \AuAu\ (solid curves), \pPb\ (dashed curves), and \pp\ (dotted curves) UPCs. The curves are $\ln ^3(\snn)$ fits to guide the eye.}
    \label{fig:sigma_vs_sqrts_1_1.5GeV}
\end{figure}

\begin{figure}[htpb!]
    \centering
    \includegraphics[width=0.6\linewidth]{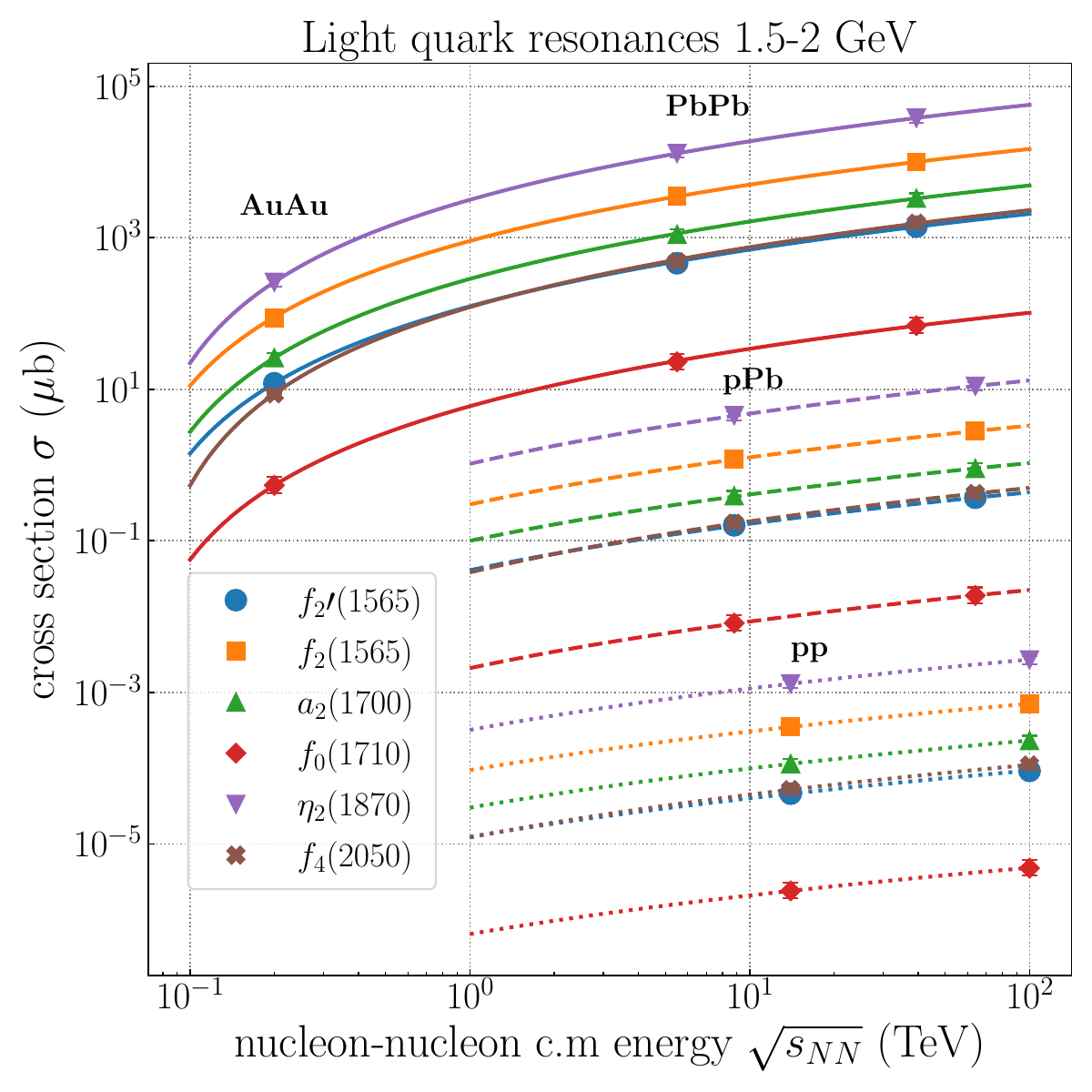}
    \caption{Cross sections for the $\gaga$ production of light mesons (with $m_\mathrm{X}= 1.5$--2.0~GeV and known $\gaga$ widths) as a function of nucleon-nucleon \cm\ energy $\sqrtsnn$, in \PbPb\ or \AuAu\ (solid curves), \pPb\ (dashed curves), and \pp\ (dotted curves) UPCs. The curves are $\ln ^3(\snn)$ fits to guide the eye.}
    \label{fig:sigma_vs_sqrts_1.5_2GeV}
\end{figure}

The photon-fusion production of pseudoscalar mesons in UPCs, via the diagram shown in Fig.~\ref{fig:gaga_diags}, is just a realization of the the well-known ``Primakoff effect''~\cite{Primakoff:1951iae}. Tables~\ref{tab:sigma_light_onia0}, \ref{tab:sigma_light_onia1}, and \ref{tab:sigma_light_onia2} in Appendix report the computed $\gaga$ cross sections for all the (pseudo) scalar/tensor resonances listed in Table~\ref{tab:light_mesons} ---grouped by mass ranges: $m_\mathrm{X}\lesssim 1$~GeV, $m_\mathrm{X} \approx 1$--1.5~GeV, and $m_\mathrm{X} = 1.5$--2~GeV, respectively--- produced in \pp, \pPb, and \PbPb\ UPCs at LHC and FCC \cm\ energies as well as p-air collisions at GZK-cutoff energies, derived using Eq.~(\ref{eq:sigma_X_master}). We have propagated into the final cross sections the parametric uncertainties from the $m_\mathrm{X}$ and $\Gamma_{\gaga}$ values of each resonance. Uncertainties due to the photon luminosities are subpercent in this mass range~\cite{Shao:2024dmk} and are neglected. 
Whenever available, we also quote the cross sections results from previous studies. Our results agree in general with older calculations (which did not quote theoretical uncertainties) with differences appearing due to the previous use of simplified (point-like, hard-sphere) photon fluxes, absence of survival probability corrections, and/or outdated diphoton widths and masses, although there are also some inconsistencies likely due to typos and/or errors in past results.
One can see, as expected from Eq.~(\ref{eq:sigma_X_master}), that for the same particle spin, all cross sections decrease with resonance mass following the $\propto\!m_\mathrm{X}^{-2}$ dependence of the photon-fusion cross section as well as the $m_\mathrm{X}^{-n}$ power-law decrease (with exponents $n=1.25$--1.8 depending on the system and \cm\ energy) of the two-photon effective luminosities (Fig.~\ref{fig:dLdWUPC}). Although, in some cases, such a generic trend is partially compensated by comparatively larger diphoton partial widths for concrete heavier resonances. The UPC cross sections results at colliders of Tables~\ref{tab:sigma_light_onia0}, \ref{tab:sigma_light_onia1}, and \ref{tab:sigma_light_onia2} in Appendix~\ref{app:lightmeson} are plotted as a function of collision energy in Figs.~\ref{fig:sigma_vs_sqrts_1GeV}, \ref{fig:sigma_vs_sqrts_1_1.5GeV}, and \ref{fig:sigma_vs_sqrts_1.5_2GeV}, respectively, showing their approximate dependence on $\ln ^3(\snn)$.

Tables~\ref{tab:sigma_light_onia0}, \ref{tab:sigma_light_onia1}, and \ref{tab:sigma_light_onia2} list also the total yields $N_\text{evts}(\gaga\to\mathrm{X})=\sigma(\gaga\to\mathrm{X})\times\LumiInt$, as well as the yields in their diphoton decay mode $N_\text{evts}(\gaga\to\mathrm{X}(\gaga))$ (obtained by multiplying the former by their $\BR(X\to\gaga)$ value), expected in UPCs at the various colliders. To obtain such numbers, we use the nominal integrated luminosities for the p-A and A-A running modes (Table~\ref{tab:1}), but only a small fraction of the \pp\ data (1~fb$^{-1}$ at the LHC, and 10~fb$^{-1}$ at the FCC-hh) that are assumed to be recorded under the low-pileup conditions needed to properly identify exclusive processes in UPCs and reconstruct such low-mass objects. Given the inherent imprecision on the actual luminosities to be integrated by the experiments, we present the  $N_\text{evts}(\gaga\to\mathrm{X})$ and $N_\text{evts}(\gaga\to\mathrm{X}(\gaga))$ values without uncertainties, so as to give the indicative order-of-magnitude of events forecast (here, as well as in all other tables of the paper).

The number of UPCs expected to exclusively produce light-quark even-spin resonances over $m_\mathrm{X}=0.14$--2.3~GeV masses is very large, ranging from millions to hundred-millions events at the LHC, and about factors of 10 to 1000 smaller for \AuAu(200~GeV) UPCs at RHIC. The number of those resonances that ``decay back'' to a pair of photons is several orders-of-magnitude smaller (except for the two lightest, $\pi^0$ and $\eta$, mesons), but some of them will have enough number of events to be also observable in such a clean decay mode. However, their small masses and the fact that photon-fusion leads to negligible transverse momentum ($\pT$) boosts, imply decays into very soft hadronic or diphoton final states that make their observation very difficult in the large-acceptance ATLAS~\cite{ATLAS:2008xda} and CMS~\cite{CMS:2008xjf} experiments, which are optimized for the reconstruction of particles with much larger $\pT$ values. Despite smaller acceptances (as well as, comparatively reduced integrated luminosities in some cases), the ALICE~\cite{ALICE:2008ngc} (in particular, the proposed ALICE-3~\cite{ALICE:2022wwr}) as well as the LHCb~\cite{LHCb:2008vvz} (in particular, the proposed LHCb upgrade II~\cite{LHCb:2018roe,LHCb:2025vou}) experiments have much better adapted detectors to reconstruct such soft decays. In the hadronic decay modes there will be potentially larger backgrounds from other exclusive processes --such as photoproduction for UPCs with ions (photon-pomeron interactions) and central-exclusive (pomeron-pomeron) processes for \pp\ collisions. On the other hand, their diphoton decays (despite being comparatively suppressed) can offer a cleaner final state to attempt their observation on top of the light-by-light continuum (by reversing the selection criteria used to measure the latter, see Section~\ref{sec:LbL}). Of course, experimental acceptance and efficiency losses, as well as selection cuts applied to remove backgrounds, will further reduce the yields (a determination of them goes beyond the scope of this paper), but the measurement of the cross sections for such light mesons in UPCs at the LHC, and the subsequent determination of their diphoton widths (which are poorly known in most cases) via Eq.~(\ref{eq:diphoton_width_from_sigmaUPC}), provides additional interesting physics cases for the future ALICE~3 and LHCb-upgrade-II experiments.

\subsection{Production of heavy quarkonium resonances}
\label{sec:heavy_onia}

Table~\ref{tab:heavy_onia} lists the relevant properties of all presently known (pseudo) scalar and tensor resonances formed by charmonium and bottomonium bound states with masses over $m_\mathrm{X}\approx 3$--10~GeV. Moreover, we also list the para-toponium $\eta_\mathrm{t}\equiv(\ttbar)_0$ state, a theoretical quasibound object formed by a top-antitop quark pair interacting via the QCD interaction for ultrashort time scales~\cite{Kuhn:1987ty}, which has been hinted at in the LHC data recently~\cite{CMS:2025kzt}. Masses are measured precisely for all the particles, but not all their two-photon widths have been (properly) experimentally determined~\cite{ParticleDataGroup:2024cfk}. 
On the one hand, starting with the lightest heavy-quarkonium state ($\etacOneS$), a recent direct measurement by BES-III finds $\Gamma(\etacOneS\to\gaga) = 11.30 \pm 1.43$~keV~\cite{BESIII:2024rex}, which is more than twice larger than the current PDG world-average for this decay, $\Gamma(\etacOneS\to\gaga) = 5.06 \pm 0.40$~keV. Similarly, the diphoton width of the $\etacTwoS$ state is also poorly known, and has presently a $\pm60\%$ experimental uncertainty. On the other hand, the $\gaga$ decays of the four $\bbbar$ resonances ($\etabOneS$, $\etabTwoS$, $\chibZero$, $\chibTwo$) remain unobserved so far. All these results highlight the issues affecting many heavy-quarkonium diphoton widths for which we use theoretical predictions as explained below. For the $\etacOneS \to\gaga$ partial width, we use the recent lattice QCD calculations of Ref.~\cite{Colquhoun:2023zbc} to avoid the aforementioned contradictory experimental results. For the $\etabOneS$ and $\etabTwoS$ cases, predictions exist for their two-photon partial widths in nonrelativistic QCD (NRQCD)~\cite{Penin:2004ay,Chung:2010vz}. Exploiting heavy-quark spin symmetry, the two-photon $\etabnS\to\gaga$ and leptonic $\UpsnS\to\ell^+\ell^-$ decay widths are proportional to the same wavefunction at the origin, and the NRQCD calculations use the theoretical ratio between them, $\Gamma(n^3\mathrm{S}_1\to\epem)/\Gamma(n^1\mathrm{S}_0\to\gaga)$, taking into account cancellations between higher-order relativistic and radiative corrections, together with the (latest) experimental values of $\Gamma(\Upsilon(\rm 1S,2S)\to\epem)=1.291\pm 0.030,\, 0.611\pm 0.050$~keV~\cite{ParticleDataGroup:2024cfk}, to obtain $\Gamma(\eta_\mathrm{b}(1\mathrm{S},2\mathrm{S})\to\gaga) = 0.493 \pm 0.092,\,0.24 \pm 0.04$~keV, respectively.
We quote this latter $\Gamma(\etabTwoS\to\gaga)$ value in Table~\ref{tab:heavy_onia}, whereas for the $\etabOneS \to\gaga$ partial width we use the more recent lattice QCD calculations with a 6\% precision~\cite{Colquhoun:2024wsj}. The diphoton widths of $\chibZero$, $\chibTwo$ are based on the range of model predictions given in Ref.~\cite{Wang:2018rjg}.

\begin{table}[htpb!]
\tabcolsep=3.5mm
\centering
\caption{List of C-even heavy-quarkonium resonances that can be produced via two-photon fusion. For each particle, we quote its $\rm J^{PC}$ quantum numbers, mass $m_\mathrm{X}$, total width $\Gamma_\mathrm{tot}$, and diphoton partial width $\Gamma_{\gaga}$ and branching fraction $\BR(\rm X\to\gaga)$, from measurements~\cite{ParticleDataGroup:2024cfk} or theoretical predictions (for $\etacOneS$, $\etabOneS$, $\etabTwoS$, $\chibZero,\chibTwo$, and $\etatnS$, see text for details), and dominant decay channels.
\label{tab:heavy_onia}}
\vspace{0.2cm}
\resizebox{\textwidth}{!}{
\begin{tabular}{lc ccccc}
\toprule
Resonance & $\rm J^{PC}$ & $m_\mathrm{X}$ (GeV) & $\Gamma_\mathrm{tot}$ (MeV) & $\Gamma_{\gaga}$ (keV) & $\BR(\rm X\to\gaga)$ & Dominant decay ($\BR$)\\ 
\midrule
$\etacOneS$ & $0^{-+}$ & \texttt{$2.9841 \pm 0.0004$} & \texttt{$30.5 \pm 0.5$} & $6.788 \pm 0.061$~\cite{Colquhoun:2023zbc} 
& $(2.23 \pm 0.06) \cdot 10^{-4}$ & $2\left(\pi^{+} \pi^{-} \pi^0\right)$ ($15.9 \pm 2.0\%$) \\

$\etacTwoS$ & $0^{-+}$ & \texttt{$3.6377 \pm 0.0009$} & \texttt{$11.8 \pm 1.6$} & $2.12 \pm 1.45$ 
& $(1.8 \pm 1.2) \cdot 10^{-4}$ & $\rm K\overline{K}\pi$ ($1.9 \pm 1.2\%$) \\
$\chicZero$ & $0^{++}$ & \texttt{$3.41471 \pm 0.00030$} & \texttt{$10.7 \pm 0.6$} & $2.18 \pm 0.16$ 
&$(2.04 \pm 0.10) \cdot 10^{-4}$ & $\pi^{+} \pi^{-} \pi^0 \pi^0$ ($3.3\pm 0.4\%$) \\
$\chicTwo$ & $2^{++}$ & \texttt{$3.55617 \pm 0.0007$} & \texttt{$1.98\pm 0.09$} & $0.578 \pm 0.035$ 
& $(2.92 \pm 0.12) \cdot 10^{-4}$ & $\jpsi\gamma$ ($19.0 \pm 0.5\%$) \\

$\etabOneS$ & $0^{-+}$ & \texttt{$9.3987 \pm 0.0020$} & \texttt{$10 ^{+5}_{-4}$} & $0.557 \pm 0.032$~\cite{Colquhoun:2024wsj} 
& $\left(5.6_{-3.2}^{+4.0}\right) \cdot 10^{-5}$ & gg (${\approx}100\%$) \\
$\etabTwoS$ & $0^{-+}$ & \texttt{$9.999 \pm 0.004$} & \texttt{$5 ^{+3}_{-2}$} & $0.24 \pm 0.04$\footnote{Result based on NRQCD calculations~\cite{Chung:2010vz} and the latest experimental data (see text).} 
&$\left(4.8 ^{+4.7}_{-3.7}\right) \cdot 10^{-5}$ & gg (${\approx}100\%$) \\
$\chibZero$ & $0^{++}$ & \texttt{$9.85944 \pm 0.00052$} & {$2.60^{+0.79}_{-0.57}$} & $0.15^{+0.05}_{-0.03}$~\cite{Wang:2018rjg} 
&$\left(5.8_{-2.1}^{+3.1}\right) \cdot 10^{-5}$ & $\Upsilon$(2S)$\gamma$ ($1.94\pm 0.27\%$) \\

$\chibTwo$ & $2^{++}$ & \texttt{$9.91221 \pm 0.00040$} & {$0.180^{+0.016}_{-0.057}$} & \texttt{$(9.3^{+1.3}_{-6.2})\cdot 10^{-3}$}~\cite{Wang:2018rjg} 
& $\left(5.2_{-4.2}^{+1.0}\right) \cdot 10^{-5}$ & $\Upsilon$(1S)$\gamma$ ($18.0\pm 1.0\%$) \\
$\etatOneS$ & $0^{-+}$ & ${\approx}343.1$ & ${\approx}2660$ & $22.6\pm 6.8$ & $1.022 \cdot 10^{-5}$ & $\rm W^+W^-\bbbar$ (${\approx}100\%$) \\
$\etatTwoS$ & $0^{-+}$ & ${\approx}344.5$ & ${\approx}2660$ & $2.8\pm 0.8$ & $0.85 \cdot 10^{-5}$ & $\rm W^+W^-\bbbar$ (${\approx}100\%$) \\
$\etatnS$ & $0^{-+}$ & ${\approx}344$ & ${\approx}2660$ & $27.2\pm 8.1$ & $1.02 \cdot 10^{-5}$ & $\rm W^+W^-\bbbar$ (${\approx}100\%$) \\
\bottomrule
\end{tabular}
}
\end{table}

We also consider the ephemeral spin-singlet para-toponium bound state, $\eta_\mathrm{t}$, formed by a top and antitop quark~\cite{Kuhn:1987ty}, which can be perturbatively described in NRQCD with a single-gluon exchange potential of the form $V(r) = -C_\mathrm{F}\alphas/r$, where $C_\mathrm{F}=4/3$ is the quark-antiquark color factor. Such a state has a mass of $m_{\eta_\mathrm{t}} \approx 2 m_\mathrm{t} + E_\text{bind} = 343.1 \pm 0.9$~GeV, obtained using the current PDG mass value of $m_\mathrm{t} = 172.5 \pm 0.45$~GeV~\cite{ParticleDataGroup:2024cfk} and $E_\text{bind} = -\frac{1}{4}m_\mathrm{t}(C_\mathrm{F}\overline{\alpha}_s)^2 \approx -1.9$~GeV, where $\overline{\alpha}_s\approx 0.16$ is the strong coupling evaluated\footnote{One can also find in the literature equivalent expressions for the toponium properties written instead as a function of the typical velocity of the top quarks in the bound state: $v_\mathrm{t}=C_\mathrm{F}\overline{\alpha}_s = 0.21$.} at the typical scale given by the toponium radius, \ie\ $\overline{\alpha}_s = \alphas(r_\text{Bohr}^{-1})$, whose numerical value can be obtained iteratively by finding the scale~$\mu$ that satisfies $\mu = C_\mathrm{F}m_\mathrm{t}\alphas(\mu)$~\cite{Fabiano:1994cz,Beneke:2005hg,Kats:2009bv}. 
Such a state has thus a Bohr radius of order $r_\text{Bohr} = (C_\mathrm{F}/2\,\overline{\alpha}_s m_\mathrm{t})^{-1} \approx 0.01$~fm. 
Toponium is extremely short-lived and its revolution time of $t \approx r_\text{Bohr} \approx 0.01$~fm~\cite{Bigi:1986jk} is of the same order of magnitude as its lifetime driven by the electroweak decay of any of its constituent quarks $\rm t,\,\overline{\rm t}\to W^+b,\,W^-\overline{b}$. The $\eta_\mathrm{t}$ decay width is thus very large, of the order of $\Gamma(\eta_\mathrm{t}) \approx 2 \Gamma_\mathrm{t} = 2.66$~GeV, using the next-to-next-to-leading-order value $\Gamma_\mathrm{t} = 1.33$~GeV of the top-quark width~\cite{Chen:2022wit}. The $\etatOneS$ diphoton width can be obtained from the standard analytic expressions for heavy-quarkonium diphoton decays~\cite{Kwong:1987ak}, and at next-to-leading-order (NLO) accuracy reads
\begin{equation}
\Gamma(\etatOneS\to\gaga) = \frac{12 q_\mathrm{t}^4 \alpha^2}{m_{\eta_\mathrm{t}}^2}|R_0(0)|^2\left[1+\frac{\alphas}{\pi}\left(\frac{\pi^2}{3}-\frac{20}{3}\right)\right],
\end{equation}
where $|R_0(0)|^2 = 4n^3/r_\text{Bohr}^3 = 4 (C_\mathrm{F}/2\,\overline{\alpha}_s m_\mathrm{t})^{3}$ is the wavefunction at the origin, $q_\mathrm{t}=2/3$ the top-quark electric charge, and $\alphas = \alphas(m_\mathrm{t})=0.096$, yielding: 
$\Gamma(\etatOneS\to\gaga) = 22.6$~keV. The diphoton widths of higher $\etatnS$ states amount to $\Gamma(\etatnS\to\gaga) = \Gamma(\etatOneS\to\gaga)/n^3$, which implies $\Gamma(\etatTwoS\to\gaga) = 2.82$~keV for the 2S state. Since toponium is a very broad pseudoresonance, it is unlikely that one can experimentally separate the different $n$S para-states, and it is more realistic to consider the photon-photon production for the sum of all of them combined, which amounts to
\begin{equation}
\hspace{-0.2cm}\Gamma(\etatnS\to\gaga) = 
\zeta(3)\frac{12 q_\mathrm{t}^4 \alpha^2}{m_{\eta_\mathrm{t}}^2}|R_0(0)|^2\left[1+\frac{\alphas}{\pi}\left(\frac{\pi^2}{3}-\frac{20}{3}\right)\right],\mbox{ with }
\zeta(3) = \sum_{n=1}^{\infty} \frac{1}{n^3} = 1.20205\mbox{ (Apéry's constant)}.
\end{equation}

\begin{figure}[htpb!]
    \centering
    \includegraphics[width=0.6\linewidth]{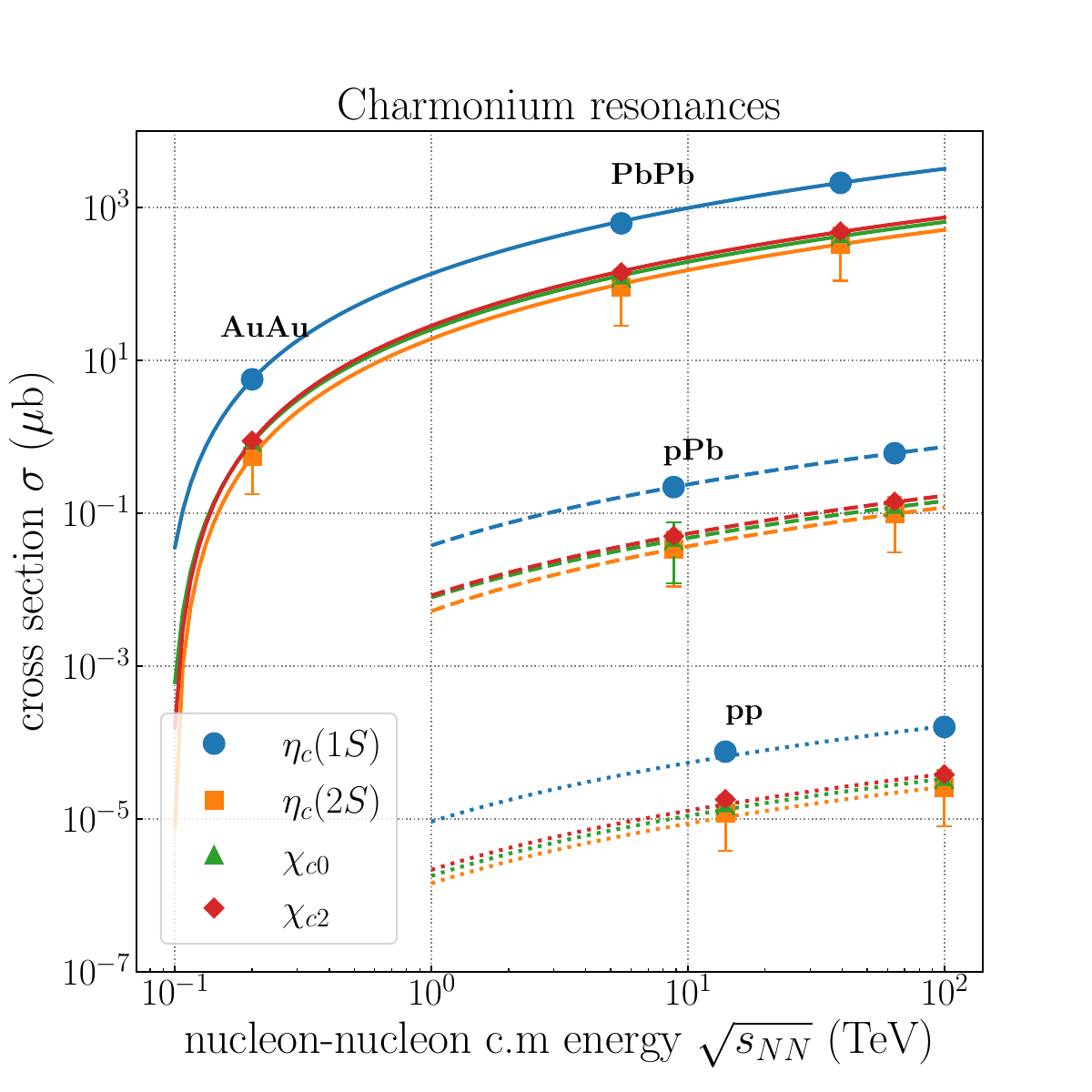}
    \caption{Cross sections for the $\gaga$ production of even-spin charmonium mesons as a function of nucleon-nucleon \cm\ energy $\sqrtsnn$, in \PbPb\ or \AuAu\ (solid curves), \pPb\ (dashed curves), and \pp\ (dotted curves) UPCs. The curves are $\ln ^3(\snn)$ fits to guide the eye.}
    \label{fig:sigma_vs_sqrts_ccbar}
\end{figure}

\begin{figure}[htpb!]
    \centering
    \includegraphics[width=.6\linewidth]{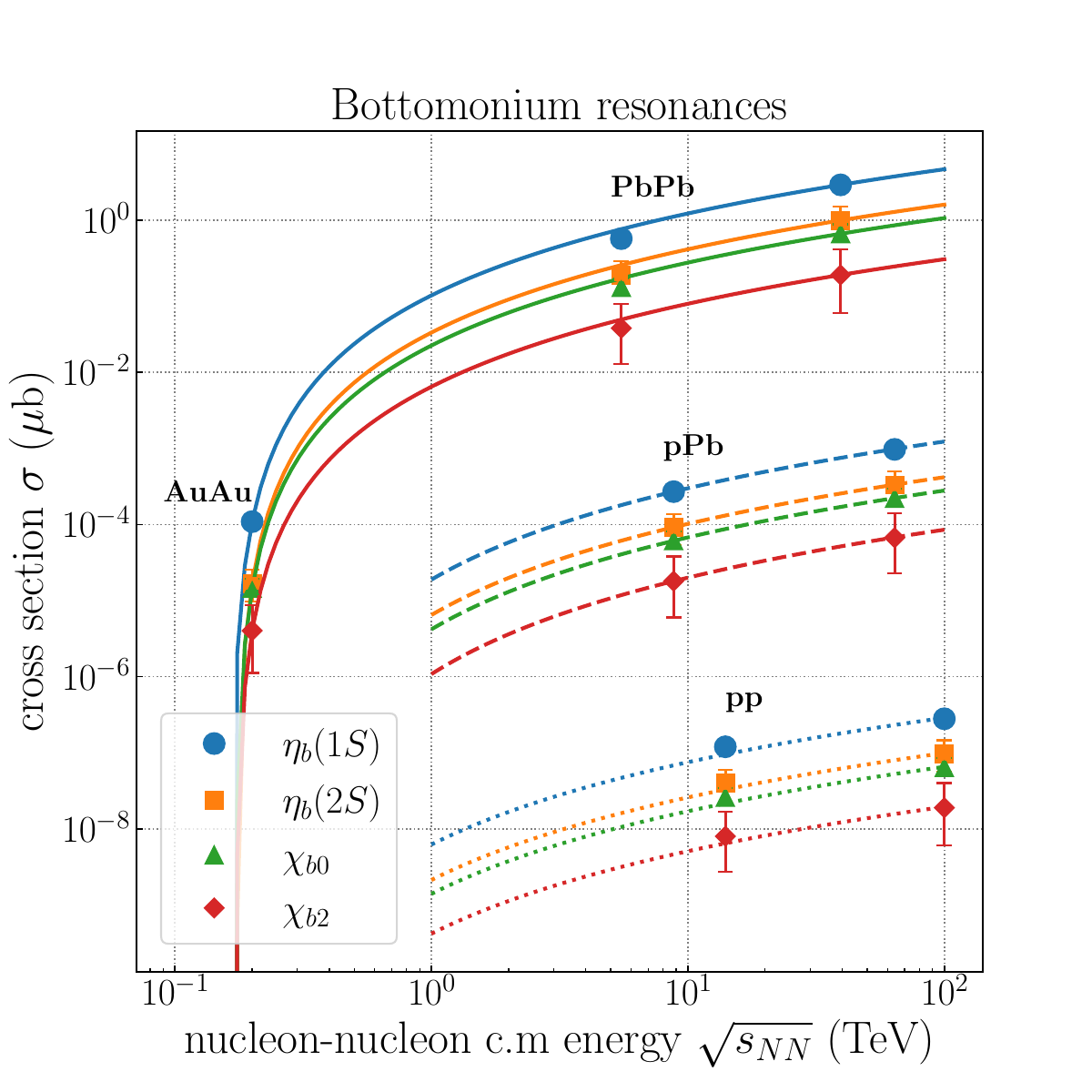}
    \caption{Cross sections for the $\gaga$ production of even-spin bottomonium mesons as a function of nucleon-nucleon \cm\ energy $\sqrtsnn$, in \PbPb\ or \AuAu\ (solid curves), \pPb\ (dashed curves), and \pp\ (dotted curves) UPCs. The curves are $\ln ^3(\snn)$ fits to guide the eye.}
    \label{fig:sigma_vs_sqrts_bbbar}
\end{figure}

\begin{figure}[htpb!]
     \centering
     \includegraphics[width=.6\linewidth]{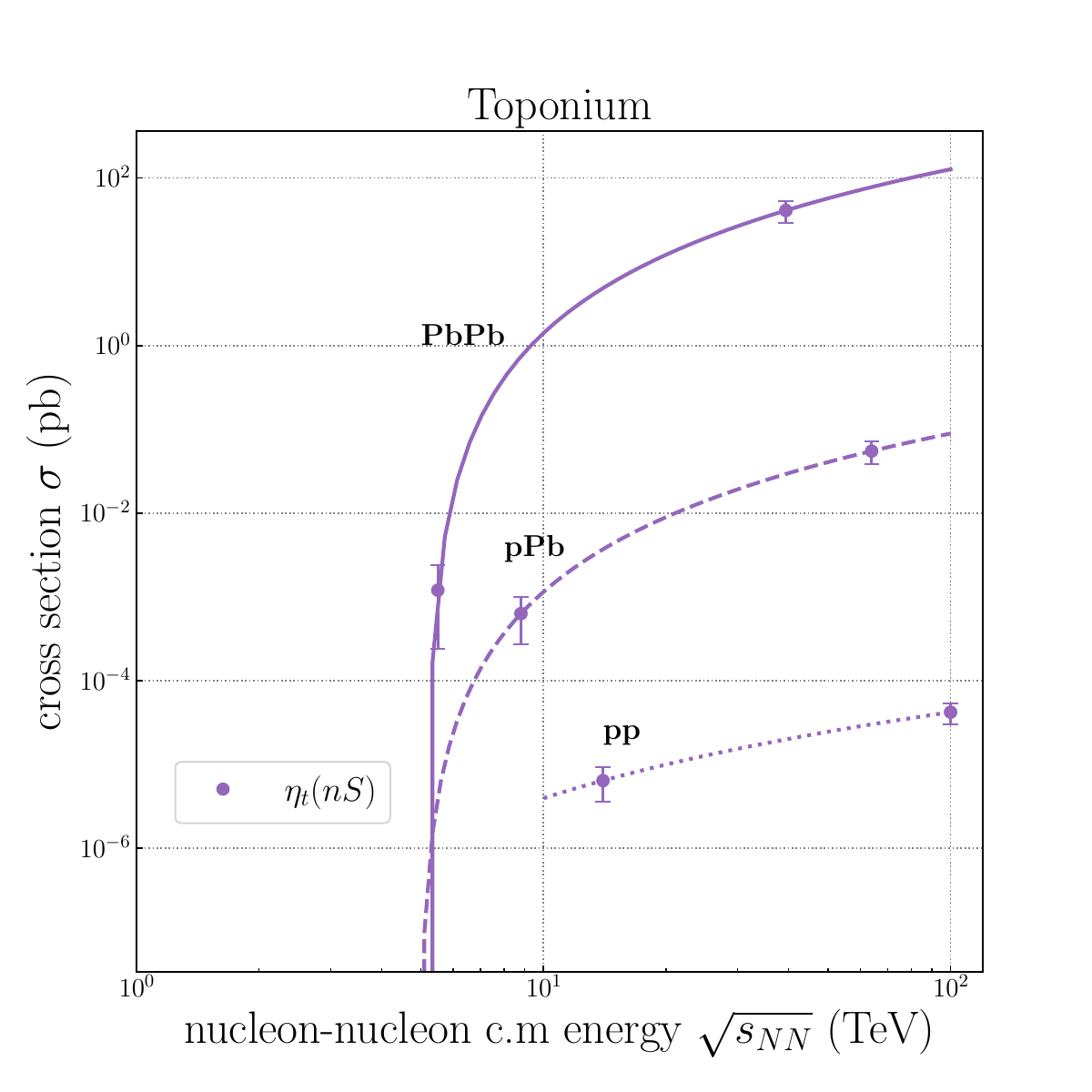}
     \caption{Cross sections for the $\gaga$ production of toponium $\etatnS$ as a function of nucleon-nucleon \cm\ energy $\sqrtsnn$, in \PbPb\ (solid curves), \pPb\ (dashed curves), and \pp\ (dotted curves) UPCs. The curves are $\ln ^3(\snn)$ fits to guide the eye.}
     \label{fig:sigma_vs_sqrts_top}
\end{figure}

Recent alternative calculations exist of the toponium diphoton width~\cite{Jiang:2024fyw,Wang:2024hzd} that, for comparable theoretical setups, give similar results within a factor of about 30\%, which we assign as theoretical uncertainty here.
Based on the diphoton widths of Table~\ref{tab:heavy_onia} and on Eq.~(\ref{eq:sigma_X_master}), we provide the theoretical predictions for the photon-fusion cross sections for even-spin charmonium, bottomonium, and toponium resonances produced in UPCs at various \cm\ energies in Tables~\ref{tab:sigma_heavy_charmonia} and~\ref{tab:sigma_heavy_bb_tt_onia} in the Appendix. Uncertainties in the cross sections are dominated by the propagated uncertainty of the corresponding $\Gamma_{\gaga}$ widths and vary between 5\% and 100\%. The cross sections are compared to a few previous estimates, with differences appearing mostly due to updated values of a few heavy-quarkonium diphoton widths. The photon-fusion cross sections for the different charmonium, bottomonium, and toponium states produced in UPCs at RHIC, LHC, and FCC-hh are plotted as a function of $\sqrtsnn$ in Figs.~\ref{fig:sigma_vs_sqrts_ccbar},~\ref{fig:sigma_vs_sqrts_bbbar}, and~\ref{fig:sigma_vs_sqrts_top}, respectively. 
The tables provide also the number of heavy-quarkonium events expected for the considered integrated luminosities at each facility, so as to assess the feasibility of their potential experimental measurement. In \PbPb\ UPCs at the LHC, we expect hundreds to thousands events with even-spin charmonium resonances produced exclusively that decay back into a pair of photons. The measurement of exclusive charmonia in this decay mode (or in their much more abundant hadronic decays) appears therefore feasible for the ALICE and LHCb detectors (on top of the light-by-light continuum, see Section~\ref{sec:LbL}). The experimental perspectives for bottomonium appear more challenging, though. First, one can see that the exclusive bottomonium production cross sections are negligible at RHIC, because the effective two-photon luminosities are too low above $m_\mathrm{X}\approx 4$~GeV at this collider (Fig.~\ref{fig:dLdWUPC}). The number of exclusive bottomonia produced in UPCs at the LHC are in the hundreds to thousands events (depending on the system and concrete particle) and their potential measurement would only be feasible in their dominant hadronic decays (although no concrete exclusive hadronic final states have been measured in their inclusive $\etabOneS, \etabTwoS\to \rm gg$ decays, yet) or in their quarkonium$\,+\,$photon decays (for $\chibZero, \chibTwo\to\Upsilon+\gamma$), as their diphoton branching fractions are too small. 

Last but not least, the rightmost row of Table~\ref{tab:sigma_heavy_bb_tt_onia} gives the photon-fusion cross sections for $\etatnS$ para-toponium, which are the smallest ones considered in this study given the very high mass of this object. The cross section for this system has large uncertainties at the LHC (see Fig.~\ref{fig:sigma_vs_sqrts_top}), because at such high masses the charged-form-factor photon fluxes are highly oscillating~\cite{Shao:2022cly} and their integration is not fully reliable. Although the largest $\gaga\to\etatnS$ cross sections (tens of pb) are reached in \PbPb\ UPCs at the FCC-hh, the beam luminosities are too low for any meaningful number of events to be produced. The only system where the para-toponium measurement can be attempted is in \pp\ collisions at the LHC (with about 40 events expected) and FCC-hh (about 1300 events to be produced) by exploiting the full dataset of 6 and 30~ab$^{-1}$ integrated luminosities to be collected, respectively, under high-pileup conditions. An observation of the $\gaga$ production of toponium could be realized in \pp\ collisions by searching for a back-to-back $\ttbar$ pair produced at rest (\ie\ with zero pair $\pT$) in coincidence with two intact protons reconstructed in very forward proton spectrometers, such as those from the CMS-TOTEM PPS system~\cite{CMS:2022hly}, whose acceptance for photon-fusion systems is optimal in the toponium mass range, $m_{\gaga}\gtrsim 350$~GeV~\cite{CMS:2018uvs,ATLAS:2020mve,CMS:2023naq}. In this case, one will have to deal with a background from the $\gaga\to\ttbar$ continuum at around threshold $m_{\ttbar} = 2m_\text{top}$, which has a cross section larger than the toponium one by about a factor of 50, $\sigma(\gaga\to\ttbar)\approx 300$~ab at the LHC~\cite{dEnterria:2009cwl,Shao:2022cly}, though over all $\ttbar$ pair masses.

\subsection{Production of exotic hadrons}
\label{sec:multiquark}

In principle, QCD permits the existence of exotic types of hadrons  ---such as multiquark (tetra-, penta-, hexa-quarks), glueballs, and hybrids ($\qqbar$ states with ``valence'' gluons) states--- that are beyond the conventional meson and baryon structure in the constituent quark model~\cite{Amsler:2004ps,Klempt:2007cp}. The discovery of the first tetraquark candidate, $\chicOne(3872)$, in 2003~\cite{Belle:2003nnu} triggered a renewed interest in hadronic spectroscopy, and multiple types of new exotic hadronic states have been observed in the last years at the LHC and B-factories~\cite{Brambilla:2019esw,Johnson:2024omq,Husken:2024rdk}. Any new exotic multiquark hadron with even spin can be produced via photon fusion provided its diphoton width is not too small. 
Since the actual existence of many of such states (with the PDG often omitting them from the summary tables unless they are confirmed by more than one experiment), as well as their exact nature (a compact multiquark system, or a hadronic molecule consisting of charge-conjugated pairs of mesons) and spectroscopic properties are often not precisely known, the study of their production via photon-photon fusion in UPCs can help confirm their quantum numbers and/or determine their diphotons widths, among others. In this section, we extend previous studies~\cite{Moreira:2016ciu,Goncalves:2018hiw, Goncalves:2021ytq, Esposito:2021ptx, Niu:2022cug, Biloshytskyi:2022dmo, Fariello:2023uvh} to consider all presently known even-spin exotic hadron states (Table~\ref{tab:multiquark}), and compute their cross sections for a large variety of colliding systems. In most of the cases, their exact spin (0 or 2) state, as well as their diphoton decays, remain experimentally unsettled. For their diphoton partial widths, we use the theoretical results of~\cite{Moreira:2016ciu} or~\cite{Esposito:2021ptx} and, given their large model dependencies, we only quote an approximate value for them. Lighter even-spin candidate exotic states (with $m_\mathrm{X}\lesssim 2.3$~GeV) are listed in Tables~\ref{tab:light_mesons} and~\ref{tab:light_mesons_unknown} discussed in Section~\ref{sec:light_onia}. 
 
\begin{table}[htpb!]
\tabcolsep=3.5mm
\centering
\caption{List of even-spin exotic hadrons producible via two-photon fusion. For each particle, we quote its $\rm J^{PC}$ quantum numbers, mass $m_\mathrm{X}$, total width $\Gamma_\mathrm{tot}$, and diphoton partial width $\Gamma_{\gaga}$ from measurements or theoretical predictions. The middle line separates light- from heavy-quark states. 
\label{tab:multiquark}}
\vspace{0.2cm}
\begin{tabular}{lc cccc}
\toprule
Resonance & $\rm J^{PC}$ & $m_\mathrm{X}$ (MeV) & $\Gamma_\mathrm{tot}$ (MeV) & $\Gamma_{\gaga}$ (eV) & Decay(s); comment \\ 
\midrule
$\rm X(2370)$ & $0^{-+}$ & $2377 \pm 9$ & $148_{-28}^{+80}$ & unknown & $\rm K\overline{K}\eta', \pi\pi\eta'$; glueball candidate~\cite{BESIII:2023wfi}\\
$\rm T^{*}_{cs0}(2900)$ & $2^{++}$ & $2892 \pm 14 \pm 15$ & $119 \pm 26 \pm 13$ & unknown & $\rm D_s\pi$; tetraquark candidate~\cite{LHCb:2020bls,LHCb:2020pxc}\\
\hline
$\chicZero(3860)$ & $0^{++}$ & $\approx 3862$ & $\approx 200$ & unknown & $\rm D^+D^-,D^0\overbar{D}^0$ \\
$\chicZero(3915)$ & $0^{++}/2^{++}$ & $3922.1 \pm 1.8$ & $20 \pm 4$ & $\approx 200$ & $\rm D^+D^-,\jpsi\omega$ \\
$\chicTwo(3930)$ & $2^{++}$ & $3922.5 \pm 1.0$ & $35.2 \pm 2.2$ & $\approx 80$ & $\rm D^+D^-,\jpsi\omega$ \\

$\rm X(3940)$ & $0^{++}/2^{++}$ & $3942_{-6}^{+7} \pm 6$ & $37_{-15}^{+26} \pm 8$ & $\approx 300$ & $\mathcal{B}(\rm D\overbar{D}^*+\mathrm{c.c.})>0.45$ \\

\multirow{2}{*}{$\rm X_0(4140)$} & $0^{++}$ & \multirow{2}{*}{$4146.8 \pm 2.5$} & \multirow{2}{*}{$19^{+8}_{-7}$} & $\approx 630$ & \multirow{2}{*}{$\rm cs\bar{c}\bar{s}$ tetraquark candidate}\\
 & $2^{++}$ & & & $\approx 500$ & \\ 

$\rm X(4350)$ & $0^{++}/2^{++}$ & $4350.6_{-5.1}^{+4.6} \pm 0.7$ & $13_{-9}^{+18} \pm 4$ & unknown & $\jpsi\phi, \gaga$ \\

$\chicZero(4500)$ & $0^{++}$ & $4474 \pm 3 \pm 3$ & $77 \pm 6^{+10}_{-8}$ & unknown & $\jpsi\phi$ \\
$\chicZero(4700)$ & $0^{++}$ & $4694 \pm 4^{+16}_{-3}$ & $87 \pm 8^{+16}_{-6}$ & unknown & $\jpsi\phi$ \\

$\rm T_{\twoccbar}(6900)$ & $0^{++}/2^{++}$ & $6899 \pm 12$ & $153 \pm 29$ & $\approx 100$~\cite{Esposito:2021ptx} & $\jpsi\jpsi$, tetra-charm candidate\\
\hline
\end{tabular}
\end{table}

\begin{figure}[htpb!]
    \centering
    \includegraphics[width=.6\linewidth]{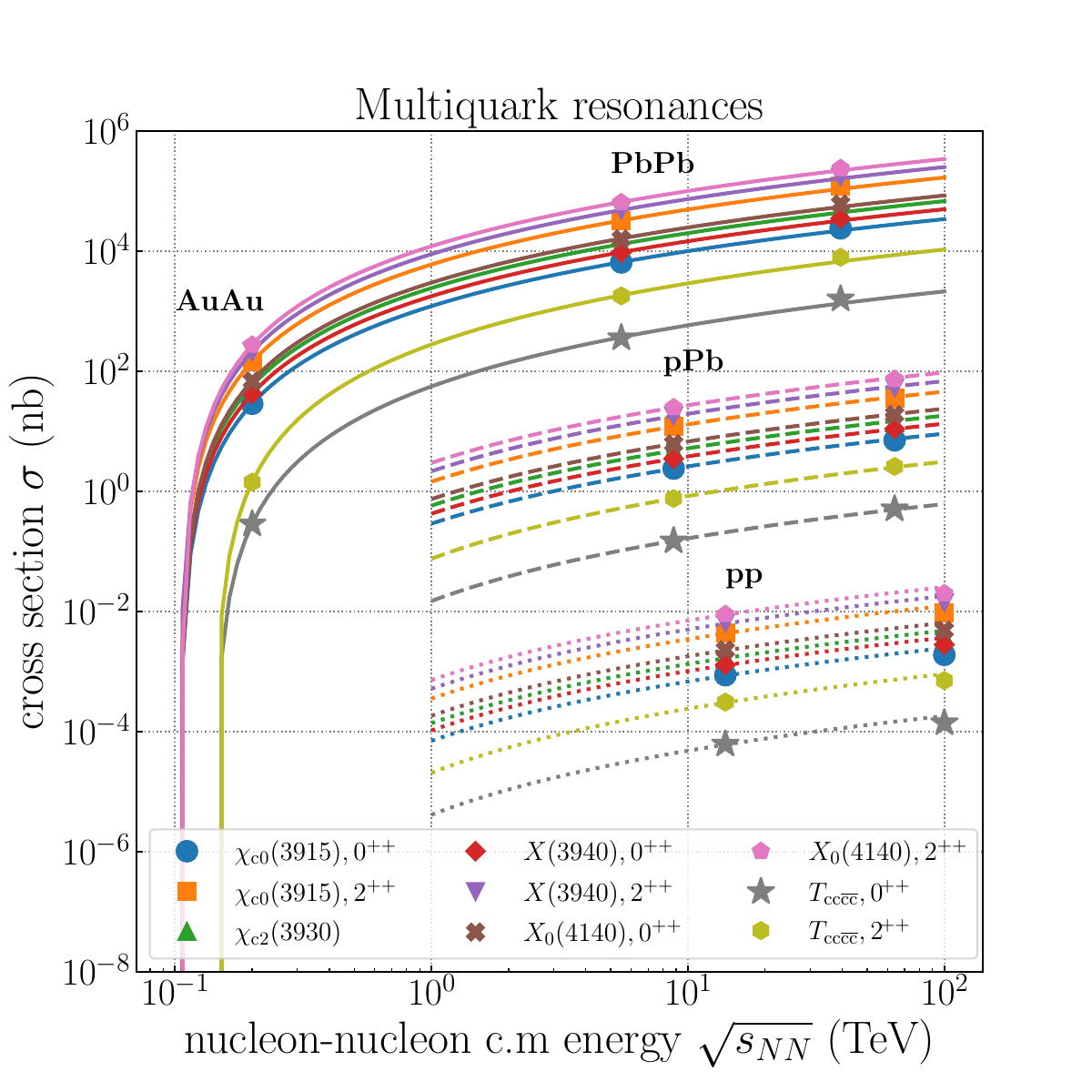}
    \caption{Cross sections for the $\gaga$ production of even-spin exotic hadrons as a function of nucleon-nucleon \cm\ energy $\sqrtsnn$, in \PbPb\ or \AuAu\ (solid curves), \pPb\ (dashed curves), and \pp\ (dotted curves) UPCs. The curves are $\ln ^3(\snn)$ fits to guide the eye. 
    \label{fig:sigma_vs_sqrts_exotic}}
\end{figure}

The calculated photon-fusion cross-sections of exotic hadrons are listed in Table~\ref{tab:sigma_multiquark}, and shown in graphical form as a function of collision energy in Fig.~\ref{fig:sigma_vs_sqrts_exotic}. The expected inclusive yields for such exotic states are relatively large, assuming that their estimated diphoton widths are correct, but their observation via diphoton decays appears unfeasible and experimental searches should be carried out in hadronic final states instead ($\rm D\overline{D}$, $\jpsi\omega$, $\jpsi\phi$,...). By searching for their $\gaga$ production in UPCs, their yields can help ascertain whether their spin is 0 or 2, as in the latter case they will be comparatively enhanced by a factor of 5 as per Eq.~(\ref{eq:sigma_X_master}). Our cross section predictions are in general consistent with those of previous works, if available, except for the $\rm T_{c c \overline{c c}}(6900)$ results of Ref.~\cite{Fariello:2023uvh} that use $\Gamma_{\gaga} = 67/45$~keV (with/without interference), which is 500 times larger than the diphoton decay width that we have adopted here. It is also worth noting that all cross sections quoted from~\cite{Goncalves:2018hiw} and~\cite{Goncalves:2021ytq} are total cross sections for the $\rm \gaga\to\chicZero(3915)\to\jpsi\omega$ and $\rm \gaga\to X(6900)\to\jpsi\jpsi$ production, respectively.

\clearpage
\section{Photon-fusion production of leptonium states}
\label{sec:leptonium}

Leptons with opposite charges ($\ellell$), such as electrons and positrons ($e^{\pm}$), muons ($\mu^{\pm}$), and tau particles ($\tau^{\pm}$), can temporarily pair up to form onium states under their QED interaction. The smallest of these pairs, known as positronium (a bound state of an electron and a positron), was first identified over 75 years ago~\cite{Deutsch:1951zza}, but its heavier siblings involving muon and tau particles (often called, respectively, dimuonium and ditauonium) have not yet been observed. The photon-photon production of leptonium states in UPCs has been considered several times in the literature~\cite{Ginzburg:1998df,Kotkin:1998hu,Baur:2001jj,Azevedo:2019hqp, dEnterria:2022ysg,Shao:2022cly,Francener:2021wzx,Dai:2024imb}. Here, we recall the basic properties of leptonium states, and present results for the production of all three species in UPCs at current and future hadron colliders.

The properties of the pure-QED leptonium systems can be straightforwardly derived from the expressions of Section~\ref{sec:QED_onium}. 
Here, we are interested in the para-leptonium systems, spin-singlet configurations where the lepton pairs have their spins aligned oppositely and are characterized by the quantum numbers $\rm J^{PC} = 0^{-+}$, which can thus be produced in two-photon collisions.  Excited tensor ($\rm J=2$) leptonium states can also be produced, but their rates are suppressed by a factor of $\rm {\sim}(2J+1)\cdot\alpha^2$ compared to the paraleptonium ones. At leading order (LO), the diphoton width of the para-leptonia ground state \(1^1S_0\) can be obtained from Eq.~(\ref{eq:Gamma_gaga_fermion}) and amounts to:
\begin{equation}
\Gamma_\mathrm{\pleptonium\to\gaga} = \frac{\alpha^5 m_\ell}{2}.
\end{equation}
The formula for the corresponding photon-photon cross section in UPCs of hadrons A and B can be obtained by plugging this expression into Eq.~(\ref{eq:sigma_X_master}). If one neglects the tiny binding, Lamb, and Breit mass corrections (Section~\ref{sec:QED_onium}), the para-leptonium production cross sections in UPCs can be simplified as,
\begin{equation}
\sigma(\mathrm{A}~\mathrm{B} \xrightarrow{\gaga}~\mathrm{A}~\pleptonium~\mathrm{B})=\left. \pi^2(2 J+1) \frac{\alpha^5}{m_{\ell}} \frac{\mathrm{d} \mathcal{L}_{\gaga}^{(\mathrm{AB})}}{\mathrm{d}m_{\gaga}}\right|_{m_{\gaga}=m_{\pleptonium}},
\label{eq:sigma_aa_leptonium}
\end{equation}
which are proportional to the ratio of fifth-power of the QED coupling over the lepton mass. For the determination of the UPC leptonium cross sections, we will not however use the LO expressions above, presented here for illustration purposes, but Eq.~(\ref{eq:sigma_X_master}) with the leptonium masses and diphoton widths computed with the highest theoretical accuracy known today.

\begin{figure}[htpb!]
\centering
\includegraphics[width = \textwidth]{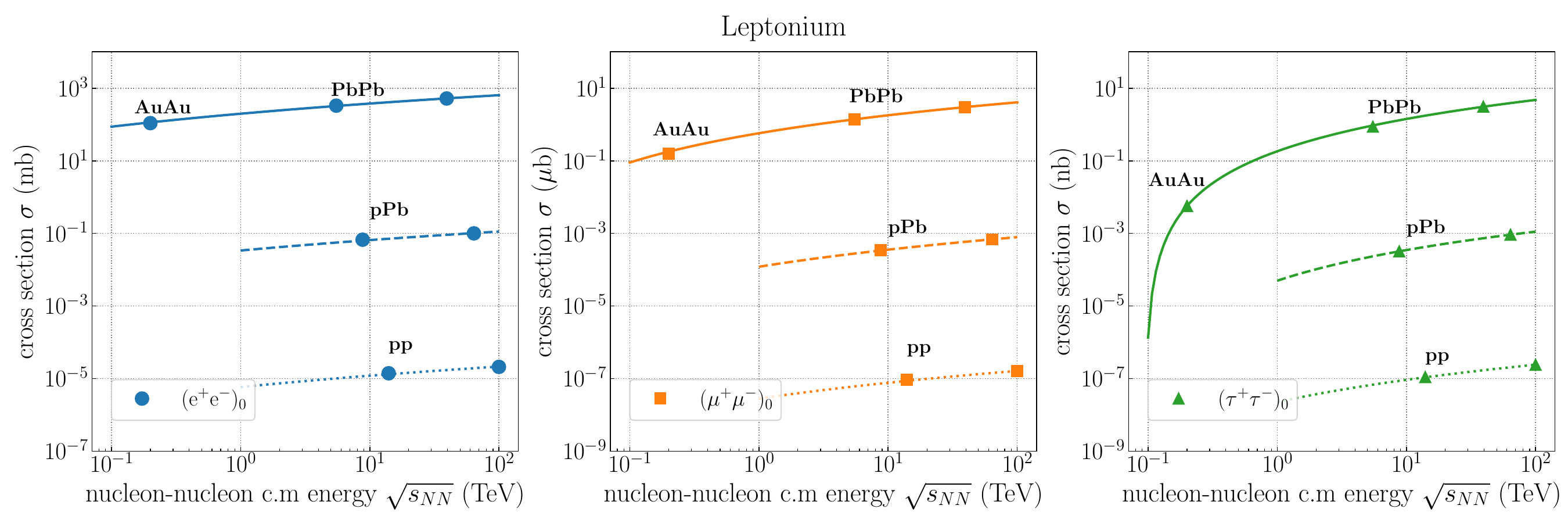}
\caption{Cross sections of positronium (left), dimuonium (center), and ditauonium (right) para-states in photon-photon fusion for various UPCs calculated in this work as a function of nucleon-nucleon \cm\ energy.\label{fig:leptonium}}
\end{figure}

In Table~\ref{tab:paralept}, the basic properties of the para-states of positronium~\cite{Czarnecki:1999mt, Czarnecki:1999uk,Kniehl:2000dh,Melnikov:2000fi}, dimuonium~\cite{Jentschura:1998vkm, Brodsky:2009gx}, and ditauonium~\cite{dEnterria:2022alo,dEnterria:2023yao} are collected. Based on the diphoton widths of each object, their corresponding $\gaga$ cross sections are 
listed in Table~\ref{tab:lep_xsection} in Appendix~\ref{app:leptonium} and plotted as a function of collision energy in Fig.~\ref{fig:leptonium}. The results computed here are in general in agreement with previous estimates, if existing, except for the \AuAu\ at $\sqrts = 0.2$~TeV case, where our calculated value for para-ditauonium, $\left(\tau^{+} \tau^{-}\right)_0$, falls between the values computed in Refs.~\cite{Dai:2024imb,Francener:2021wzx}. The production cross sections, and associated yields, are very large for positronium and dimuonium, whereas they are very small for the heaviest true-tauonium system. However, the observation of the production of any of the three para-leptonium ground states in UPCs appears unfeasible or very challenging. On the one hand, the fact that positronium is extremely light (leading to a pair of ultrasoft, 0.5~MeV, decay photons) and that the $(\tautau)_0\to\gaga$ decay fully overlaps with the much more probable $\chicTwo\to\gaga$ final state~\cite{dEnterria:2022ysg}, precludes the experimental observation of both leptonium states. On the other hand, dimuonium could only be observed if the experimental detectors are able to reconstruct two soft decay photons with $\mathcal{O}(100$ MeV) transverse momentum, given that the $(\mumu)_0$ production via quasireal photon-fusion leads to very small transverse boosts.



\begin{table}[htbp]
\tabcolsep=1.4mm
\centering
\caption{Main properties of para-leptonium ground states $\pleptonium$: quantum numbers $\rm J^{PC}$, mass $m_{\pleptonium}$, binding energy $E_{n=1}$, Bohr radius $r_\text{Bohr}$, lifetime $(\tau)$, total width $\Gamma_\mathrm{tot}$, diphoton width $\Gamma_{\gaga}$ from Eq.~(\ref{eq:Gamma_gaga_fermion}), and dominant decay branching ratio.
\label{tab:paralept}}
\vspace{0.2cm}
\begin{tabular}{lc ccccccc}
\toprule
$\pleptonium$ state & $\rm J^{PC}$ & $m_\mathrm{X}$ (MeV) & $E_{n=1}$ (keV)& $r_\text{Bohr}$ (fm)& $\tau$~(fs) & $\Gamma_\mathrm{tot}$ (meV) & $\Gamma_{\gaga}$ (meV) & Dominant decay ($\BR$) \\
\midrule
$\ppositronium$ & $0^{-+}$ & 1.02991& $-6.859\cdot 10^{-3}$ & $106 \cdot 10^3$ & $125.2\cdot 10^3$ & $5.257 \cdot 10^{-3}$ & $5.257 \cdot 10^{-3}$ 
& $\gaga$ (${\approx}100\%$) \\ 
$\pdimuonium$ & $0^{-+}$ & $211.316$ & $-1.407$ & 512 & 595.4 
& $1.105$ & 1.105 
& $\gaga$ (${\approx}100\%$) \\ 
$\pditauonium$ & $0^{-+}$ & $3553.6962 \pm 0.240$ & $-23.655$ & 30.4& 27.60 & 23.84 & 18.533 & $\gaga$ (77.72\%) \\ 
\bottomrule
\end{tabular}
\end{table}

\section{Photon-fusion production of QED hadronium states}
\label{sec:hadronium}

In this section, we discuss the photon-fusion production of even-spin systems formed by two identical hadrons of opposite charge, bound by their Coulomb interaction, which we refer to as QED ``hadronium''. The hadrons must be charged and long-lived enough so as to be able to form a bound state before decaying individually via the weak interaction (if they decay strongly, they would be too short-lived to bind together). These implies the following list of six charged pseudoscalar mesons: $\rm \pi^\pm, K^\pm, D^\pm, D_s^\pm, B^\pm, B_c^\pm$, 
(whose shortest lifetimes among them are $\tau\approx 500$~fs for the $\rm D_s^\pm$ and $\rm B_c^\pm$ mesons, and $\tau\approx 1000,\,1600$~fs for the $\rm D^\pm$ and $\rm B^\pm$ mesons), plus the following list of eight charged baryons: p, $\Sigma^\pm$, $\Xi^\pm$, $\Omega^\pm$, $\Lambda^\pm_\mathrm{c}$, $\Xi_\mathrm{c}^\pm$, $\Xi_\mathrm{b}^\pm$, $\Omega_\mathrm{b}^\pm$ (where the proton is stable, and the rest of baryons have lifetimes $\tau \approx 10^{-10}$--$10^{-13}$~s). On the one hand, the ``mesonium'' atoms are constituted by pairs of bosons and, therefore, are scalars with $\rm J^{PC} = 0^{++}$ quantum numbers, given that the pair parity P and C-parity combine as $(-1)^l$. On the other, QED-baryonium states are formed by opposite-charge identical fermions, and can be in ortho- or para-states, like leptonium, and we consider only the latter $\rm J^{PC} = 0^{-+}$ pseudoscalar cases, which we denote as $\baryonium$ and are producible in photon-photon collisions. We discuss here the following six QED-mesonium systems: pionium, kaonium, D$_{\mathrm{(s)}}^\pm$-onium, and B$_{\mathrm{(s)}}^\pm$-onium, which we denote by $\rm A_{\rm 2h} \equiv (h^+h^-)$, where `A' stands for an (exotic) atom and `h' its hadronic constituent. To our knowledge, the $\rm \DsDs=(D_\mathrm{s}^+$D$_\mathrm{s}^-)$ and $\rm \BcBc=(B_\mathrm{c}^+$B$_\mathrm{c}^-)$ onium states have not been considered before in the literature, whereas the other dimeson systems have been previously studied theoretically and/or experimentally~\cite{Palfrey:1961kt, Dumbrajs:1985ay,Wycech:1993ci, Afanasev:1993zp, Kerbikov:1995ge, Jallouli:1997ux, Ivanov:1998wx, Labelle:1998gh, Hammer:1999up, Colangelo:2001df, Gasser:2001un, DIRAC:2003kif, Krewald:2003ab, Suebka:2004zi, DIRAC:2005hsg, Zhang:2006ix,  Gasser:2007zt, Gasser:2009wf, Yan:2009zzb, Klevansky:2011hi, DIRAC:2015lmc, Afanasyev:2017xpx, DIRAC:2018xvz,Zhang:2020mpi, Shi:2021hzm}. Among the QED-baryonium atoms, only protonium (also known as antiprotonic hydrogen) has been thoroughly investigated~\cite{Batty:1989gg,Carbonell:1989cs,Augsburger:1999yt,Klempt:2002ap, ATHENA:2006clk,Doser:2022tlg}, whereas the rest of heavier systems have not been experimentally or theoretically studied to our knowledge.

The QED hadronium states are bound predominately by their Coulomb force (photon exchange) and have relatively large Bohr radii, of the order of $r_\mathrm{Bohr}\approx10$--400~fm (Tables~\ref{tab:mesonium} and~\ref{tab:baryonium}), that are (much) larger than the range of the strong interaction $\mathcal{O}(1$~fm). 
They should not be confused with hadronic molecules\footnote{In particular, the term ``baryonium'' is used most often to refer to a baryon-antibaryon (not necessarily charged) system bound by pion exchange.}, such as some of the objects discussed in Section~\ref{sec:multiquark}, which are bound primarily by the strong interaction (gluon or pion exchanges), have much smaller radii, and much shorter lifetimes~\cite{Hanhart:2007wa}. Nonetheless, it may occur that a hadronium state first forms through its long-range QED interaction that subsequently, before decaying, evolves into a shorter-lived QCD-bound hadronic molecule. While their binding is electromagnetic, hadronium states decay mostly through their strong interaction, and their two-photon partial decay width 
is very small, \ie\ their lifetimes are fully dominated by QCD effects. The production of such systems provides an interesting testbed for the study of low-energy hadron-hadron interactions, as modeled by chiral perturbation theory (ChPT), nonperturbative lattice QCD (LQCD) and dispersion relation analysis~\cite{Gasser:2007zt,Gasser:2009wf}. In addition, the understanding of $A_\mathrm{2h}$ atoms can also provide valuable information on loosely bound (molecular-like) tetraquark states~\cite{Zhang:2020mpi}.
Dimeson $\Apipi$ and $\AKK$~\cite{Dumbrajs:1985ay,Wycech:1993ci,Kerbikov:1995ge} atoms can be produced by colliding oppositely charged meson pairs with low relative momentum~\cite{Afanasyev:2017xpx}, %
and dedicated experiments, such as the CERN Proton Synchrotron Dimeson Relativistic Atom Complex (DIRAC)~\cite{DIRAC:2003kif}, have produced the lightest $\Apipi$ atom. Similarly, experiments at the CERN Low Energy Antiproton Ring (LEAR) over the 1982--1996 period studied antiprotonic atoms via nucleon-antinucleon scattering at low energies, and in particular protonium (in states of high angular momenta $l$) via antiproton stopping in liquid hydrogen~\cite{Batty:1989gg,Augsburger:1999yt}.

Tables~\ref{tab:mesonium} and~\ref{tab:baryonium} list the properties of the even-spin mesonium and baryonium QED atoms, respectively, determined from the expressions of Section~\ref{sec:QED_onium}. Single hadron masses are from the latest PDG values~\cite{ParticleDataGroup:2024cfk}. The  diphoton widths $\Gamma_{\gaga}$ are obtained from Eqs.~(\ref{eq:Gamma_gaga_boson}) and (\ref{eq:Gamma_gaga_fermion}) for mesonium and baryonium systems, respectively. The dominant strong-interaction decays of the QED hadronium atoms (listed in the last column) are estimated by simple inspection of the valence quark content of each annihilating pair and requiring the conservation of C and P quantum numbers, although their total width (or, equivalently, lifetime) has not been explicitly computed for about half of the systems\footnote{For those cases, we give an order-of-magnitude for $\Gamma_{\text{had}}$ determined as follows. 
Since the dominant decay occurs via the strong interaction when the hadrons annihilate, 
the hadronic decay width is proportional to their squared wavefunction at the origin multiplied by the annihilation cross section times the relative velocity of the bound hadrons $v\propto \alpha$, \ie\
\begin{equation}
\Gamma_{\text{had}} \approx |\psi(0)|^2 \cdot \mean{\sigma_\text{ann} v} \propto \frac{(\alpha m_\mathrm{h})^3}{m_\mathrm{h}^2} \alpha \propto \alpha^4 m_\mathrm{h},
\end{equation}
where the approximate scaling is derived using Eq.~(\ref{eq:para_WF}), and knowing that $\sigma_\text{ann}\propto 1/m_\mathrm{h}^2$ (because annihilation happens at short distances comparable to the hadron Compton wavelength $\lambda \approx 1/m_\mathrm{h}$). 
Since the hadronic decay width scales as $m_h$, we can approximate the total widths of heavy baryonia from the $\Gamma_{\text{had}}$ values already determined for lighter hadronium systems.}. 

\begin{table}[htbp!]
\tabcolsep=1mm
\centering
\caption{Main properties of QED mesonium states $A_{\rm 2h}$, with $\rm h = \pi^\pm, K^\pm, D^\pm_{\mathrm{(s)}}, B^\pm_{\mathrm{(c)}}$. For each ground state ($n=1$), we list its $\rm J^{PC}$ quantum numbers, constituent hadron mass $m_{\rm h^{\pm}}$, atom mass $m_\mathrm{X}$ from Eq.~(\ref{eq:m_para}), QED binding energy (from $E_{n=1} = 2m_{\rm h}-m_\mathrm{X}$), Bohr radius $r_\text{Bohr}$ from Eq.~(\ref{eq:Rbohr}), lifetime $\tau$ and total width $\Gamma_\mathrm{tot}$ (if known), diphoton width $\Gamma_{\gaga}$ from Eq.~(\ref{eq:Gamma_gaga_boson}), and typical hadronic decays (and branching fraction for the pionium case).
\label{tab:mesonium}}
\vspace{0.2cm}
\resizebox{\textwidth}{!}{
\begin{tabular}{lc cccccccc}
\toprule
Mesonium & $\rm J^{PC}$ & $m_{\rm h^{\pm}}$ (MeV)& $m_\mathrm{X}$ (MeV) & $E_{n=1}$ (keV) & $r_\text{Bohr}$ (fm) & $\tau$ (fs)& $\Gamma_\mathrm{tot}$ (eV) & $\Gamma_{\gaga}$ (meV) & Typical decays ($\BR$)\\
\midrule
$\Apipi$ & $0^{++}$ & $139.57039 \pm 0.00018$ & $279.140 \pm 0.00036$ & $-1.858$ & 387 & $3.15_{-0.26}^{+0.28}$
& $(208.6_{-17.2}^{+18.5}) \cdot 10^{-3}$ & 0.873\footnote{Including higher-order QED and QCD corrections.} & $\pi^0\pi^0$ (99.6\%) \\

$\AKK$ & $0^{++}$ & $493.677 \pm 0.013$ & $987.347 \pm 0.026$ & $-6.576$ 
& 110 & $(2.2 \pm 0.9) \cdot 10^{-3}$ 
& $300\pm120$ & 2.56 & $\rm \pi\pi, \pi^0\eta$ \\ 

$\DD$ & $0^{++}$ & $1869.66 \pm 0.05$ & $3739.32 \pm 0.10$ & $-24.90$ & 28.88 & $(0.36^{+0.28}_{-0.12})\cdot10^{-3}$ & $1800_{-600}^{+1400}$
& 9.67 & $\rm D^0 \overbar{D}^0$ \\ 
$\DsDs$ & $0^{++}$ & $1968.35 \pm 0.14$ & $3936.70 \pm 0.28$ & $-26.20$ & 27.43 & $\mathcal{O}(10^{-4})$ & $\mathcal{O}(2000)$ & $10.18$ & $\rm \eta\eta$ \\ 
$\BB$ & $0^{++}$ & $5279.34 \pm 0.12$ & $10558.68 \pm 0.24$ & $-70.28$ & 10.23 & $\mathcal{O}(10^{-4})$ & $\mathcal{O}(5000)$ & $27.31$ & $\rm \pi\pi, \eta\eta$  \\ 
$\BcBc$ & $0^{++}$ & $6274.47 \pm 0.32$ & $12548.94 \pm 0.64$ & $-83.53$ & 8.60 & $\mathcal{O}(10^{-5})$ & $\mathcal{O}(6000)$ & $32.46$ & $\rm D\overline{D}$ \\ 
\bottomrule
\end{tabular}}
\end{table}

The first key observation is that the hadronium Bohr radii, derived using Eq.~(\ref{eq:Rbohr}) with reduced $\mu = m_\mathrm{X}/2$ mass for symmetric states, is (much) larger than the strong interaction range. For this reason, QCD effects do not change drastically the structure of the lightest hadronium bound-state spectra although they shift and broaden the purely QED energy levels. Both effects can be related to the S-wave hadron-hadron scattering length $a_0$, a quantity of fundamental importance in studies of low-energy QCD interactions. 
The binding energies and diphoton widths quoted in Tables~\ref{tab:mesonium} and~\ref{tab:baryonium} are derived from the QED expressions alone (unless otherwise stated, as for the pionium case).
Modifications of the binding energy of 1S hadronium atoms due to QCD effects can be estimated with the formula~\cite{Trueman:1961zza}
\begin{equation}
    \Delta E_{n=1}^\text{QCD} \approx -4\frac{|a_0|}{r_\text{Bohr}}E_{n=1}\,,
\end{equation}
which is approximately valid for $|a_0|\ll r_\text{Bohr}$. Estimates of the QCD-induced modifications of the hadronium binding energies amount to a few percent for pionium and kaonium~\cite{Krewald:2003ab,Yan:2009zzb}, protonium~\cite{Batty:1989gg,Augsburger:1999yt,Klempt:2002ap} and D-onium~\cite{Shi:2021hzm}, although they should arguably be larger for the heaviest charm and bottom hadronium atoms considered here. 

The lightest system of Table~\ref{tab:mesonium} is pionium, discovered in 1993 at the 70-GeV Serpukhov proton-synchrotron in proton collisions on a Ta target~\cite{Afanasev:1993zp}, and further studied at the CERN DIRAC experiment~\cite{DIRAC:2003kif, DIRAC:2005hsg, DIRAC:2018xvz}. Its lifetime, predicted to be $\tau=(2.90 \pm 0.10)$~fs by ChPT~\cite{Colangelo:2001df}, has been experimentally determined with an 8\% precision ($\tau = 3.15_{-0.26}^{+0.28}$~fs) through a measurement of the S-wave $\pi\pi$ scattering length difference~\cite{DIRAC:2018xvz}. Pionium decays mostly into a pair of its lighter neutral counterparts via the $\pi^+\pi^- \to\pi^0\pi^0$ charge-exchange process~\cite{Palfrey:1961kt,Jallouli:1997ux,Ivanov:1998wx, Gasser:2001un} and, to a much lesser extent, into two photons with a 0.36\% branching ratio. For the latter partial decay width, the pure Coulomb width of $\Gamma_{\gaga}=0.722$~meV from Eq.~(\ref{eq:Gamma_gaga_boson}), is increased by about 21\% to $0.722\cdot[1+0.174+0.033] = 0.873$~meV by including higher-order QED (vacuum polarization)~\cite{Hammer:1999up} and chiral expansion~\cite{Gasser:2007zt} QCD corrections that are of $\mathcal{O}(1+0.132+0.004)$ and $\mathcal{O}(1+0.042+0.029)$ sizes, respectively.
For the kaonium atom, contrary to the pionium case where charge exchange dominates, the $\rm \AKK\to K^0\overbar{K}^0$ decay is forbidden because the neutral kaon is heavier than the charged one. Thus, the principal strong decay modes are $\AKK \to\pipifree, \pi^0\eta$ that proceed via strange quark annihilation. The kaonium lifetime has been calculated under different assumptions~\cite{Wycech:1993ci,Krewald:2003ab, Zhang:2006ix,Klevansky:2011hi} and found to be 3 orders of magnitude smaller than the pionium one. In Table~\ref{tab:mesonium}, we use the value $\tau = (2.2 \pm 0.9) \cdot 10^{-3}$~fs (and associated total width $\Gamma = 1/\tau\approx 300$~eV) from Ref.~\cite{Klevansky:2011hi}. 

The lightest heavy-quark QED mesonium system is the $\DD$ bound state, which decays dominantly via $\rm \DD \to D^0 \overbar{D}^0$ charge exchange thanks to the neutral D being slightly lighter than its charged counterparts. Searches for such an exotic atom have been proposed in the $\rm D^0 \overbar{D}^0$ invariant mass distribution in high energy interactions~\cite{Shi:2021hzm} (where this exotic atom is dubbed ``dionium'').
For its total decay width, we adopt the $\Gamma_{\mathrm{tot}} = \Gamma_\mathrm{D^0 \overbar{D}^0}$ estimate derived in Ref.~\cite{Shi:2021hzm} using lattice inputs for the $\rm D \overline{D}$ strong interactions~\cite{Prelovsek:2020eiw}. Its heavier sibling, the $\DsDs$ system, decays predominantly into $\eta\eta$ with charm-anticharm quark annihilation. The $\DsDs$ total width could be theoretically derived following the same approach used by Ref.~\cite{Shi:2021hzm} plus the LQCD calculations of Ref.~\cite{Prelovsek:2020eiw}, but such an exercise goes beyond the scope of this work. At variance with the $\DD$ case, the $\BB$ mesonium atom cannot decay via charge exchange as the $\rm B^0$ mesons are slightly heavier than the charged ones. Its dominant decays are $\BB\to \pi\pi,\;\eta\eta$ following bottom-antibottom annihilation. The heaviest QED mesonium system is $\BcBc$, which decays more dominantly into $\rm D \overline{D}$ pairs also after bottom-antibottom quark annihilation.

We now turn to the QED baryonium systems listed in Table~\ref{tab:baryonium}, of which only protonium has been studied in detail, starting in the 1990s at the CERN LEAR through antiproton stopping in hydrogen, followed by atomic cascade of the highly excited states via X-rays emission, and final \ppbar\ annihilation and production of mesons. More recently\footnote{Also, the BES~III Collaboration has observed a pseudoscalar meson X(1880) in the mass spectrum of the decay of charmonium into a photon plus three pairs of charged pions $\jpsi\to\gamma3(\pi^+\pi^-)$~\cite{BESIII:2023vvr}, which appears consistent with protonium, although being much broader than the QED state discussed here, it is rather a molecular $\protonium$ state bound by the QCD interaction.}, highly excited ($n\approx 68$) protonium has been measured by the ATHENA experiment via cold antiproton interactions with molecular hydrogen ions in vacuum, $\rm \overline{p} + H^{+}_{2}$, in a Penning trap at the CERN antiproton decelerator~\cite{ATHENA:2006clk}. The 1S ground state of protonium decays mostly into a pair of neutral pions and $\eta$ mesons. The total width of the 1S$_0$ ground state was found to be: $\Gamma_\text{tot} = 1096.7 \pm 42.3$~eV~\cite{Augsburger:1999yt}, which is $10^5$ times larger than its partial diphoton decay width obtained via Eq.~(\ref{eq:Gamma_gaga_fermion}). The heavier even-spin QED baryonium systems considered here include those formed by pairs of light-quark (u,\,d,\,s) baryons --- such as the $\Sigmaonium$, $\Xionium$, and $\Omegaonium$ --- as well as by pairs of charmed and/or bottom baryons: $\Lambdaconium$, $\Xiconium$, $\Xibonium$, and $\Omegabonium$, with generic properties listed in Table~\ref{tab:baryonium}. We indicatively list some possible two-body decays from baryon-antibaryon annihilation at rest, 
but their large mass allows for multimeson decays too. None of these theoretical systems (which have increasing mass and decreasing Bohr radii) has been explicitly investigated previously, as far as we can tell. 

\begin{table}[htbp!]
\tabcolsep=1.mm
\centering
\caption{Main properties of QED baryonium para-states $\baryonium$, with $\rm h = p, \Sigma^\pm, \Xi^\pm, \Omega^\pm, \Lambda^\pm_\mathrm{c}, \Xi_\mathrm{c}^\pm, \Xi_\mathrm{b}^\pm, \Omega_\mathrm{b}^\pm$. For each ground state ($n=1$), we list its $\rm J^{PC}$ quantum numbers, constituent hadron mass $m_{\rm h^{\pm}}$, atom mass $m_\mathrm{X}$ from Eq.~(\ref{eq:m_para}), QED binding energy (from $E_{n=1} = 2m_{\rm h}-m_\mathrm{X}$), Bohr radius $r_\text{Bohr}$ from Eq.~(\ref{eq:Rbohr}), lifetime $\tau$ and total width $\Gamma_\mathrm{tot}$ (assessed as explained in the text), diphoton width $\Gamma_{\gaga}$ from Eq.~(\ref{eq:Gamma_gaga_fermion}), and typical hadronic decays ('mult.' stands for multiple).
\label{tab:baryonium}}
\vspace{0.2cm}
\resizebox{\textwidth}{!}{
\begin{tabular}{lc cccccccc}
\toprule
Baryonium & $\rm J^{PC}$ & $m_{\rm h^{\pm}}$ (MeV)& $m_\mathrm{X}$ (MeV) & $E_{n=1}$ (keV) & $r_\text{Bohr}$ (fm) & $\tau$ (fs)& $\Gamma_\mathrm{tot}$ (eV) & $\Gamma_{\gaga}$ (meV) & Typical decays \\
\midrule
$\protonium$ & $0^{-+}$ & $938.272088$ & $1876.5317$ & $-12.49$ & 57.54 & $(0.60\pm 0.02)\cdot 10^{-3}$ & $1096.7 \pm 42.3$ & 9.71 & $\rm \pi^0\pi^0, \eta\eta$ \\ 
$\Sigmaonium$ & $0^{-+}$ & $1189.37 \pm 0.07$ & $2378.72 \pm 0.14$ & $-15.84$ & 45.40 & $\mathcal{O}(10^{-3})$ & $\mathcal{O}(1000)$ & 12.31 & $\rm \pi^0\pi^0, \eta\eta,\Sigma^0\Sigma^0$ \\ 
$\Xionium$ & $0^{-+}$ & $1321.71 \pm 0.07$ & $2643.40 \pm 0.14$ & $-17.60$ & 40.85 & $\mathcal{O}(10^{-3})$ & $\mathcal{O}(1000)$ & 13.68 & $\rm \pi\pi, \eta\eta, KK$ \\ 
$\Omegaonium$ & $0^{-+}$ & $1672.45 \pm 0.29$ & $3344.88 \pm 0.58$ & $-22.3$ & 32.3 & $\mathcal{O}(10^{-4})$ & $\mathcal{O}(2000)$ & 17.3 & $\rm \eta\eta, \eta'\eta'$\\ 
$\Lambdaconium$ & $0^{-+}$ & $2286.46 \pm 0.14$ & $4572.89 \pm 0.28$ & $-30.4$ & 23.6 & $\mathcal{O}(10^{-4})$ & $\mathcal{O}(2000)$ & 23.7 & $\rm D^0 \overbar{D^0},...$ \\ 
$\Xiconium$ & $0^{-+}$ & $2467.71 \pm 0.23$ & $4935.39 \pm 0.46$ & $-32.8$ & 21.9 & $\mathcal{O}(10^{-4})$ & $\mathcal{O}(2000)$ & 25.5 & $\rm \pi\pi, \eta\eta, KK,...$ \\ 
$\Xibonium$ & $0^{-+}$ & $5797.0 \pm 0.6$ & $11593.9 \pm 1.2$ & $-77.2$ & 9.3 & $\mathcal{O}(10^{-4})$ & $\mathcal{O}(6000)$ & 60.0 & mult.\ mesons \\ 
$\Omegabonium$ & $0^{-+}$ & $6045.8 \pm 0.8$ & $12091.5 \pm 1.6$ & $-80.5$ & 8.9 & $\mathcal{O}(10^{-4})$ & $\mathcal{O}(6000)$ & 62.5 & mult.\ mesons \\ 
\bottomrule
\end{tabular}
}
\end{table}

\begin{figure}[htpb!]
\centering
\includegraphics[width=0.495\linewidth]{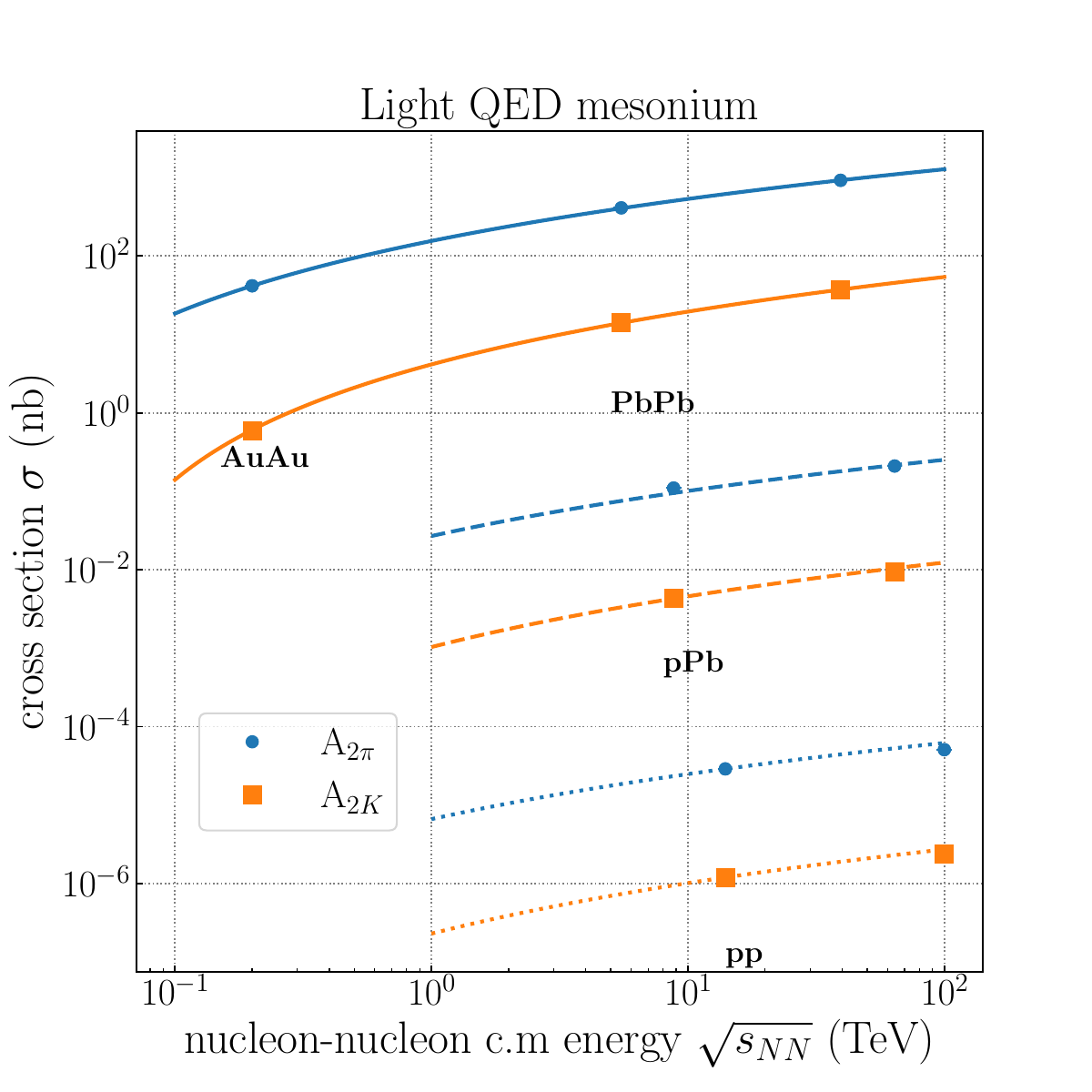}
\includegraphics[width=0.495\linewidth]{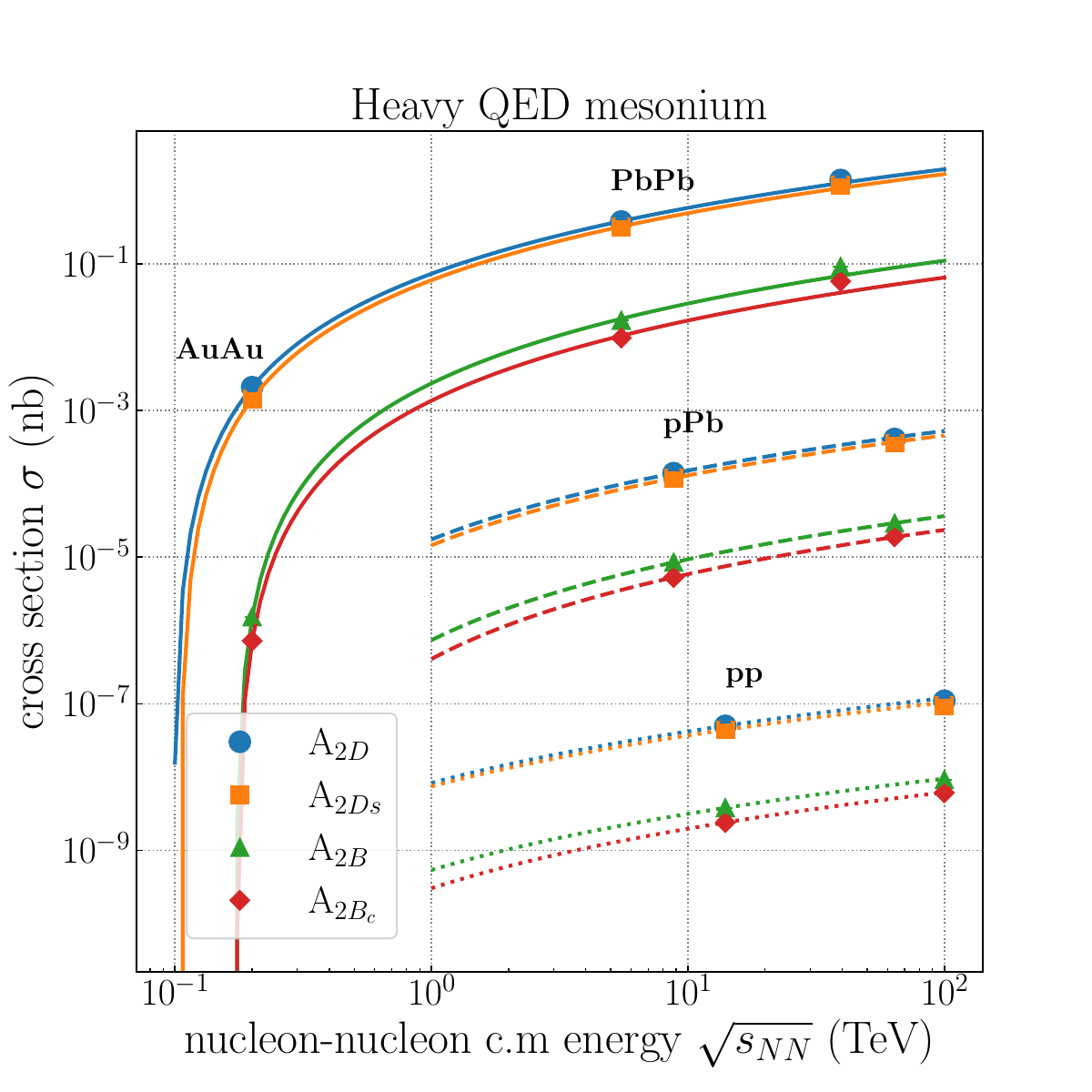}
\caption{Cross sections for the $\gaga$ production of even-spin QED mesonium states as a function of nucleon-nucleon \cm\ energy $\sqrtsnn$, in \PbPb\ or \AuAu\ (solid curves), \pPb\ (dashed curves), and \pp\ (dotted curves) UPCs. The curves are $\ln ^3(\snn)$ fits to guide the eye. The left and right panels show the results for light-quark ($\Apipi$, $\AKK$) and heavy-quark ($\DD$, $\DsDs$, $\BB$, $\BcBc$) states, respectively, whose properties are listed in Table~\ref{tab:mesonium}.}
\label{fig:sigma_vs_sqrts_mesonium}
\end{figure}

To our knowledge, the photon-fusion production of QED hadronium states has never been considered before in the literature, although such exotic systems should be theoretically producible through this channel given their nonzero diphoton widths. We estimate here their cross sections in UPCs through our ``master formula'' Eq.~(\ref{eq:sigma_X_master}). The relevant tables for this section can be found in Appendix~\ref{app:qedhadronium}. The QED mesonium cross sections and expected yields in UPCs at different colliders are listed in Table~\ref{tab:mesonium_xsection}, and the cross sections as a function of \cm\ energy are plotted in Fig.~\ref{fig:sigma_vs_sqrts_mesonium}. Whereas the cross sections appear too low to be visible above backgrounds in p-p and p-Pb UPCs, they are not that small in Pb-Pb UPCs for the case of light-quark dimeson systems, where a few thousands pionium and hundreds kaonium events are expected with the nominal LHC integrated luminosity. The ALICE (or ALICE-3) and LHCb experiments could have a chance to measure the lightest of such exotic QED atoms in their dominant hadronic decay modes. However, the $\gaga$ production of heavy-quark QED mesonium systems ($\DD, \DsDs, \BB, \BcBc$) features much smaller cross sections and is only potentially visible in UPCs at the FCC-hh.

\begin{figure}[htpb!]
\centering
\includegraphics[width=0.48\linewidth]{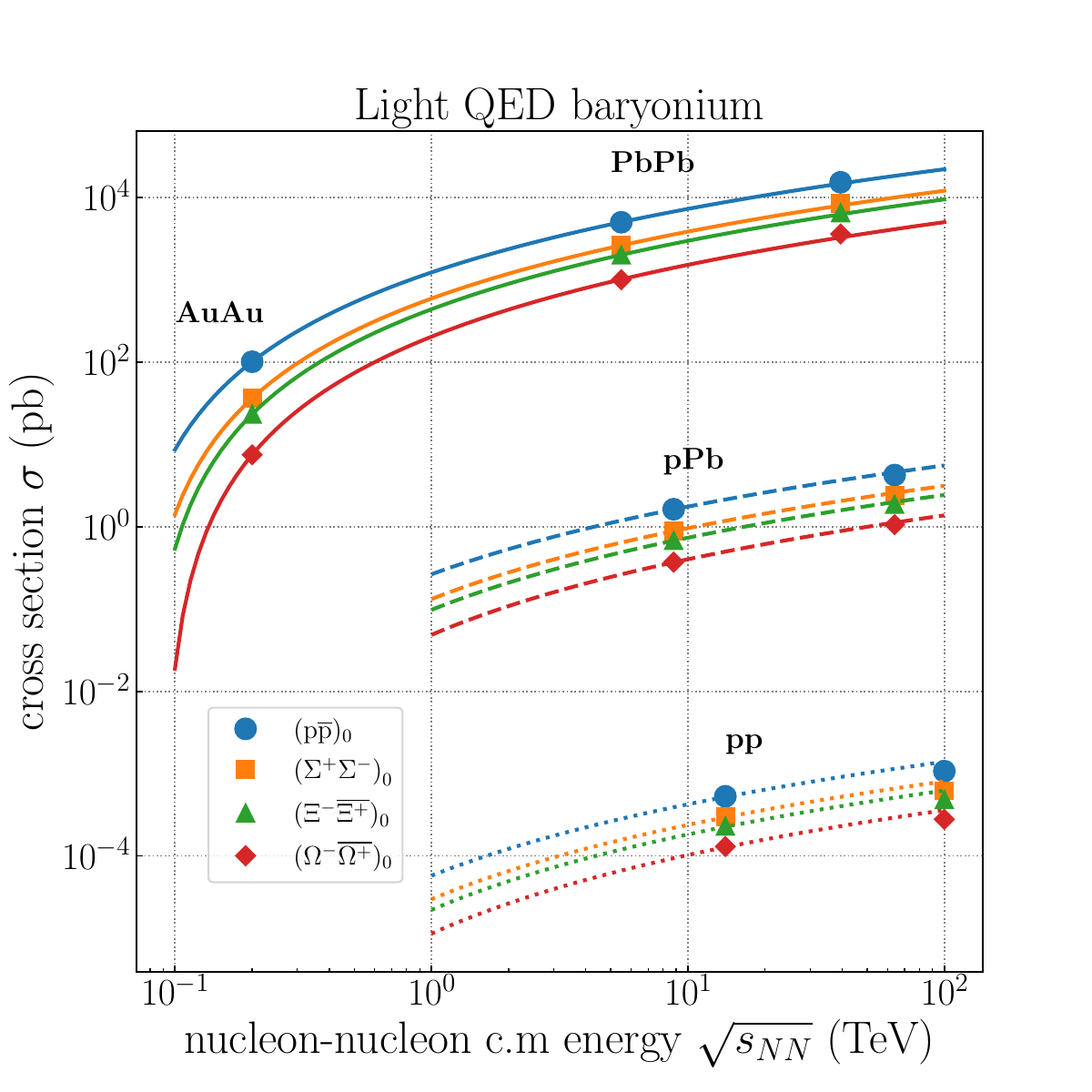}
\includegraphics[width=0.48\linewidth]{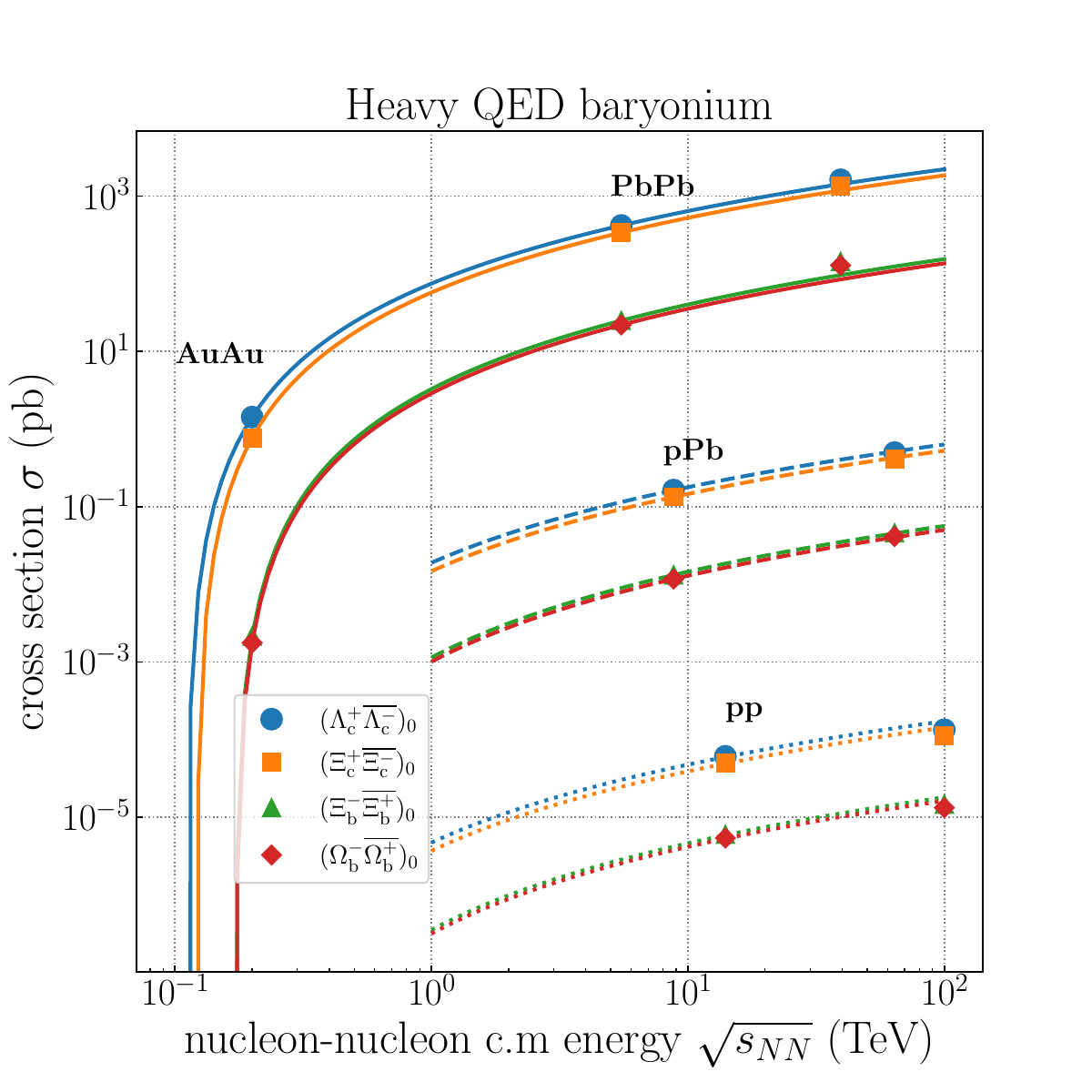}
\caption{Cross sections for the $\gaga$ production of even-spin QED baryonium states as a function of nucleon-nucleon \cm\ energy $\sqrtsnn$, in \PbPb\ or \AuAu\ (solid curves), \pPb\ (dashed curves), and \pp\ (dotted curves) UPCs. The curves are $\ln ^3(\snn)$ fits to guide the eye. The left and right panels show the results for light-quark ($\protonium$ , $\Sigmaonium$, $\Xionium$, $\Omegaonium$) and heavy-quark ($\Lambdaconium$, $\Xiconium$, $\Xibonium$, $\Omegabonium$) baryonium states, respectively, whose properties are listed in Table~\ref{tab:baryonium}.}
\label{fig:sigma_vs_sqrts_baryonium}
\end{figure}

The QED baryonium cross sections and expected yields in UPCs at the various colliders are listed in Table~\ref{tab:baryonium_xsection}, and shown graphically in Fig.~\ref{fig:sigma_vs_sqrts_baryonium} as a function of \cm\ energy. Cross sections appear too low to be visible above backgrounds in p-p and p-Pb UPCs, but light-quark baryonium systems appear producible in Pb-Pb UPCs, where a few tens of $\protonium$, $\Sigmaonium$, $\Xionium$, $\Omegaonium$ events are expected with the nominal LHC integrated luminosity. The ALICE (or ALICE-3) and LHCb experiments could attempt a measurement of such exotic QED atoms in their dominant hadronic decay modes, although backgrounds are likely very large. The $\gaga$ production of QED baryonium systems with charm or bottom quarks features much smaller cross sections and would be potentially visible only in \PbPb\ UPCs at the FCC-hh energies.

\section{Total \texorpdfstring{$\gaga$}{gamma-gamma} even-spin resonance cross sections in UPCs}

In Table~\ref{tab:sigma_sum} we collect the sum of all photon-photon resonance cross sections computed in this work (Tables~\ref{tab:sigma_light_onia0}--\ref{tab:baryonium_xsection}) for UPCs at RHIC/LHC/FCC/GZK-cutoff energies, and compare their values to the total hadronic cross section for each system. The latter have been computed with the Glauber model of Ref.~\cite{Loizides:2017ack} using the parametrization of the inelastic nucleon-nucleon cross section $\sigma_\mathrm{inel,had}(\rm NN\to X)$ vs.\ $\sqrts$ of Ref.~\cite{dEnterria:2020dwq}. The photon-fusion cross sections are dominated by the sum of the lightest even-spin systems that have the largest individual cross sections. In general, the photon-fusion cross sections represent a very small fraction of the p-p (about 1 part in 1 $10^{6}$) and proton-nucleus (1 part in $10^5$) inclusive hadronic cross sections, but the production of even-spin hadron and leptonium systems amounts to about 2.5--6\% 
of the Pb-Pb hadronic inelastic cross sections at the LHC and FCC, and are clearly not negligible.

\begin{table}[htpb!]
\tabcolsep=1.0mm
\centering
\caption{Sum of all photon-fusion cross sections in UPCs for hadronic resonances, $\sum_i\sigma(\gaga\to\mathrm{X}_i)$, and leptonium systems $\sum_i\sigma(\gaga\to(\ell^+_i \ell^-_i)_0)$, obtained here for all considered colliding systems (Tables~\ref{tab:sigma_light_onia0}--\ref{tab:baryonium_xsection}) compared to the total inclusive hadronic cross section, $\sigma(\rm A B)_\text{had}$, computed with a Glauber MC model~\cite{Loizides:2017ack} using the $\sigma_\mathrm{inel,had}(\rm NN \to X)$ vs.\ $\sqrtsnn$ parametrization of Ref.~\cite{dEnterria:2020dwq}.
\label{tab:sigma_sum}}
\vspace{0.2cm}
\resizebox{\textwidth}{!}{%
\begin{tabular}{lc ccccccc}\toprule
colliding system & \AuAu & \PbPb & \pPb & \pp & \PbPb & \pPb & \pp & p-air \\
$\sqrtsnn$ & 0.2~TeV & 5.5~TeV & 8.8~TeV & 14~TeV & 39.4~TeV & 62.8~TeV & 100~TeV & 400~TeV\\\hline
$\sum_i\sigma(\gaga\to\mathrm{X}_i)$ (mb) & $13.1 \pm 0.56$ & $189 \pm 8.8$ & $0.056 \pm 0.003$ & $(1.57\pm 0.075)\cdot 10^{-5} $ & $479 \pm 23$ & $0.12 \pm 0.006$ & $(2.96\pm 0.14)\cdot 10^{-5} $ & $(1.79\pm 0.087)\cdot 10^{-3} $ \\
$\sum_i\sigma(\gaga\to(\ell^+_i\ell^-_i)_0)$ (mb) & $109$ & $328 $ & $0.0675$ & $1.42\cdot 10^{-5}$ & $516$ & $0.10$ & $2.06\cdot 10^{-5}$ & $1.15\cdot 10^{-3}$ \\
$\sigma_\text{inel,had}(\rm A B\to X)$ (mb) & $6840\pm 140$ & $7640\pm 150$ & $2130 \pm 30$ & $79.2\pm 1.9$ & $7930\pm 160$ & $2300\pm 50$ & $107.5\pm 6.5$ & $570\pm20$ \\
\hline
\end{tabular}
}
\end{table}

\section{Two-photon even-spin backgrounds to LbL scattering in Pb-Pb UPCs at the LHC}
\label{sec:LbL}

\begin{figure}[htpb!]
  \centering
  \includegraphics[width=0.26\linewidth]{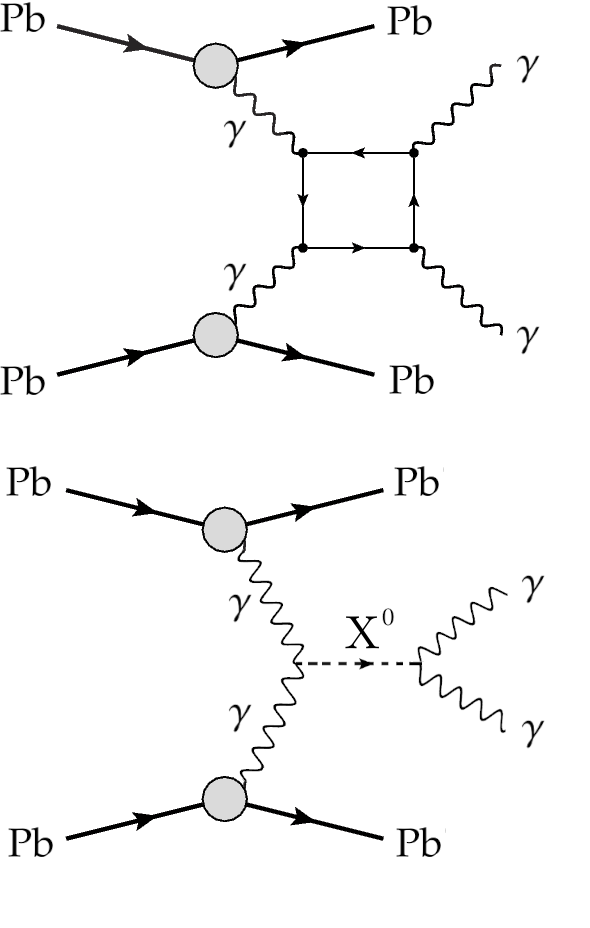}
  \includegraphics[width=0.735\linewidth]{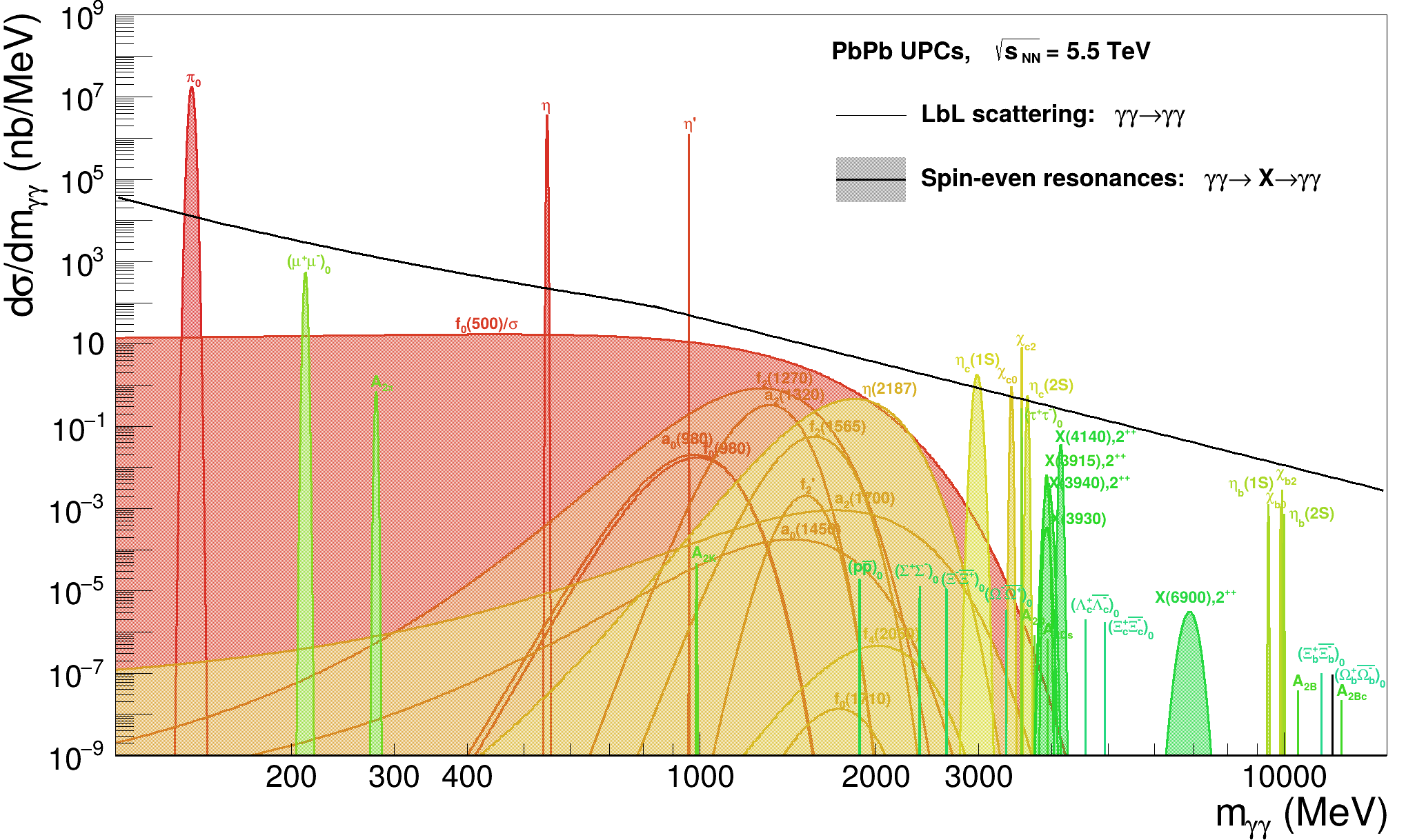}
  \caption{Left: Diagrams for LbL scattering (top) and exclusive two-photon even-spin particle production decaying into two photons (bottom) in \PbPb\ UPCs. Right: Exclusive diphoton mass distribution over $m_{\gaga}\approx0.1$--15~GeV for \PbPb(5.5~TeV) UPCs showing the light-by-light continuum  (\gammaUPC$\,+\,$\ttt{LbL@NLO}, dashed black curve) and all diphoton even-spin resonances (filled colored areas) over this mass range (note that the curves are not stacked on top of each other, and that the width of the 19 narrowest resonances has been arbitrarily set to 1-MeV for visibility purposes).}
  \label{fig:LbL}
\end{figure}

The work of~\cite{dEnterria:2013zqi} proposed to exploit the very large quasireal photon fluxes available in \PbPb\ UPCs at the LHC to measure and study the elastic $\gaga\to\gaga$ process, also known as light-by-light (LbL) scattering, that had remained experimentally unobserved because of its very small elementary cross section (proportional to the fourth power of the QED coupling, $\alpha^4 \approx 3 \cdot 10^{-9}$). Following the analysis strategy outlined in Ref.~\cite{dEnterria:2013zqi}, both the ATLAS and CMS experiments have measured the LbL process at the LHC~\cite{ATLAS:2017fur,CMS:2018erd,ATLAS:2019azn,CMS:2024bnt} for diphoton masses above $m_{\gaga} = 5$~GeV, with cross sections consistent (albeit with relatively large experimental uncertainties) with the theoretical prediction at NLO accuracy in QCD and QED~\cite{AH:2023ewe,AH:2023kor}. 
The LbL scattering proceeds via virtual box diagrams containing charged particles, as depicted in the top left diagram of Fig.~\ref{fig:LbL}. Whereas the contributions from charged leptons and heavy-quark boxes are well controlled theoretically, the nonperturbative light-quark hadronic contributions that dominate the cross section at lower diphoton masses are much more uncertain~\cite{Bern:2001dg}. As a matter of fact, the same hadronic virtual contributions to LbL scattering (also known as HLbL) are among the leading sources of uncertainty in the calculations of QCD corrections to the anomalous magnetic moment of the muon $(g-2)_\mu$~\cite{Colangelo:2015ama,Cappiello:2021vzi, Hoferichter:2024vbu}, whose measured value~\cite{Muong-2:2021ojo} appears in contradiction with standard model predictions based on data-driven dispersive approaches~\cite{Aoyama:2020ynm}. Measuring LbL scattering at lower diphoton masses in UPCs, in the region $m_{\gaga}\approx 0.1$--5~GeV currently unexplored experimentally, would thus provide valuable complementary input on the the HLbL contributions and their interplay with the resonant even-spin hadronic resonances similarly produced via photon fusion (bottom left diagram of Fig.~\ref{fig:LbL}). In addition, both LbL and hadronic diphoton resonances constitute backgrounds in searches for new even-spin particles, such as pseudoscalar axion-like~\cite{Knapen:2016moh,dEnterria:2021ljz} or graviton-like~\cite{dEnterria:2023npy} particles, identically produced via photon-fusion.
The ALICE and LHCb experiments are well placed to attempt such a measurement, given that the ATLAS and CMS experiments have poorer reconstruction capabilities at such low diphoton masses. 

Studies of LbL at low masses in UPCs at the LHC have been previously presented in Refs.~\cite{Klusek-Gawenda:2019ijn,Jucha:2023hjg}, but only a few background hadron diphoton states were considered. We include here all diphoton resonances discussed in this paper, and compare their cross sections to the LbL continuum computed with \gammaUPC$\,+\,$\ttt{LbL@NLO}~\cite{AH:2023ewe,AH:2023kor}. 
The right panel of Fig.~\ref{fig:LbL} shows the expected diphoton mass distribution in \PbPb(5.5~TeV) UPCs from the LbL continuum (dashed black curve) and from all even-spin diphoton resonances (filled colored areas) considered in this work. For visibility purposes, the width of the narrowest resonances has been arbitrarily set to 1-MeV, and the exotic hadron resonances cross sections plotted assume J$^\text{CP}=2^{++}$ (which yield larger yields than the scalar case).
The LbL curve has been obtained at LO QED and QCD accuracy, including the contributions from light-quarks boxes, which are not well-defined perturbatively, using $m_u = m_d = 0$ and $m_s \approx 95$ MeV for the up/down and strange quark masses, respectively. A full theoretical calculation would require to properly consider also interferences between the two diagrams of Fig.~\ref{fig:LbL} (left) at each relevant mass point, but this goes beyond the scope of this paper where we want to show the relative size of both contributions in different diphoton mass ranges, and emphasize the interest of such an experimental measurement. 
One can see that, in the absence of any selection cuts, the only even-spin resonances that would stand out clearly above the LbL continuum (provided a good experimental diphoton mass resolution is achieved) are the $\pi^0$, $\eta$, $\eta'$, and the charmonium mesons. On the one hand, if the main goal of the measurement is to measure or constrain HLbL scattering, the absence of significant resonant backgrounds is good news. On the other hand, if the aim is to identify other diphoton-decaying even-spin particles, one would need to apply appropriate event selection criteria through multivariate analysis techniques (\eg\ 
exploiting kinematic and angular properties of the single and double photons to separate scalar/tensor resonances from the box-mediated LbL process), so as to identify any potential resonant excess of events above the 
smooth LbL continuum. Nonetheless, there will remain many resonances with diphoton yields orders-of-magnitude smaller than the LbL background, whose measurement could only be potentially attempted through other more probable (hadronic) decay channels.

\section{Summary}
\label{sec:summ}

The cross sections for the single exclusive production of (pseudo)scalar and (pseudo)tensor hadrons, as well as of even-spin QED bound states formed by pairs of opposite-charge leptons or hadrons, have been estimated for photon-fusion processes in ultraperipheral collisions (UPCs) of proton-proton, proton-nucleus, and nucleus-nucleus at the RHIC, LHC and FCC colliders, as well as in proton-air interactions at the highest energies reached by cosmic-rays impinging on Earth. The UPC cross sections have been computed in the equivalent photon approximation with realistic photon fluxes from the charged form factors of proton, lead, gold, and nitrogen ions. The production of four types of even-spin systems has been considered: quarkonium (spin-0,\,2,\,4 meson bound states, from the lightest $\pi^0$ meson up to toponium), exotic mesons (including candidate multiquark states), leptonium (positronium, dimuonium, and ditauonium), and hadronium QED atoms (including pionium, kaonium, and protonium, plus dimeson/dibaryon onium systems with strange and/or heavy quarks). 
The production cross sections for about 50 such even-spin composite particles have been computed. To our knowledge, those are the first calculations of the UPC production cross sections for about half of these particles, including several light-quark resonances, exotic hadronic states, QED-hadronium systems, and para-toponium. 
Compared to other existing previous works, our study uses improved photon-photon UPC luminosities, propagates theoretical uncertainties to the production cross sections, and also includes predictions for future colliders, such as the FCC-hh, as well for cosmic-rays interactions on the Earth's atmosphere at the highest (GZK cutoff) energies observed.\\

We find, first, that the number of UPCs producing the lightest even-spin light-quark resonances (with masses over the $m_\mathrm{X}\approx 0.135$--2.1~GeV range) reaches the millions to hundred-millions events at the LHC. The ALICE (in particular, the proposed ALICE-3) as well as the LHCb (in particular, the proposed LHCb upgrade II) experiments should be able to reconstruct many of these resonances in their decays into soft hadronic or diphoton final states. Such measurements would allow to shed light on the properties (quantum numbers, diphoton widths, quark/gluon composition,...) of some of the least well-known of such states.
Similarly, one expects hundreds to thousands of events with even-spin charmonium resonances (with $m_\mathrm{X}\approx 2.95$--3.6~GeV masses) exclusively produced in UPCs at the LHC that decay back into a pair of photons. The measurements of exclusive charmonia in this decay mode (or in their much more abundant hadronic decays) appear also feasible for the ALICE and LHCb detectors, and 
can help determine their diphoton widths, which are either not known (as is the case for the scalar $\chicOne$ and tensor $\chicTwo$ states) or for which contradictory results exist today (as is the case for the pseudoscalar $\etacOneS$ and $\etacTwoS$ states). The number of exclusive bottomonia produced in UPCs at the LHC lies in the hundreds to thousands events (depending on the system and concrete state) and their potential measurement would only be possible in their hadronic decays, 
as their diphoton partial widths are too small. The heaviest particle known today is the quasibound state formed by a top-antitop quark pair (toponium). An observation of para-toponium could be attempted in \pp\ UPCs at the HL-LHC (where about 40 events are expected) and at FCC-hh (with about 1300 events expected) by exploiting the whole dataset of 6 and 30~ab$^{-1}$ integrated luminosities to be collected under high-pileup conditions. Such a measurement would require the search for a back-to-back $\ttbar$ pair produced at rest (\ie\ with zero pair $\pT$) in coincidence with two intact protons reconstructed in very forward proton spectrometers, such as those from the CMS-TOTEM PPS system, whose acceptance for such a heavy system is very large.\\

We have also studied the production of pure-QED para-leptonium systems formed by a pair of opposite-charged leptons. The UPC cross sections and associated yields are very large for positronium and dimuonium, whereas they are very small for the heaviest (true tauonium) system. The observation of paraleptonium production in UPCs appears, however, unfeasible either because the diphoton final state is ultra soft (positronium), or likely too soft (dimuonium) to be reconstructed, or because it is swamped by the decays of more abundant diphoton resonances in the same mass range (in the ditauonium case). Lastly, we have studied for the first time the two-photon production of even-spin systems formed by two identical hadrons of opposite charge, bound by their Coulomb interaction, which we refer to as QED ``hadronium''. We discussed the properties and production cross sections of six QED-mesonium scalar systems: pionium ($\Apipi$), kaonium ($\AKK$), D$_{\mathrm{(s)}}^\pm$-onium, and B$_{\mathrm{(s)}}^\pm$-onium; as well as eight QED-baryonium para-atoms formed by opposite-charge pairs of p, $\Sigma^\pm$, $\Xi^\pm$, $\Omega^\pm$, $\Lambda^\pm_\mathrm{c}$, $\Xi_\mathrm{c}^\pm$, $\Xi_\mathrm{b}^\pm$, $\Omega_\mathrm{b}^\pm$. Whereas the cross sections appear too low to be visible above backgrounds in p-p and p-Pb UPCs, they are not that small in Pb-Pb UPCs for the case of light-quark systems, where a few thousands of $\Apipi$ and hundreds of $\AKK$ events, as well as a few tens of protonium $\protonium$, $\Sigmaonium$, $\Xionium$, and $\Omegaonium$ events, are expected with the nominal LHC integrated luminosity. At the LHC, the ALICE and LHCb experiments could venture a measurement of the lightest of such exotic QED atoms in their dominant hadronic decay modes. However, the $\gaga$ production of the heavy-quark QED hadronium systems in UPCs features much smaller cross sections, which are only potentially producible at the FCC-hh.\\

Last but not least, we have computed the differential cross section for light-by-light (LbL) scattering, $\gaga\to\gaga$, in \PbPb\ UPCs at the LHC in the low-mass range, $m_{\gaga} = 0.1$--15~GeV, and compared it to the expected contributions from all diphoton resonances discussed in this study. The only even-spin resonances that would stand out clearly above the LbL continuum (provided a good experimental diphoton mass resolution is achieved) are the $\pi^0$, $\eta$, $\eta'$, and (partially) $\chicTwo$ mesons. Identifying other diphoton-decaying particles (as well as new axion- or graviton-like particles) would require applying appropriate event selection criteria in multivariate analyses, or attempting their reconstruction through other more probable decay channels.\\

We hope that the results reported in this work can help motivate upcoming experimental, and further theoretical, studies of multiple even-spin particles and exotic QED atoms, which either remain unobserved or whose properties are poorly known, as well as of low-mass LbL scattering, in UPCs at the LHC and future hadron colliders.

\vspace{0.5cm}
\paragraph*{Acknowledgments.---} 
K.K. gratefully acknowledges support from the Amherst College Charles Houston Hamilton Internship Program. 
We want to warmly thank Hua-Sheng~Shao and Nicolas Cr\'epet for common work with the \gammaUPC\ code used to produce many of the results shown in this study (additional thanks to H.S.~Shao for providing the low-mass light-by-light continuum prediction with the \gammaUPC$\,+\,$\ttt{LbL@NLO} code). We are grateful to H.S.~Shao and Spencer~Klein for useful feedback on a previous version of this manuscript, and to Guang-Zhi Xu for clarifications on the toponium wavefunction.

\clearpage
\appendix
\renewcommand\thesubsection{\Alph{subsection}}
\section*{Appendix: Photon-fusion cross sections and yields tables}

This appendix collects in tabulated form all the even-spin photon-fusion cross sections, $\sigma(\gaga \rightarrow \mathrm{X})$, and yields, $N_\text{evts}(\gaga \rightarrow \mathrm{X})$, computed in this paper for all colliding systems and c.m.\ energies. Whenever previous calculations exist for a given system, we also include their results for comparison purposes.

\subsection{Light-quark meson resonances}
\label{app:lightmeson}

\begin{table}[htbp!]
\centering
\tabcolsep=2.mm
\caption{Photon-fusion cross sections $\sigma(\gaga\to\mathrm{X})$, total yields $N_\text{evts}(\gaga\to\mathrm{X})$, and yields $N_\text{evts}(\gaga\to\mathrm{X}(\gaga))$ in the diphoton decay mode, for the production of light-quark spin-0,-2 resonances with masses $m_\mathrm{X}\lesssim 1$~GeV and known $\gaga$ decay widths (Table~\ref{tab:light_mesons}) in UPCs for various colliding systems at RHIC, LHC, and FCC \cm\ energies. Previously derived cross sections (if available) are also listed for reference. The last row gives the corresponding cross sections in proton-air collisions at GZK-cutoff energies. The '---' symbol indicates missing predictions from older works.
\label{tab:sigma_light_onia0}}
\vspace{0.2cm}
\begin{tabular}{lc cccc}
\toprule
System, $\sqrtsnn$, $\LumiInt$ & {Ref.} & {$\pi^0$} & {$\eta$} & $f_0(500)$ & {$\eta'$} \\
\midrule 
\multicolumn{6}{l}{\hspace{-0.cm}\AuAu, 0.2~TeV, 10~nb$^{-1}$:}\\ 
$\sigma(\gaga\to\mathrm{X})$ & Eq.~(\ref{eq:sigma_X_master}) & $6.1 \pm 0.1$ mb & $1.6 \pm 0.06$ mb & $2.3\pm0.5$ mb & $1.1 \pm 0.04$ mb \\
 & \cite{Krauss:1997vr,Baur:2001jj,Bertulani:2001zk} & 5.72, 5.0, 4.94 mb & 1.29, 0.85, 1.00 mb & --- & 0.99, 0.59, 0.75 mb \\

  & \cite{Nystrand:1998hw, Baltz:2009jk} & 4.7 mb, --&  0.88, 1.05 mb & --- &  0.64, 0.72 mb \\
$N_\text{evts}(\gaga\to\mathrm{X})$ & & $6.1\times 10^7$ & $1.6\times 10^7$ & $2.3\times 10^7$ & $1.1 \times 10^7$ \\
$N_\text{evts}(\gaga\to\mathrm{X}(\gaga))$ & & $6.1\times 10^7$ & $6.2\times 10^6$ & $15$ & $2.6\times 10^5$ \\
\midrule 
\multicolumn{6}{l}{\hspace{-0.cm}\PbPb, 5.5~TeV, 10 nb$^{-1}$:}\\ 
$\sigma(\gaga\to\mathrm{X})$ & Eq.~(\ref{eq:sigma_X_master}) & $45 \pm 1$ mb & $23 \pm 1$ mb & $29\pm 6$ mb & $26 \pm 1$ mb \\
 & \cite{Krauss:1997vr,Baur:2001jj,Bertulani:2001zk} & 43, 46, 28\footnote{Result for \PbPb(5.02~TeV) UPCs.} mb & 19.9, 20, 16 mb & --- & 24.8, 25, 21 mb \\

  & \cite{Fariello:2023uvh, Baltz:2009jk} &  38\footnote{Result for \PbPb(5.02~TeV) UPCs.} mb, -- &  17.3$^a$, 18.8 mb & --- &  21.8$^a$, 21.9 mb \\
$N_\text{evts}(\gaga\to\mathrm{X})$ & & $4.5\times 10^8$ & $2.3\times 10^8$ & $2.9\times 10^8$ & $2.6 \times 10^8$ \\
$N_\text{evts}(\gaga\to\mathrm{X}(\gaga))$ & & $4.4\times 10^8$ & $8.9\times 10^7$ & $180$ & $6.0 \times 10^6$ \\
\midrule
\multicolumn{6}{l}{\hspace{-0.cm}\pPb, 8.8~TeV, 1 pb$^{-1}$:}\\
$\sigma(\gaga\to\mathrm{X})$ & Eq.~(\ref{eq:sigma_X_master}) & $11.1 \pm 0.2$~$\mu$b & $6.4 \pm 0.2$~$\mu$b & $8.1\pm 1.8$~$\mu$b & $7.9 \pm 0.3$~$\mu$b \\
$N_\text{evts}(\gaga\to\mathrm{X})$ & & $1.1\times 10^7$ & $6.4\times 10^6$ & $8.1\times 10^6$ & $7.9\times 10^6$ \\
$N_\text{evts}(\gaga\to\mathrm{X}(\gaga))$ & & $1.1\times 10^7$ & $2.5\times 10^6$ & $5$ & $1.8 \times 10^5$ \\
\midrule
\multicolumn{6}{l}{\hspace{-0.cm}\pp, 14~TeV, 1 fb$^{-1}$:}\\
$\sigma(\gaga\to\mathrm{X})$ & Eq.~(\ref{eq:sigma_X_master}) & $2.8 \pm 0.1$ nb & $1.8 \pm 0.1$ nb & $2.2 \pm 0.5$ nb & $2.3 \pm 0.1$ nb \\ 
$N_\text{evts}(\gaga\to\mathrm{X})$ & & $2.8\times 10^6$ & $1.8\times 10^6$ & $2.2\times 10^6$ & $2.3\times 10^6$ \\
$N_\text{evts}(\gaga\to\mathrm{X}(\gaga))$ & & $2.8\times 10^6$ & $7.1\times 10^5$ & $1.3$ & $5.4 \times 10^4$ \\
\midrule
\multicolumn{6}{l}{\hspace{-0.cm}\PbPb, 39.4~TeV, 110 nb$^{-1}$:}\\
$\sigma(\gaga\to\mathrm{X})$ & Eq.~(\ref{eq:sigma_X_master}) & $93 \pm 1.4$ mb & $56 \pm 2$ mb & $70\pm 15$ mb & $70 \pm 2$ mb \\
$N_\text{evts}(\gaga\to\mathrm{X})$ & & $1.0\times 10^{10}$ & $6.1\times 10^9$ & $7.7\times 10^9$ & $7.7\times 10^9$ \\
$N_\text{evts}(\gaga\to\mathrm{X}(\gaga))$ & & $1.0\times 10^{10}$ & $2.4\times 10^9$ & $4.6\times 10^3$ & $1.8 \times 10^8$ \\
\midrule 
\multicolumn{6}{l}{\hspace{-0.cm}\pPb, 62.8~TeV, 29 pb$^{-1}$:}\\
$\sigma(\gaga\to\mathrm{X})$ & Eq.~(\ref{eq:sigma_X_master}) & $21.0 \pm 0.3$~$\mu$b & $13 \pm 0.5$~$\mu$b & $17 \pm 3.6$~$\mu$b & $18 \pm 0.6$~$\mu$b \\
$N_\text{evts}(\gaga\to\mathrm{X})$ & & $6.1\times 10^8$ & $3.9\times 10^8$ & $4.9\times 10^8$ & $5.1\times 10^8$ \\
$N_\text{evts}(\gaga\to\mathrm{X}(\gaga))$ & & $6.0 \times 10^8$ & $1.5\times 10^8$ & $290$ & $1.2 \times 10^7$ \\
\midrule
\multicolumn{6}{l}{\hspace{-0.cm}\pp, 100~TeV, 10 fb$^{-1}$:}\\
$\sigma(\gaga\to\mathrm{X})$ & Eq.~(\ref{eq:sigma_X_master}) & $4.8 \pm 0.1$ nb & $3.3 \pm 0.12$ nb & $4.1\pm0.9$ nb & $4.5 \pm 0.14$ nb \\
$N_\text{evts}(\gaga\to\mathrm{X})$ & & $4.8\times 10^7$ & $3.3\times 10^7$ & $4.1\times 10^7$ & $4.5\times 10^7$ \\
$N_\text{evts}(\gaga\to\mathrm{X}(\gaga))$ & & $4.8\times 10^7$ & $1.3\times 10^7$ & $25$ & $1.0 \times 10^6$ \\
\midrule
\multicolumn{6}{l}{\hspace{-0.cm}p-air, 400~TeV:}\\
$\sigma(\gaga\to\mathrm{X})$ & Eq.~(\ref{eq:sigma_X_master}) & $280\pm 4$ nb & $200\pm 7$ nb & $246^{+55}_{-54}$ nb & $270\pm 9$ nb \\
\bottomrule
\end{tabular}
\end{table}

\begin{table}[htbp!]
\centering
\tabcolsep=1.mm
\caption{Photon-fusion cross sections $\sigma(\gaga\to\mathrm{X})$, total yields $N_\text{evts}(\gaga\to\mathrm{X})$, and yields $N_\text{evts}(\gaga\to\mathrm{X}(\gaga))$ in the diphoton decay mode, for the production of light-quark spin-0,-2 resonances with masses $m_\mathrm{X}\approx 1$--1.5~GeV and known $\gaga$ decay widths (Table~\ref{tab:light_mesons}) in UPCs for various colliding systems at RHIC, LHC, and FCC \cm\ energies. Previously derived cross sections (if available) are also listed for reference. The last row gives the corresponding cross sections in proton-air collisions at GZK-cutoff energies. The '---' symbol indicates missing predictions from older works.
\label{tab:sigma_light_onia1}}
\vspace{0.2cm}
\resizebox{\textwidth}{!}{%
\begin{tabular}{lc cccccc}
\toprule
System, $\sqrtsnn$, $\LumiInt$ & {Ref.} & $f_0(980)$ & $a_0(980)$ & $f_2(1270)$ & $a_2(1320)$ & $a_0(1450)$ \\
\midrule 
\multicolumn{7}{l}{\hspace{-0.cm}\AuAu, 0.2~TeV, 10~nb$^{-1}$:}\\ 
$\sigma(\gaga\to\mathrm{X})$ & Eq.~(\ref{eq:sigma_X_master}) & $66^{+25}_{-14}$~$\mu$b & $71\pm 24$~$\mu$b & $920 \pm 180$~$\mu$b & $310 \pm 20$~$\mu$b & $190\pm 10$~$\mu$b \\
 & \cite{Krauss:1997vr, Baur:2001jj, Bertulani:2001zk} & 91~$\mu$b, --, -- & --- & 680, 410, 545~$\mu$b & 250, 140, 195~$\mu$b & --- \\

 & \cite{Nystrand:1998hw, Baltz:2009jk} & 75~$\mu$b, -- & --- & 514, 550~$\mu$b & 155~$\mu$b, -- & --- \\
 $N_\text{evts}(\gaga\to\mathrm{X})$ & & $6.6\times 10^5$ & $7.1\times 10^5$ & $9.2\times 10^6$ & $3.1\times 10^6$ & $1.9\times 10^6$ \\
 $N_\text{evts}(\gaga\to\mathrm{X}(\gaga))$ & & $2$ & $2$ & $130$ & $30$ & $35$ \\
\midrule 
\multicolumn{7}{l}{\hspace{-0.cm}\PbPb, 5.5~TeV, 10 nb$^{-1}$:}\\ 
$\sigma(\gaga\to\mathrm{X})$ & Eq.~(\ref{eq:sigma_X_master}) & $1.5^{+0.6}_{-0.3}$ mb & $1.7\pm 0.6$ mb & $28 \pm 5.3$ mb & $9.5 \pm 0.6$ mb & $6.4\pm 0.3$ mb \\
 & \cite{Krauss:1997vr, Baur:2001jj, Bertulani:2001zk} & 2.50~mb, --, -- & --- & 24.7, 25, 22 mb & 9.54, 7.7, 8.2 mb & --- \\

  & \cite{Fariello:2023uvh, Baltz:2009jk} &  --- & --- &  22.4\footnote{Result for \PbPb(5.02~TeV) UPCs.}, 23.4 mb &  8.3$^a$ mb, -- & --- \\
$N_\text{evts}(\gaga\to\mathrm{X})$ & & $1.5\times 10^7$ & $1.7\times 10^7$ & $2.8\times 10^8$ & $9.5\times 10^7$ & $6.4\times 10^7$ \\
 $N_\text{evts}(\gaga\to\mathrm{X}(\gaga))$ & & $45$ & $50$ & $3900$ & $890$ & $1100$ \\
\midrule
\multicolumn{7}{l}{\hspace{-0.cm}\pPb, 8.8~TeV, 1 pb$^{-1}$:}\\
$\sigma(\gaga\to\mathrm{X})$ & Eq.~(\ref{eq:sigma_X_master}) & $470^{+180}_{-100}$ nb & $510\pm 170$ nb & $9.3 \pm 1.8$~$\mu$b & $3.2\pm 0.2$~$\mu$b & $2.2\pm 0.1$~$\mu$b \\
$N_\text{evts}(\gaga\to\mathrm{X})$ & & $4.7\times 10^5$ & $5.1\times 10^5$ & $9.3\times 10^6$ & $3.2\times 10^6$ & $2.2\times 10^6$ \\
 $N_\text{evts}(\gaga\to\mathrm{X}(\gaga))$ & & $1.5$ & $1.5$ & $130$ & $30$ & $40$ \\
\midrule
\multicolumn{7}{l}{\hspace{-0.cm}\pp, 14~TeV, 1 fb$^{-1}$:}\\
$\sigma(\gaga\to\mathrm{X})$ & Eq.~(\ref{eq:sigma_X_master}) & $140^{+50}_{-30}$ pb & $150 \pm 50$ pb & $2.6 \pm 0.5$ nb & $910 \pm 55$ pb & $630\pm 26$ pb \\
$N_\text{evts}(\gaga\to\mathrm{X})$ & & $1.4\times 10^5$ & $1.5\times 10^5$ & $2.6\times 10^6$ & $9.1\times 10^5$ & $6.3\times 10^5$ \\
 $N_\text{evts}(\gaga\to\mathrm{X}(\gaga))$ & & $0.5$ & $0.5$ & $40$ & $10$ & $10$ \\
\midrule
\multicolumn{7}{l}{\hspace{-0.cm}\PbPb, 39.4~TeV, 110 nb$^{-1}$:}\\
$\sigma(\gaga\to\mathrm{X})$ & Eq.~(\ref{eq:sigma_X_master}) & $4.1^{+1.6}_{-0.9}$ mb & $4.5 \pm 1.5$ mb & $78 \pm 15$ mb & $27 \pm 2$ mb & $18\pm 1$ mb \\
$N_\text{evts}(\gaga\to\mathrm{X})$ & & $4.6\times 10^8$ & $4.9\times 10^8$ & $8.6\times 10^9$ & $3.0\times 10^9$ & $2.0\times 10^9$ \\
 $N_\text{evts}(\gaga\to\mathrm{X}(\gaga))$ & & $1300$ & $1500$ & $1.2\times 10^5$ & $2.8 \times 10^4$ & $3.6 \times 10^4$ \\
\midrule 

\multicolumn{7}{l}{\hspace{-0.cm}\pPb, 62.8~TeV, 29 pb$^{-1}$:}\\
$\sigma(\gaga\to\mathrm{X})$ & Eq.~(\ref{eq:sigma_X_master}) & $1.1^{+0.4}_{-0.2}$~$\mu$b & $1.1\pm 0.4$~$\mu$b & $21 \pm 4$~$\mu$b & $7.2 \pm 0.4$~$\mu$b & $5.0\pm 0.2$~$\mu$b \\
$N_\text{evts}(\gaga\to\mathrm{X})$ & & $3.1\times 10^7$ & $3.3\times 10^7$ & $6.1\times 10^8$ & $2.1\times 10^8$ & $1.4\times 10^8$ \\
 $N_\text{evts}(\gaga\to\mathrm{X}(\gaga))$ & & $100$ & $100$ & $85\times 10^3$ & $2000$ & $2600$ \\
\midrule

\multicolumn{7}{l}{\hspace{-0.cm}\pp, 100~TeV, 10 fb$^{-1}$:}\\
$\sigma(\gaga\to\mathrm{X})$ & Eq.~(\ref{eq:sigma_X_master}) & $265^{+100}_{-55}$ pb & $280\pm 95$ pb & $5.2 \pm 1.0$ nb & $1.8\pm 0.1$ nb & $1.24\pm 0.05$ nb \\
$N_\text{evts}(\gaga\to\mathrm{X})$ & & $2.6\times 10^6$ & $2.8\times 10^6$ & $5.2\times 10^7$ & $1.8\times 10^7$ & $1.2\times 10^7$ \\
 $N_\text{evts}(\gaga\to\mathrm{X}(\gaga))$ & & $10$ & $10$ & $720$ & $170$ & $220$ \\
\midrule
\multicolumn{7}{l}{\hspace{-0.cm}p-air, 400~TeV:}\\
$\sigma(\gaga\to\mathrm{X})$ & Eq.~(\ref{eq:sigma_X_master})& $16^{+6}_{-3}$ nb & $17 \pm 6$ nb & $320 \pm 60$ nb & $110 \pm 7$ nb & $76\pm 3$ nb \\
\bottomrule
\end{tabular}
}
\end{table}

\begin{table}[htbp!]
\centering
\tabcolsep=1.mm
\caption{Photon-fusion cross sections $\sigma(\gaga\to\mathrm{X})$, total yields $N_\text{evts}(\gaga\to\mathrm{X})$, and yields $N_\text{evts}(\gaga\to\mathrm{X}(\gaga))$ in the diphoton decay mode, for the production of light-quark spin-0,-2 resonances with masses $m_\mathrm{X}\approx 1.5$--2~GeV and known $\gaga$ decay widths (Table~\ref{tab:light_mesons}) in UPCs for various colliding systems at RHIC, LHC, and FCC \cm\ energies. Previously derived cross sections (if available) are also listed for reference. The last row gives the corresponding cross sections in proton-air collisions at GZK-cutoff energies. The '---' symbol indicates missing predictions from older works.
\label{tab:sigma_light_onia2}}
\vspace{0.2cm}
\begin{tabular}{lc cccccc}
\toprule
System, $\sqrtsnn$, $\LumiInt$ & {Ref.} & $f_2'(1525)$ & $f_2(1565)$ & $a_2(1700)$ & $f_0(1710)$ & $\eta_2(1870)$ & $f_4(2050)$ \\
\midrule 
\multicolumn{7}{l}{\hspace{-0.cm}\AuAu, 0.2~TeV, 10~nb$^{-1}$:}\\ 
$\sigma(\gaga\to\mathrm{X})$ & Eq.~(\ref{eq:sigma_X_master}) & $12\pm 1.3$~$\mu$b & $88 \pm 18$~$\mu$b & $26\pm 4.4$~$\mu$b & $540^{+150}_{-110}$ nb & $264 \pm 34$~$\mu$b & $8.7\pm 0.17$~$\mu$b\\
 & \cite{Krauss:1997vr,Baur:2001jj} & --, 6.6~$\mu$b & --- & --- & --- & --- & 22.1\footnote{Using $\Gamma_{\gaga} = 1.4$ keV, which is about 10 times larger than the value we used.}~$\mu$b, --- \\

  & \cite{Nystrand:1998hw, Baltz:2009jk} & 7, 6.9~$\mu$b & --- & --- & --- & --- & 22.1\footnote{Using $\Gamma_{\gaga} = 1.4$ keV, which is about 10 times larger than the value we used.}~$\mu$b, --- \\
$N_\text{evts}(\gaga\to\mathrm{X})$ & & $1.2\times 10^5$ & $8.8\times 10^5$ & $2.6\times 10^5$ & $5.4\times 10^3$ & $2.6\times 10^6$ & $8.7\times 10^3$ \\
$N_\text{evts}(\gaga\to\mathrm{X}(\gaga))$ & & $0.1$ & $5$ & $0.2$ & $10^{-3}$ & $50$ & $0.05$ \\
\midrule 
\multicolumn{7}{l}{\hspace{-0.cm}\PbPb, 5.5~TeV, 10 nb$^{-1}$:}\\ 
$\sigma(\gaga\to\mathrm{X})$ & Eq.~(\ref{eq:sigma_X_master}) & $465 \pm 50$~$\mu$b & $3.5\pm 0.7$ mb & $1.1 \pm 0.2$ mb & $23_{-4.7}^{+6.4}$~$\mu$b & $13\pm 1.6$ mb & $480 \pm 10$~$\mu$b\\
 & \cite{Krauss:1997vr,Baur:2001jj, Baltz:2009jk} & --, 450, 380~$\mu$b & --- & --- & --- & --- & 1.6$^a$ mb, --- \\
$N_\text{evts}(\gaga\to\mathrm{X})$ & & $4.6\times 10^6$ & $3.5\times 10^7$ & $1.1\times 10^7$ & $2.3\times 10^5$ & $1.3\times 10^8$ & $4.8\times 10^6$ \\
$N_\text{evts}(\gaga\to\mathrm{X}(\gaga))$ & & $5$ & $180$ & $10$ & $0.05$ & $2500$ & $3$ \\
\midrule
\pPb, 8.8~TeV, 1 pb$^{-1}$: & & & & & \\
$\sigma(\gaga\to\mathrm{X})$ & Eq.~(\ref{eq:sigma_X_master}) & $0.16 \pm 0.02$~$\mu$b & $1.2\pm 0.2$~$\mu$b & $390\pm 65$ nb & $8.2_{-1.7}^{+2.3}$ nb & $4.5 \pm 0.58$~$\mu$b & $175\pm 3$ nb\\
$N_\text{evts}(\gaga\to\mathrm{X})$ & & $1.6\times 10^5$ & $1.2\times 10^6$ & $3.9\times 10^5$ & $8200$ & $4.5\times 10^6$ & $1.8\times 10^5$ \\
$N_\text{evts}(\gaga\to\mathrm{X}(\gaga))$ & & $0.2$ & $6$ & $0.3$ & $10^{-3}$ & $90$ & $0.1$ \\
\midrule
\pp, 14~TeV, 1 fb$^{-1}$: & & & & & \\
$\sigma(\gaga\to\mathrm{X})$ & Eq.~(\ref{eq:sigma_X_master}) & $46 \pm 5$ pb & $350\pm 70$ pb & $110\pm 20$ pb & $2.4_{-0.5}^{+0.7}$ pb & $1.3 \pm 0.17$ nb & $52 \pm 1$ pb\\ 
$N_\text{evts}(\gaga\to\mathrm{X})$ & & $4.6\times 10^4$ & $3.5\times 10^5$ & $1.1\times 10^5$ & $2400$ & $1.3\times 10^6$ & $5.2\times 10^4$ \\
$N_\text{evts}(\gaga\to\mathrm{X}(\gaga))$ & & $0.05$ & $2$ & $0.1$ & $5\times 10^{-4}$ & $30$ & $0.03$ \\
\midrule
\multicolumn{7}{l}{\hspace{-0.cm}\PbPb, 39.4~TeV, 110 nb$^{-1}$:}\\ 
$\sigma(\gaga\to\mathrm{X})$ & Eq.~(\ref{eq:sigma_X_master}) & $1.4 \pm 0.1$ mb & $10\pm 2$ mb & $3.3\pm 0.5$ mb & $69.3_{-14}^{+19}$~$\mu$b & $38 \pm 5$ mb & $1.6\pm {0.03}$ mb\\
$N_\text{evts}(\gaga\to\mathrm{X})$ & & $1.5\times 10^8$ & $1.1\times 10^9$ & $3.6\times 10^8$ & $7.6\times 10^6$ & $4.2\times 10^9$ & $1.7\times 10^8$ \\
$N_\text{evts}(\gaga\to\mathrm{X}(\gaga))$ & & $140$ & $6000$ & $290$ & $2$ & $8 \times 10^4$ & $100$ \\
\midrule 

\pPb, 62.8~TeV, 29 pb$^{-1}$: & & & & & \\
$\sigma(\gaga\to\mathrm{X})$ & Eq.~(\ref{eq:sigma_X_master}) & $370 \pm 40$ nb & $2.8\pm 0.6$~$\mu$b & $900\pm 150$ nb & $19 _{-3.9}^{+5.3}$ nb & $11\pm 1.4$~$\mu$b & $420\pm 8$ nb\\
$N_\text{evts}(\gaga\to\mathrm{X})$ & & $1.1\times 10^7$ & $8.1\times 10^7$ & $2.6\times 10^7$ & $5.5\times 10^5$ & $3.1\times 10^8$ & $1.2\times 10^7$ \\
$N_\text{evts}(\gaga\to\mathrm{X}(\gaga))$ & & $10$ & $430$ & $20$ & $0.1$ & $6100$ & $7$ \\
\midrule

\multicolumn{7}{l}{\hspace{-0.cm}\pp, 100~TeV, 10 fb$^{-1}$:}\\ 
$\sigma(\gaga\to\mathrm{X})$ & Eq.~(\ref{eq:sigma_X_master}) & $92\pm 10$ pb & $700\pm 140$ pb & $280\pm 47$ pb & $4.8_{-1.0}^{+1.3}$ pb & $2.7 \pm 0.4$ nb & $110\pm 2$ pb\\
$N_\text{evts}(\gaga\to\mathrm{X})$ & & $9.2\times 10^5$ & $7.0\times 10^6$ & $2.3\times 10^6$ & $4.8\times 10^4$ & $2.7\times 10^7$ & $1.1\times 10^6$ \\
$N_\text{evts}(\gaga\to\mathrm{X}(\gaga))$ & & $1$ & $40$ & $0.2$ & $0.01$ & $540$ & $0.6$ \\
\midrule
\multicolumn{7}{l}{\hspace{-0.cm}p-air, 400~TeV:}\\ 
$\sigma(\gaga\to\mathrm{X})$ & Eq.~(\ref{eq:sigma_X_master}) & $5.7\pm 0.6$ nb & $43 \pm 8.6$ nb & $14\pm 2.4$ nb & $300_{-61}^{+82}$ pb & $164\pm 21$ nb & $6.9 \pm 0.1$ nb\\
\bottomrule
\end{tabular}
\end{table}

\clearpage
\subsection{Heavy quarkonium resonances}
\label{app:heavyquarkonium}

\begin{table}[htbp!]
\centering
\tabcolsep=2.3mm
\caption{Photon-fusion cross sections $\sigma(\gaga\to\mathrm{X})$, total yields $N_\text{evts}(\gaga\to\mathrm{X})$, and yields $N_\text{evts}(\gaga\to\mathrm{X}(\gaga))$ in the diphoton decay mode, for the production of all known even-spin charmonium resonances (Table~\ref{tab:heavy_onia}) in UPCs for various colliding systems at RHIC, LHC, and FCC \cm\ energies. Previously derived cross sections (if available) are also listed for reference. The last row gives the corresponding cross sections in proton-air collisions at GZK-cutoff energies. 
The '---' symbol indicates missing predictions from older works.
\label{tab:sigma_heavy_charmonia}}
\vspace{0.2cm}
\resizebox{\textwidth}{!}{
\begin{tabular}{lc cccc}
\toprule
System, $\sqrtsnn$, $\LumiInt$ & {Ref.} & $\etacOneS$ & $\etacTwoS$ & $\chicZero$ & $\chicTwo$ \\
\midrule 
\multicolumn{6}{l}{\hspace{-0.cm}\AuAu, 0.2~TeV, 10~nb$^{-1}$:}\\ 
$\sigma(\gaga\to\mathrm{X})$ & Eq.~(\ref{eq:sigma_X_master}) & $5.7 \pm 0.05$~$\mu$b & $0.56 \pm 0.38$~$\mu$b & $0.85 \pm 0.04$~$\mu$b & $0.88\pm 0.05$~$\mu$b \\
& \cite{Krauss:1997vr,Baur:2001jj,Bertulani:2001zk} & 3.66, 1.8, 3.3~$\mu$b & --- & 1.36, 0.38, 0.63~$\mu$b & --, 0.17, 0.59~$\mu$b \\

& \cite{Nystrand:1998hw, Baltz:2009jk} & 2, 2.9~$\mu$b & --- & --- & --- \\
$N_\text{evts}(\gaga\to\mathrm{X})$ & & $5.6\times 10^4$ & $5600$ & $8500$ & $8800$ \\
$N_\text{evts}(\gaga\to\mathrm{X}(\gaga))$ & & $10$ & $1$ & $2$ & $2.5$ \\
\midrule 
\multicolumn{6}{l}{\hspace{-0.cm}\PbPb, 5.5~TeV, 10 nb$^{-1}$:}\\ 
$\sigma(\gaga\to\mathrm{X})$ & Eq.~(\ref{eq:sigma_X_master})& $0.62 \pm 0.01$ mb & $91\pm 63$~$\mu$b & $0.12 \pm 0.01$ mb & $0.14 \pm 0.01$ mb \\
 & \cite{Krauss:1997vr,Baur:2001jj,Bertulani:2001zk} & 0.56, 0.54, 0.61 mb & --- & 0.29, 0.17, 0.16 mb & --, 0.085, 0.15 mb \\
 & \cite{dEnterria:2022ysg,Shao:2022cly, Fariello:2023uvh, Baltz:2009jk} & 0.46, 0.39, 0.43\footnote{Result for \PbPb(5.02~TeV) UPCs.}, 0.57 mb & 95, 80 , 90$^a$~$\mu$b, -- & 0.12, 0.10, 0.11$^a$ mb, -- & 0.13, 0.11, 0.12$^a$ mb, -- \\
$N_\text{evts}(\gaga\to\mathrm{X})$ & & $6.2\times 10^6$ & $9.1\times 10^5$ & $1.2\times 10^6$ & $1.4\times 10^6$ \\
$N_\text{evts}(\gaga\to\mathrm{X}(\gaga))$ & & $1400$ & $160$ & $240$ & $400$ \\
\midrule

\multicolumn{6}{l}{\hspace{-0.cm}\pPb, 8.8~TeV, 1 pb$^{-1}$:}\\
$\sigma(\gaga\to\mathrm{X})$ & Eq.~(\ref{eq:sigma_X_master}) & $220 \pm 2$ nb & $33.6 \pm 23.0$ nb & $44 \pm 3.2$ nb & $50 \pm 3.1$ nb \\
 & \cite{Shao:2022cly,dEnterria:2022ysg} & 180, 160 nb & 38, 33.2 nb & 49, 43 nb & 53, 46 nb \\
$N_\text{evts}(\gaga\to\mathrm{X})$ & & $2.2\times 10^5$ & $3.4\times 10^4$ & $4.4\times 10^4$ & $5.0\times 10^4$ \\
$N_\text{evts}(\gaga\to\mathrm{X}(\gaga))$ & & $50$ & $5$ & $10$ & $15$ \\
\midrule

\multicolumn{6}{l}{\hspace{-0.cm}\pp, 14~TeV, 1 fb$^{-1}$:}\\
$\sigma(\gaga\to\mathrm{X})$ & Eq.~(\ref{eq:sigma_X_master}) & $75.8\pm 6.8$ pb & $12.0 \pm 8.2$ pb & $15.3 \pm 1.1$ pb & $17.7 \pm 1.1$ pb \\
 & \cite{Shao:2022cly,dEnterria:2022ysg} & 56, 50 pb & 12, 10.5 pb & 15, 13.7 pb & 17, 15 pb \\
$N_\text{evts}(\gaga\to\mathrm{X})$ & & $7.6\times 10^4$ & $1.2\times 10^4$ & $1.5\times 10^4$ & $1.8\times 10^4$ \\
$N_\text{evts}(\gaga\to\mathrm{X}(\gaga))$ & & $20$ & $2$ & $3$ & $5$ \\
\midrule

\multicolumn{6}{l}{\hspace{-0.cm}\PbPb, 39.4~TeV, 110 nb$^{-1}$:}\\
$\sigma(\gaga\to\mathrm{X})$ & Eq.~(\ref{eq:sigma_X_master})& $2.1 \pm 0.02$ mb & $0.33 \pm 0.22$ mb & $0.42 \pm 0.03$ mb & $0.49 \pm 0.03$ mb \\
 & \cite{Shao:2022cly} & 1.6 mb & 0.33 mb & 0.43 mb & 0.47 mb \\
$N_\text{evts}(\gaga\to\mathrm{X})$ & & $2.3\times 10^8$ & $3.6\times 10^7$ & $4.6\times 10^7$ & $5.3\times 10^7$ \\
$N_\text{evts}(\gaga\to\mathrm{X}(\gaga))$ & & $5.2\times 10^4$ & $6500$ & $9500$ & $1.6\times 10^4$ \\
\midrule 
\multicolumn{6}{l}{\hspace{-0.cm}\pPb, 62.8~TeV, 29 pb$^{-1}$:}\\
$\sigma(\gaga\to\mathrm{X})$ & Eq.~(\ref{eq:sigma_X_master}) & $610\pm 60$ nb & $98 \pm 67$ nb & $130 \pm 1$ nb & $144 \pm 5$ nb\\
 & \cite{Shao:2022cly} & 460 nb & 100 nb & 130 nb & 140 nb \\
$N_\text{evts}(\gaga\to\mathrm{X})$ & & $1.8\times 10^7$ & $2.8\times 10^6$ & $3.6\times 10^6$ & $4.2\times 10^6$ \\
$N_\text{evts}(\gaga\to\mathrm{X}(\gaga))$ & & $4000$ & $510$ & $740$ & $1200$ \\
\midrule

\multicolumn{6}{l}{\hspace{-0.cm}\pp, 100~TeV, 10 fb$^{-1}$:}\\
$\sigma(\gaga\to\mathrm{X})$ & Eq.~(\ref{eq:sigma_X_master})& $160\pm 1.4$ pb & $26\pm 18$ pb & $33.0 \pm 2.4$ pb & $37.9 \pm 2.3$ pb \\
 & \cite{Shao:2022cly} & 120 nb & 26 pb &33 pb & 37 pb \\
$N_\text{evts}(\gaga\to\mathrm{X})$ & & $1.6\times 10^6$ & $2.58\times 10^5$ & $3.28\times 10^5$ & $3.79\times 10^5$ \\
$N_\text{evts}(\gaga\to\mathrm{X}(\gaga))$ & & $360$ & $45$ & $70$ & $10$ \\
\midrule
\multicolumn{6}{l}{\hspace{-0.cm}p-air, 400~TeV:}\\
$\sigma(\gaga\to\mathrm{X})$ & Eq.~(\ref{eq:sigma_X_master}) & $8.01\pm 0.59$ nb & $1.71 \pm 1.12$ nb & $2.09\pm 0.09$ nb & $2.33\pm 0.08$ nb \\
\bottomrule
\end{tabular}
}
\end{table}

\begin{table}[htbp!]
\centering
\tabcolsep=2.9mm
\caption{Photon-fusion cross sections $\sigma(\gaga\to\mathrm{X})$, total yields $N_\text{evts}(\gaga\to\mathrm{X})$, and yields $N_\text{evts}(\gaga\to\mathrm{X}(\gaga))$ in the diphoton decay mode, for the production of all known even-spin bottomonium and toponium resonances (Table~\ref{tab:heavy_onia}) in UPCs for various colliding systems at RHIC, LHC, and FCC \cm\ energies. Previously derived cross sections (if available) are also listed for reference. The last row gives the corresponding cross sections in proton-air collisions at GZK-cutoff energies. 
The '---' symbol indicates either missing predictions from older works or negligible expected cross sections or yields.
\label{tab:sigma_heavy_bb_tt_onia}}
\vspace{0.2cm}
\begin{tabular}{lc ccccc}
\toprule
System, $\sqrtsnn$, $\LumiInt$ & {Ref.} & $\etabOneS$ & $\etabTwoS$ & $\chibZero$ & $\chibTwo$ & $\etatnS$ \\ 
\midrule 
\multicolumn{7}{l}{\hspace{-0.cm}\AuAu, 0.2~TeV, 10~nb$^{-1}$:}\\ 
$\sigma(\gaga\to\mathrm{X})$ & Eq.~(\ref{eq:sigma_X_master}) & $110\pm 10$ pb & $16.8 \pm 2.8$ pb & $14.5^{+4.8}_{-2.9}$ pb & $4.0^{+0.6}_{-2.7}$ pb & --- \\
 & \cite{Krauss:1997vr} & 20 pb & --- & --- & --- & --- \\
$N_\text{evts}(\gaga\to\mathrm{X})$ & & $1$ & $0.2$ & $0.15$ & $0.05$ & --- \\
$N_\text{evts}(\gaga\to\mathrm{X}(\gaga))$ & & --- & --- & --- & --- & --- \\
\midrule 
\multicolumn{7}{l}{\hspace{-0.cm}\PbPb, 5.5~TeV, 10 nb$^{-1}$:}\\ 
$\sigma(\gaga\to\mathrm{X})$ & Eq.~(\ref{eq:sigma_X_master})& $570\pm30$ nb & $190\pm 30$ nb & $130^{+40}_{-30}$ nb & $38^{+5}_{-25}$ nb & $1.2^{+1.2}_{-0.96}$ fb\\
& \cite{Krauss:1997vr,Baur:2001jj,Shao:2022cly} & --, 410, 500~nb & --, 320, 190 nb & --, 15, 130~nb & --, 20, 38~nb & --- \\
$N_\text{evts}(\gaga\to\mathrm{X})$ & & $5700$ & $1900$ & $1300$ & $380$ & $1\times 10^{-5}$ \\
$N_\text{evts}(\gaga\to\mathrm{X}(\gaga))$ & & $0.3$ & $0.1$ & $0.1$ & $0.02$ & --- \\
\midrule 
\multicolumn{7}{l}{\hspace{-0.cm}\pPb, 8.8~TeV, 1 pb$^{-1}$:}\\ 
$\sigma(\gaga\to\mathrm{X})$ & Eq.~(\ref{eq:sigma_X_master}) & $270\pm 20$ pb & $92 \pm 15$ pb & $61_{-12}^{+20}$ pb & $18_{-12}^{+3}$ pb & $0.63 \pm 0.36$ fb \\
 & \cite{Shao:2022cly} & 270 pb & 106 pb & 70 pb & 21 pb & --- \\
$N_\text{evts}(\gaga\to\mathrm{X})$ & & $270$ & $90$ & $60$ & $20$ & $6\times 10^{-4}$ \\
$N_\text{evts}(\gaga\to\mathrm{X}(\gaga))$ & & $0.01$ & $5\times 10^{-3}$ & $3\times 10^{-3}$ & $10^{-3}$ & --- \\
\midrule
\multicolumn{7}{l}{\hspace{-0.cm}\pp, 14~TeV, 1 fb$^{-1}$ (6~ab$^{-1}$ for $\etatnS$):}\\ 
$\sigma(\gaga\to\mathrm{X})$ & Eq.~(\ref{eq:sigma_X_master}) & $116\pm 7$ fb & $40 \pm 7$ fb & $26_{-5}^{+9}$ fb & $8.0_{-5.4}^{+1.1}$ fb & $6.4\pm 2.8$ ab \\
 & \cite{Shao:2022cly} & 100 fb & 40 fb & 26 fb & 8.0 fb & --- \\
$N_\text{evts}(\gaga\to\mathrm{X})$ & & $120$ & $40$ & $25$ & $10$ & $40$ \\ 
$N_\text{evts}(\gaga\to\mathrm{X}(\gaga))$ & & $6\times 10^{-3}$ & $2 \times 10^{-3}$ & $2\times 10^{-3}$ & $4\times 10^{-4}$ & --- \\
\midrule
\multicolumn{7}{l}{\hspace{-0.cm}\PbPb, 39.4~TeV, 110 nb$^{-1}$:}\\ 
$\sigma(\gaga\to\mathrm{X})$ & Eq.~(\ref{eq:sigma_X_master})& $2.88\pm 0.17$~$\mu$b & $1.0 \pm 0.2$~$\mu$b & $0.66^{+0.22}_{-0.13}$~$\mu$b & $0.20^{+0.03}_{-0.13}$~$\mu$b & $40\pm 12$ pb \\ 
 & \cite{Shao:2022cly} & 2.5 $\mu$b & 1.0 $\mu$b & 0.66 $\mu$b & 0.19 $\mu$b & --- \\
$N_\text{evts}(\gaga\to\mathrm{X})$ & & $3.2\times 10^5$ & $1.1\times 10^5$ & $7.2\times 10^4$ & $2.2\times 10^4$ & 4.5 \\ 
$N_\text{evts}(\gaga\to\mathrm{X}(\gaga))$ & & $20$ & $5$ & $4$ & $1$ & --- \\
\midrule
\multicolumn{6}{l}{\hspace{-0.cm}\pPb, 62.8~TeV, 29 pb$^{-1}$:}\\
$\sigma(\gaga\to\mathrm{X})$ & Eq.~(\ref{eq:sigma_X_master}) & $0.97\pm{0.56}$ nb & $0.33 \pm 0.06$ nb & $0.22^{+0.07}_{-0.04}$ nb & $67^{+9}_{-45}$ pb & $55\pm 17$ fb\\ 
 & \cite{Shao:2022cly} & 0.83 nb & 0.33 nb & 0.22 nb & 67 pb & --- \\
$N_\text{evts}(\gaga\to\mathrm{X})$ & & $2.8\times 10^4$ & $9700$ & $6400$ & $1900$ & 1.6 \\ 
$N_\text{evts}(\gaga\to\mathrm{X}(\gaga))$ & & $2$ & $50$ & $0.4$ & $0.1$ & --- \\
\midrule
\multicolumn{7}{l}{\hspace{-0.cm}\pp, 100~TeV, 10 fb$^{-1}$  (30~ab$^{-1}$ for $\etatnS$):}\\ 
$\sigma(\gaga\to\mathrm{X})$ & Eq.~(\ref{eq:sigma_X_master}) & $0.28\pm{0.16}$ pb & $97 \pm 16$ fb & $64^{+21}_{-13}$ fb & $19^{+3}_{-13}$ fb & $40\pm12$ ab \\
 & \cite{Shao:2022cly} & 0.24 pb & 96 fb & 63 fb & 19 fb & --- \\
$N_\text{evts}(\gaga\to\mathrm{X})$ & & $2800$ & $970$ & $640$ & $190$ & $1300$ \\ 
$N_\text{evts}(\gaga\to\mathrm{X}(\gaga))$ & & $0.15$ & $0.05$ & $0.05$ & $0.01$ & 0.01\\
\midrule
p-air, 400~TeV: & & & & \\
$\sigma(\gaga\to\mathrm{X})$ & Eq.~(\ref{eq:sigma_X_master}) & $18.0\pm 1.0$ pb & $6.3\pm 1.0$ pb & $4.2_{-0.8}^{+1.4}$~pb & $1.3_{-0.8}^{+0.2}$ pb & $3.4\pm 1.0$ fb \\
\bottomrule
\end{tabular}
\end{table}

\clearpage
\subsection{Exotic heavy hadrons}
\label{app:exotichadrons}

\begin{table}[htbp!]
\centering
\tabcolsep=2.5mm
\caption{Photon-fusion cross sections $\sigma(\gaga\to\mathrm{X})$, total yields $N_\text{evts}(\gaga\to\mathrm{X})$, and yields $N_\text{evts}(\gaga\to\mathrm{X}(\gaga))$ in the diphoton decay mode, for the production of spin-0,-2 exotic heavy-quark hadronic states (Table~\ref{tab:multiquark}) in UPCs for various colliding systems at RHIC, LHC, and FCC \cm\ energies. Previously derived cross sections (if available) are also listed for reference. The last row gives the corresponding cross sections in proton-air collisions at GZK-cutoff energies. The '---' symbol indicates either missing predictions from older works or negligible expected cross sections or yields.
\label{tab:sigma_multiquark}}
\vspace{0.2cm}
\resizebox{\textwidth}{!}{%
\begin{tabular}{lc ccccccccc}
\toprule
System, $\sqrtsnn$, $\LumiInt$ & {Ref.} & \multicolumn{2}{c}{$\chicZero(3915)$} & $\chicTwo(3930)$ & \multicolumn{2}{c}{$\rm X(3940)$} & \multicolumn{2}{c}{$\rm X_0(4140)$} & \multicolumn{2}{c}{$\rm T_{c c \overline{c c}}(6900)$} \\
& & $0^{++}$ & $2^{++}$ & & $0^{++}$ & $2^{++}$ & $0^{++}$ & $2^{++}$ & $0^{++}$ & $2^{++}$ \\
\midrule 
\AuAu, 0.2~TeV, 10~nb$^{-1}$: & & & & & \\
$\sigma(\gaga\to\mathrm{X})$ & Eq.~(\ref{eq:sigma_X_master}) & $28.8$ nb & $144$ nb & $57.6$ nb & $42.2$~nb & $211$~nb & $69.2$~nb & $275$~nb & $290 $~pb & $1.43 $ nb\\
$N_\text{evts}(\gaga\to\mathrm{X})$ & & $290$ & $1400$ & $580$ & $420$ & $2100$ & $690$ & $2700$ & $2.9$ & $14$ \\
$N_\text{evts}(\gaga\to\mathrm{X}(\gaga))$ & & $10^{-3}$ & $0.01$ & $10^{-3}$ & $10^{-3}$ & $0.02$ & $0.02$ & $0.1$ & -- & -- \\
\midrule

\PbPb, 5.5~TeV, 10 nb$^{-1}$: & & & & & \\
$\sigma(\gaga\to\mathrm{X})$ & Eq.~(\ref{eq:sigma_X_master}) & $6.5$~$\mu$b & $32$~$\mu$b & $13$~$\mu$b & $9.5$~$\mu$b & $48$~$\mu$b & $16$~$\mu$b & $65$~$\mu$b & $360 $~nb & $1.8$~$\mu$b\\
& \cite{Moreira:2016ciu}\,~ 
& 6.7~$\mu$b & --- & 13.9~$\mu$b & 10.8~$\mu$b & 44.2~$\mu$b & --- & --- & --- & --- \\
 & \cite{Goncalves:2018hiw}\footnote{Results for Pb-Pb(5.02~TeV) UPCs.} & 1.5--2.8~$\mu$b & 2.2--4.0~$\mu$b & --- & --- & --- & --- & --- & --- & ---\\
 & \cite{Goncalves:2021ytq}$^{a}$ & --- & --- & --- & --- & --- & --- & --- & 171 nb & 206 nb\\
 & \cite{Fariello:2023uvh}$^{a}$ & 6.0~$\mu$b & --- & 12.4~$\mu$b & 9.7~$\mu$b & 39.6~$\mu$b & --- & --- & 238/160~$\mu$b
 & --- \\
$N_\text{evts}(\gaga\to\mathrm{X})$ & & $6.5\times 10^{4}$ & $3.2\times 10^5$ & $1.3\times 10^5$ & $9.5\times 10^4$ & $4.8\times 10^5$ & $1.6\times 10^5$ & $6.5\times 10^5$ & $3.6\times 10^3$ & $1.8\times 10^4$ \\
$N_\text{evts}(\gaga\to\mathrm{X}(\gaga))$ & & $1$ & $3$ & $0.3$ & $1$ & $4$ & $5$ & $20$ & -- & $0.01$ \\
\midrule

\pPb, 8.8~TeV, 1 pb$^{-1}$: & & & & & \\
$\sigma(\gaga\to\mathrm{X})$ & Eq.~(\ref{eq:sigma_X_master}) & $2.4$ nb & $12$ nb & $4.8$ nb & $3.5$ nb & $18$~nb & $6.2$~nb & $25$~nb & $0.15$~nb & $0.76$~nb \\
  \hspace{0.6cm} 
 & \cite{Moreira:2016ciu}\,~ & 2.8 nb & --- & 5.7 nb & 4.5 nb & 18.3 nb & --- & --- & --- & --- \\
 & \cite{Goncalves:2018hiw}\footnote{Results for p-Pb(8.1~TeV) UPCs.} & 0.56--1.1 nb & 0.84--1.6 nb & --- & --- & --- & --- & --- & --- & ---\\
 & \cite{Goncalves:2021ytq}$^{b}$ & --- & --- & --- & --- & --- & --- & --- & 76.3 pb & 92.4 pb\\
$N_\text{evts}(\gaga\to\mathrm{X})$ & & $2.4\times 10^{3}$ & $1.2\times 10^4$ & $4.8\times 10^3$ & $3.5\times 10^3$ & $1.8\times 10^4$ & $6.2\times 10^3$ & $2.5\times 10^4$ & $150$ & $760$ \\
$N_\text{evts}(\gaga\to\mathrm{X}(\gaga))$ & & $0.02$ & $0.1$ & $0.01$ & $0.03$ & $0.1$ & $0.2$ & $1$ & -- & -- \\
\midrule
\pp, 14~TeV, 1 fb$^{-1}$: & & & & & \\
$\sigma(\gaga\to\mathrm{X})$ & Eq.~(\ref{eq:sigma_X_master}) & $0.87$ pb & $4.4$ pb & $1.7$ pb & $1.3$~pb & $6.4$~pb & $2.3$~pb & $9.0$~pb & $62$ fb & $310$ fb\\
& \cite{Moreira:2016ciu}\,~ & 0.86 pb & --- & 1.8 pb & 1.5 pb & 5.7 pb & --- & --- & --- & --- \\
 & \cite{Goncalves:2018hiw}\footnote{Results for p-p(13~TeV) UPCs.} & 0.18--0.33 fb & 0.27--0.49 fb & --- & --- & --- & --- & --- & --- & ---\\
 & \cite{Goncalves:2021ytq}$^{c}$ & --- & --- & --- & --- & --- & --- & --- & 26.3 fb & 31.9 fb\\
$N_\text{evts}(\gaga\to\mathrm{X})$ & & $870$ & $4.4\times 10^3$ & $1.7\times 10^3$ & $1.3\times 10^3$ & $6.4\times 10^3$ & $2.3\times 10^3$ & $9\times 10^3$ & $62$ & $310$ \\
$N_\text{evts}(\gaga\to\mathrm{X}(\gaga))$ & & $0.01$ & $0.05$ & $10^{-3}$ & $0.01$ & $0.05$ & $0.1$ & $0.2$ & -- & -- \\
\midrule
\PbPb, 39.4~TeV, 110 nb$^{-1}$: & & & & & \\
$\sigma(\gaga\to\mathrm{X})$ & Eq.~(\ref{eq:sigma_X_master}) & $24$~$\mu$b & $120$~$\mu$b & $47$~$\mu$b & $35$~$\mu$b & $170$~$\mu$b & $61$~$\mu$b & $240$~$\mu$b & $1.6$~$\mu$b & $7.9$~$\mu$b \\
& \cite{Moreira:2016ciu} & 24.5~$\mu$b & --- & 50.5~$\mu$b & 39.6~$\mu$b & 162~$\mu$b & --- & --- & --- & --- \\
 & \cite{Fariello:2023uvh} & 20.1~$\mu$b & --- & 41.7~$\mu$b & 32.5~$\mu$b & 133~$\mu$b & --- & --- & 912/612~$\mu$b
 & --- \\
$N_\text{evts}(\gaga\to\mathrm{X})$ & & $2.6 \times 10^6$ & $1.3\times 10^7$ & $5.2\times 10^6$ & $3.8\times 10^6$ & $1.9 \times 10^7$ & $6.8 \times 10^6$ & $2.7 \times 10^7$ & $1.7 \times 10^5$ & $8.7 \times 10^5$ \\
$N_\text{evts}(\gaga\to\mathrm{X}(\gaga))$ & & $26$ & $130$ & $10$ & $30$ & $160$ & $220$ & $710$ & $0.1$ & $1$ \\
\midrule

\pPb, 62.8~TeV, 29 pb$^{-1}$: & & & & & \\
$\sigma(\gaga\to\mathrm{X})$ & Eq.~(\ref{eq:sigma_X_master}) & $7.1$ nb & $36$ nb & $14$ nb & $11$~nb & $53$~nb & $19$~nb & $75$~nb & $0.5$~nb & $2.6$~nb \\
 & \cite{Moreira:2016ciu} & 7.0 nb & --- & 14.5 nb & 11.3 nb & 46.3 nb & --- & --- & --- & --- \\ 
$N_\text{evts}(\gaga\to\mathrm{X})$ & & $2.1 \times 10^5$ & $10^6$ & $4.1\times 10^5$ & $3\times 10^5$ & $1.5 \times 10^6$ & $5.4 \times 10^5$ & $2.1 \times 10^6$ & $1.5 \times 10^4$ & $7.4 \times 10^4$ \\
$N_\text{evts}(\gaga\to\mathrm{X}(\gaga))$ & & $2$ & $10$ & $1$ & $2.5$ & $10$ & $20$ & $50$ & $0.01$ & $0.05$ \\
\midrule

\pp, 100~TeV, 10 fb$^{-1}$: & & & & & \\
$\sigma(\gaga\to\mathrm{X})$ & Eq.~(\ref{eq:sigma_X_master}) & $1.9$ pb & $9.5$ pb & $3.8$ pb & $2.8$~pb & $14$~pb & 4.9 pb & 20 pb & 0.14 pb & 0.7 pb \\
& \cite{Moreira:2016ciu} & 1.8 pb & --- & 3.6 pb & 2.8 pb & 11.6 pb & --- & --- & --- & --- \\
$N_\text{evts}(\gaga\to\mathrm{X})$ & & $1.9 \times 10^4$ & $9.5\times 10^4$ & $3.8\times 10^4$ & $2.8\times 10^4$ & $1.4 \times 10^5$ & $4.9 \times 10^4$ & $2 \times 10^5$ & $1400$ & $7100$ \\
$N_\text{evts}(\gaga\to\mathrm{X}(\gaga))$ & & $0.2$ & $1$ & $0.1$ & $0.2$ & $1$ & $2$ & $5$ & -- & -- \\
\midrule

p-air, 400~TeV: & & & & & \\
$\sigma(\gaga\to\mathrm{X})$ & Eq.~(\ref{eq:sigma_X_master}) & 120~pb & 600 pb & 240 pb & 180 pb & 880 pb & 310 pb & 1.2 nb & 9.1 pb & 45 pb\\
\bottomrule
\end{tabular}
}
\end{table}

\clearpage
\subsection{Leptonium states}
\label{app:leptonium}

\begin{table}[htbp!]
\tabcolsep=2.1mm
\centering
\caption{Photon-fusion cross sections $\sigma(\gaga\to\mathrm{X})$, total yields $N_\text{evts}(\gaga\to\mathrm{X})$, and yields $N_\text{evts}(\gaga\to\mathrm{X}(\gaga))$ in the diphoton decay mode, for the production of paraleptonium states (Table~\ref{tab:paralept}) in UPCs for various colliding systems at RHIC, LHC, and FCC \cm\ energies. The last row lists the corresponding cross sections in proton-air collisions at GZK-cutoff energies. The '---' symbol indicates either missing predictions from older works or negligible expected cross sections or yields.
\label{tab:lep_xsection}}
\vspace{0.2cm}
\begin{tabular}{lc ccc}
\toprule
System, $\sqrtsnn$, $\LumiInt$ & {Ref.} & {$\ppositronium$} & {$\pdimuonium$} & {$\pditauonium$} \\
\midrule
\multicolumn{5}{l}{\hspace{-0.cm}\AuAu, 0.2~TeV, 10~nb$^{-1}$:}\\
$\sigma(\gaga\to\mathrm{X})$ & Eq.~(\ref{eq:sigma_X_master}) & 109 mb & 159 nb & 5.69 pb \\
& \cite{Baur:2001jj,Ginzburg:1998df} & --- & 150, 150 nb & --- \\
&\cite{Azevedo:2019hqp,Francener:2021wzx,Dai:2024imb} &  --, 112.1, 136 mb &  160\footnote{Result for \PbPb(0.2~TeV) UPCs.},150, 200  nb & --, 3.8, 9.68 pb\\
$N_\text{evts}(\gaga\to\mathrm{X})$ & & $1.1\times 10^9$ & $1600$ & $0.06$ \\
$N_\text{evts}(\gaga\to\mathrm{X}(\gaga))$ & & $1.1\times 10^9$ & $1600$ & $0.04$ \\
\midrule
\multicolumn{5}{l}{\hspace{-0.cm}\PbPb, 5.5~TeV, 10 nb$^{-1}$:}\\
$\sigma(\gaga\to\mathrm{X})$ & Eq.~(\ref{eq:sigma_X_master}) & 328 mb & 1.36~$\mu$b & 0.873 nb \\
& \cite{Baur:2001jj,Ginzburg:1998df,Kotkin:1998hu,Azevedo:2019hqp} & --, --, 110~mb, -- & 1.35, 1.35, --, 1.24$^b$~$\mu$b & -- \\
& \cite{Shao:2022cly,dEnterria:2022ysg,Francener:2021wzx,Dai:2024imb} & --, --, 333\footnote{Result for \PbPb(5.02~TeV) UPCs.}, 4010\footnote{This value an order-of-magnitude larger than other estimates, and it is likely a typo of the paper.} mb & --, --, 1.30$^b$, 1.59~$\mu$b & 0.86, 0.74, 0.833$^b$, 1.08 nb \\
$N_\text{evts}(\gaga\to\mathrm{X})$ & & $3.3\times 10^9$ & $1.4\times 10^{4}$ & $9$ \\
$N_\text{evts}(\gaga\to\mathrm{X}(\gaga))$ & & $3.3\times 10^9$ & $1.4\times 10^{4}$ & $7$ \\
\midrule
\multicolumn{5}{l}{\hspace{-0.cm}\pPb, 8.8~TeV, 1 pb$^{-1}$:}\\  
$\sigma(\gaga\to\mathrm{X})$ & Eq.~(\ref{eq:sigma_X_master}) & 67.5~$\mu$b & 351 pb & 0.356 pb \\
& \cite{Shao:2022cly,dEnterria:2022ysg} & --- & --- & 0.35, 0.31 pb \\
$N_\text{evts}(\gaga\to\mathrm{X})$ & & $6.8\times 10^7$ & $350$ & $0.3$ \\
$N_\text{evts}(\gaga\to\mathrm{X}(\gaga))$ & & $6.8\times 10^7$ & $350$ & $0.2$ \\
\midrule
\multicolumn{5}{l}{\hspace{-0.cm}\pp, 14~TeV, 1 fb$^{-1}$:}\\  
$\sigma(\gaga\to\mathrm{X})$ & Eq.~(\ref{eq:sigma_X_master}) & 14.3 nb & 92.0 fb & 0.113 fb \\
& \cite{Shao:2022cly,dEnterria:2022ysg} & --- & --- & 0.11, 0.11 fb \\
$N_\text{evts}(\gaga\to\mathrm{X})$ & & $1.4\times 10^7$ & $92$ & $0.1$ \\
$N_\text{evts}(\gaga\to\mathrm{X}(\gaga))$ & & $1.4\times 10^7$ & $92$ & $0.1$ \\
\midrule
\multicolumn{5}{l}{\hspace{-0.cm}\PbPb, 39.4~TeV, 110 nb$^{-1}$:}\\
$\sigma(\gaga\to\mathrm{X})$ & Eq.~(\ref{eq:sigma_X_master}) & $516$ mb & 2.97~$\mu$b & 3.11 nb \\
 & \cite{Shao:2022cly,Francener:2021wzx,Azevedo:2019hqp} & --, 538, -- mb & --, 2.95, 2.74~$\mu$b & 3.1, 3.14 nb \\
$N_\text{evts}(\gaga\to\mathrm{X})$ & & $5.7\times 10^{10}$ & $3.3\times 10^5$ & $340$ \\
$N_\text{evts}(\gaga\to\mathrm{X}(\gaga))$ & & $5.7\times 10^{10}$ & $3.3\times 10^5$ & $270$ \\
\midrule

\multicolumn{5}{l}{\hspace{-0.cm}\pPb, 62.8~TeV, 29 pb$^{-1}$:}\\  
$\sigma(\gaga\to\mathrm{X})$ & Eq.~(\ref{eq:sigma_X_master}) & 102~$\mu$b & 682 pb & 0.924 pb \\
 & \cite{Shao:2022cly} & --- & --- & 0.91 pb \\ 
$N_\text{evts}(\gaga\to\mathrm{X})$ & & $3.0\times 10^9$ & $2.0\times 10^4$ & $27$ \\
$N_\text{evts}(\gaga\to\mathrm{X}(\gaga))$ & & $3.0\times 10^9$ & $2.0\times 10^4$ & $20$ \\
\midrule 
\multicolumn{5}{l}{\hspace{-0.cm}\pp, 100~TeV, 10 fb$^{-1}$:}\\
$\sigma(\gaga\to\mathrm{X})$ & Eq.~(\ref{eq:sigma_X_master}) & 20.6 nb & 0.16 pb & 0.24 fb \\
 & \cite{Shao:2022cly} & --- & --- & 0.24 fb \\ 
$N_\text{evts}(\gaga\to\mathrm{X})$ & & $2.1\times 10^8$ & $1600$ & $2.5$ \\
$N_\text{evts}(\gaga\to\mathrm{X}\to\gaga)$ & & $2.1\times 10^8$ & $1600$ & $2$ \\
\midrule
\multicolumn{5}{l}{\hspace{-0.cm}p-air, 400~TeV:}\\
$\sigma(\gaga\to\mathrm{X})$ & Eq.~(\ref{eq:sigma_X_master}) & $1.15$~$\mu$b & $9.52$ pb & $15.4$ fb \\
\bottomrule
\end{tabular}
\end{table}

\subsection{QED hadronium states}
\label{app:qedhadronium}

\begin{table}[htbp!]
\tabcolsep=4.5mm
\centering
\caption{Photon-fusion cross sections $\sigma(\gaga\to\mathrm{X})$, total yields $N_\text{evts}(\gaga\to\mathrm{X})$, and yields $N_\text{evts}(\gaga\to\mathrm{X}(\gaga))$ in the diphoton decay mode, for the production of QED mesonium states (Table~\ref{tab:mesonium}) in UPCs for various colliding systems at RHIC, LHC, and FCC \cm\ energies. The last row lists the corresponding cross sections in proton-air collisions at GZK-cutoff energies. The '---' symbol indicates either missing predictions from older works or negligible expected cross sections or yields.
\label{tab:mesonium_xsection}}
\vspace{0.2cm}
\begin{tabular}{lc cccccc}
\toprule
System, $\sqrtsnn$, $\LumiInt$ & Ref. & $\Apipi$ & $\AKK$ & $\DD$ & $\DsDs$ & $\BB$ & $\BcBc$ \\
\midrule
\multicolumn{8}{l}{\hspace{-0.cm}\AuAu, 0.2~TeV, 10~nb$^{-1}$:} \\ 
$\sigma(\gaga\to\mathrm{X})$ & Eq.~(\ref{eq:sigma_X_master}) & 42 nb & 0.6 nb & 2.1 pb & 1.5 pb & 1.5 ab & 0.7 ab\\
$N_\text{evts}(\gaga\to\mathrm{X})$ & & $420$ & $5.9$ & 0.02 & 0.01 & -- & -- \\
\midrule
\multicolumn{8}{l}{\hspace{-0.cm}\PbPb, 5.5~TeV, 10 nb$^{-1}$:} \\ 
$\sigma(\gaga\to\mathrm{X})$ & Eq.~(\ref{eq:sigma_X_master}) & 410 nb & 14 nb & 0.38 nb & 0.33 nb & 17 pb & 10 pb\\
$N_\text{evts}(\gaga\to\mathrm{X})$ & & $4100$ & $100$ & $4$ & $3$ & $0.2$ & $0.1$ \\
\midrule
\multicolumn{8}{l}{\hspace{-0.cm}\pPb, 8.8~TeV, 1 pb$^{-1}$:}  \\ 
$\sigma(\gaga\to\mathrm{X})$ & Eq.~(\ref{eq:sigma_X_master}) & 110 pb & 4.3 pb & 140 fb & 120 fb & 8.5 fb & 5.3 fb\\
$N_\text{evts}(\gaga\to\mathrm{X})$ & & $100$ & $4$ & $0.1$ & $0.1$ & -- & -- \\
\midrule
\multicolumn{8}{l}{\hspace{-0.cm}\pp, 14~TeV, 1 fb$^{-1}$:} \\ 
$\sigma(\gaga\to\mathrm{X})$ & Eq.~(\ref{eq:sigma_X_master}) & 28.8 fb & 1.2 fb & 50 ab & 44 ab & 3.8 ab & 2.4 ab\\
$N_\text{evts}(\gaga\to\mathrm{X})$ & & $30$ & $1$ & -- & -- & -- & -- \\
\midrule
\multicolumn{8}{l}{\hspace{-0.cm}\PbPb, 39.4~TeV, 110 nb$^{-1}$:} \\
$\sigma(\gaga\to\mathrm{X})$ & Eq.~(\ref{eq:sigma_X_master}) & 920 nb & 37 nb & 1.36 nb & 1.2 nb & 92 pb & 58 pb\\
$N_\text{evts}(\gaga\to\mathrm{X})$ & & $1.0\times 10^5$ & $4100$ & $100$ & $100$ & $10$ & $6$ \\
\midrule
\multicolumn{8}{l}{\hspace{-0.cm}\pPb, 62.8~TeV, 29 pb$^{-1}$:} \\ 
$\sigma(\gaga\to\mathrm{X})$ & Eq.~(\ref{eq:sigma_X_master}) & 215 nb & 9.4 pb & 0.4 pb & 0.36 pb & 29 fb & 19 fb\\
$N_\text{evts}(\gaga\to\mathrm{X})$ & & $6200$ & $270$ & $10$ & $10$ & $1$ & $0.5$ \\
\midrule
\multicolumn{8}{l}{\hspace{-0.cm} \pp, 100~TeV, 10 fb$^{-1}$:} \\ 
$\sigma(\gaga\to\mathrm{X})$ & Eq.~(\ref{eq:sigma_X_master}) & 50 fb & 2.4 fb & 0.1 fb & 95 ab & 9 ab & 6 ab \\
$N_\text{evts}(\gaga\to\mathrm{X})$ & & $500$ & $25$ & $1$ & $1$ & $0.1$ & $0.05$ \\
\midrule
\multicolumn{8}{l}{\hspace{-0.cm}p-air, 400~TeV:} \\
$\sigma(\gaga\to\mathrm{X})$ & Eq.~(\ref{eq:sigma_X_master}) & 3.0 pb & 0.14 pb & 6.8 ab & 6.0 ab & 0.6 ab & 0.4 ab
\\
\bottomrule
\end{tabular}
\end{table}

\begin{table}[htbp!]
\tabcolsep=3mm
\centering
\caption{Photon-fusion cross sections $\sigma(\gaga\to\mathrm{X})$ and total yields $N_\text{evts}(\gaga\to\mathrm{X})$ 
for the production of QED baryonium states (Table~\ref{tab:baryonium}) in UPCs for various colliding systems at RHIC, LHC, and FCC \cm\ energies. The last row lists the corresponding cross sections in proton-air collisions at GZK-cutoff energies. The '---' symbol indicates either missing predictions from older works or negligible expected cross sections or yields.
\label{tab:baryonium_xsection}}
\vspace{0.2cm}
\resizebox{\textwidth}{!}{
\begin{tabular}{lc cccccccc}
\toprule
System, $\sqrtsnn$, $\LumiInt$ & Ref. & $\protonium$ & $\Sigmaonium$ & $\Xionium$ & $\Omegaonium$ & $\Lambdaconium$ & $\Xiconium$ & $\Xibonium$ & $\Omegabonium$ \\
\midrule
\multicolumn{8}{l}{\hspace{-0.cm}\AuAu, 0.2~TeV, 10~nb$^{-1}$:} \\ 
$\sigma(\gaga\to\mathrm{X})$ & Eq.~(\ref{eq:sigma_X_master}) & 0.10 nb & 36.3 pb & 23.3 pb & 7.5 pb & 1.4 pb & 0.8 pb & 2.2 fb & 1.8 fb\\
$N_\text{evts}(\gaga\to\mathrm{X})$ & & $1$ & 0.4 & 0.2 & 0.1 & 0.01 & $0.01$ & -- & -- \\
\midrule
\multicolumn{8}{l}{\hspace{-0.cm}\PbPb, 5.5~TeV, 10 nb$^{-1}$:} \\ 
$\sigma(\gaga\to\mathrm{X})$ & Eq.~(\ref{eq:sigma_X_master}) & 5.0 nb & 2.6 nb & 2.0 nb & 1.0 nb & 0.42 nb & 0.34 nb & 25 pb & 22 pb \\
$N_\text{evts}(\gaga\to\mathrm{X})$ & & $50$ & 25 & 20 & 10 & 4 & 3 & 0.2 & 0.2  \\
\midrule
\multicolumn{8}{l}{\hspace{-0.cm}\pPb, 8.8~TeV, 1 pb$^{-1}$:}  \\ 
$\sigma(\gaga\to\mathrm{X})$ & Eq.~(\ref{eq:sigma_X_master}) & 1.63 pb & 0.9 pb & 0.7 pb & 0.37 pb & 0.16 pb & 0.13 pb & 13 fb & 12 fb \\
$N_\text{evts}(\gaga\to\mathrm{X})$ & & 1.5 & 1 & 1 & 0.5 & 0.1 & 0.1 & 0.01 & 0.01  \\
\midrule
\multicolumn{8}{l}{\hspace{-0.cm}\pp, 14~TeV, 1 fb$^{-1}$:} \\ 
$\sigma(\gaga\to\mathrm{X})$ & Eq.~(\ref{eq:sigma_X_master}) & 0.53 fb & 0.30 fb & 0.23 fb & 0.13 fb & 60 ab & 50 ab & 6 ab & 5.5 ab \\
$N_\text{evts}(\gaga\to\mathrm{X})$ & & $0.5$ & 0.3 & 0.2 & 0.1 & 0.05 & 0.05 & -- & --  \\
\midrule
\multicolumn{8}{l}{\hspace{-0.cm}\PbPb, 39.4~TeV, 110 nb$^{-1}$:} \\
$\sigma(\gaga\to\mathrm{X})$ & Eq.~(\ref{eq:sigma_X_master}) & 15.3 nb & 8.47 nb & 6.51 nb & 3.6 nb & 1.63 nb & 1.34 nb & 0.14 nb & 0.13 nb \\
$N_\text{evts}(\gaga\to\mathrm{X})$ & & $1700$ & 930 & 720 & 400 & 180 & 150 & 15 & 15 \\
\midrule
\multicolumn{8}{l}{\hspace{-0.cm}\pPb, 62.8~TeV, 29 pb$^{-1}$:} \\ 
$\sigma(\gaga\to\mathrm{X})$ & Eq.~(\ref{eq:sigma_X_master}) & 4.24 pb & 2.4 pb & 1.9 pb & 1.06 pb & 0.5 pb & 0.4 pb & 46 fb & 41 fb\\
$N_\text{evts}(\gaga\to\mathrm{X})$ & & $120$ & 70 & 54 & 30 & 15 & 10 & 1 & 1 \\
\midrule
\multicolumn{8}{l}{\hspace{-0.cm} \pp, 100~TeV, 10 fb$^{-1}$:} \\ 
$\sigma(\gaga\to\mathrm{X})$ & Eq.~(\ref{eq:sigma_X_master}) & 1.1 fb & 0.62 fb & 0.48 fb & 0.28 fb & 0.13 fb & 0.12 fb & 14.7 ab & 13.3 ab \\
$N_\text{evts}(\gaga\to\mathrm{X})$ & & 10 & 5 & 5 & 3 & 1 & 1 & 0.1 & 0.1 \\
\midrule
\multicolumn{8}{l}{\hspace{-0.cm}p-air, 400~TeV:} \\
$\sigma(\gaga\to\mathrm{X})$ & Eq.~(\ref{eq:sigma_X_master}) & 66.4 fb & 38.6 fb & 30.3 fb & 17.5 fb & 8.5 fb & 7.1 fb & 0.96 fb & 0.87 fb \\
\bottomrule
\end{tabular}}
\end{table}

\clearpage

\bibliographystyle{myutphys}
\bibliography{refs.bib}



\end{document}